\documentclass[11pt]{article}
\usepackage[margin=22mm]{geometry}
\usepackage[T1]{fontenc}
\usepackage[utf8]{inputenc}
\usepackage{amsmath,amssymb,bm,braket,graphicx}
\usepackage[dvipsnames]{xcolor}
\usepackage[numbers,sort&compress]{natbib}
\usepackage{hyperref}
\hypersetup{colorlinks=true,linkcolor=blue,citecolor=blue,urlcolor=blue}
\usepackage{caption,float}
\usepackage[expansion=false]{microtype}
\usepackage{adjustbox}
\newcommand{\Tr}{\operatorname{Tr}}
\renewcommand{\Re}{\operatorname{Re}}
\renewcommand{\Im}{\operatorname{Im}}
\newcommand{\bh}[1]{\hat{\pmb{#1}}}
\newcommand{\bs}[1]{\pmb{#1}}
\newcommand{\Del}[1]{}

\newenvironment{widetext}{}{}
\newcommand{\acknowledgements}[1]{\section*{Acknowledgements}#1}

\newcounter{firstbib}
\newcommand{\figseries}{}

\newcommand{\eqseries}{}

\newcommand{\doi}[1]{doi: \href{https://doi.org/#1}{#1}}
\begin{document}
\begin{center}{\Large\bfseries Quantum simulation of the Dicke model in a two-dimensional ion crystal}\par\vspace{8pt}{\small Bryce Bullock$^{1,*}$, Sean R. Muleady$^{2,3,*}$, Jennifer F. Lilieholm$^1$, Yicheng Zhang$^4$, Arghavan Safavi-Naini$^{5,6}$, Robert J. Lewis-Swan$^4$, John J. Bollinger$^1$, Ana Maria Rey$^{7,8}$ and Allison L. Carter$^1$}\end{center}
\begin{center}\small
$^1$NIST, Boulder, Colorado, USA. $^2$QuICS, NIST and University of Maryland, USA. $^3$JQI, NIST and University of Maryland, USA. $^4$University of Oklahoma, USA. $^5$University of Amsterdam, Netherlands. $^6$QuSoft, Netherlands. $^7$JILA, NIST and University of Colorado, Boulder, USA. $^8$Center for Theory of Quantum Matter, University of Colorado, Boulder, USA.\\
$^*$These authors contributed equally. Correspondence: sean.muleady@gmail.com; arey@jilau1.colorado.edu.
\end{center}
\begin{abstract}
Quantum many-body systems driven far from equilibrium can exhibit phase transitions, chaos and entanglement, but probing these phenomena in large, nearly closed systems remains challenging~\cite{marino_dynamical_2022,lewis-swan_unifying_2019}. The Dicke model, which couples an ensemble of two-level systems collectively to a single bosonic mode, provides a fundamental framework for exploring these phenomena in light--matter interactions~\cite{dicke_coherence_1954,hepp_superradiant_1973}. Here we show that a two-dimensional crystal of approximately 100 trapped ions realizes this model and accesses integrable dynamical phase transitions, signatures of chaotic dynamics and fluctuation-driven spin--phonon excitation growth. Adiabatic elimination of the vibrational mode reveals ferromagnetic and paramagnetic dynamical spin phases, whereas strong spin--phonon coupling produces irregular trajectories and local relaxation consistent with quantum thermalization. Following a quench from an unstable fixed point, correlated excitation growth precedes generalized vacuum Rabi collapses and revivals; calibrated calculations identify two-mode squeezing 2.6 dB below the standard quantum limit, supported by reconstruction from spin measurements. By connecting collective dynamical instabilities to experimentally accessible signatures of quantum correlations, these results establish large ion crystals as a scalable platform for investigating non-equilibrium light--matter dynamics and generating resources for quantum-enhanced sensing~\cite{barberena_fast_2024,zhang_harnessing_2025}.
\end{abstract}

\section{Introduction}

Understanding how classical dynamical instabilities and chaos persist or break down as quantum correlations develop is a central question in non-equilibrium many-body physics. Quantum simulators offer a way to address this question through unitary dynamics, which can produce dynamical phase transitions, entanglement growth and thermalization without an external bath. However, accessing these phenomena  experimentally remains challenging because the exponential growth of Hilbert space makes coherent control increasingly difficult as system size grows.

Spin-boson models provide a natural setting for this exploration. Relevant throughout condensed-matter physics, quantum optics, and molecular and chemical dynamics, they couple discrete quantum spins to bosonic modes that can become highly occupied and mediate relaxation, transport, and collective behavior. A paradigmatic example is the Dicke model~\cite{dicke_coherence_1954,hepp_superradiant_1973}, in which a collective spin couples to a single bosonic mode, supporting a crossover between integrable and chaotic dynamics~\cite{lewis-swan_exploiting_2024,lewis-swan_unifying_2019,emary_chaos_2003,emary_quantum_2003,brandes_excited-state_2013,klinder_dynamical_2015}. Despite the model's broad relevance, experiments have yet to access its quantum dynamics at scale~\cite{black_observation_2003,domokos_mechanical_2003,baumann_dicke_2010,nagy_dicke-model_2010,brennecke_real-time_2013,klinder_dynamical_2015,leonard_supersolid_2017,kollar_supermode-density-wave-polariton_2017,landini_formation_2018,morales_two-mode_2019,yan_superradiant_2023,yoshihara_superconducting_2017,forn-diaz_ultrastrong_2019,todorov_ultrastrong_2010}, with features such as pair production, two-mode spin–boson squeezing, and generalized vacuum Rabi collapses and revivals remaining elusive.

\begin{figure*}[t]
\centering
\includegraphics[width = 0.9\textwidth]{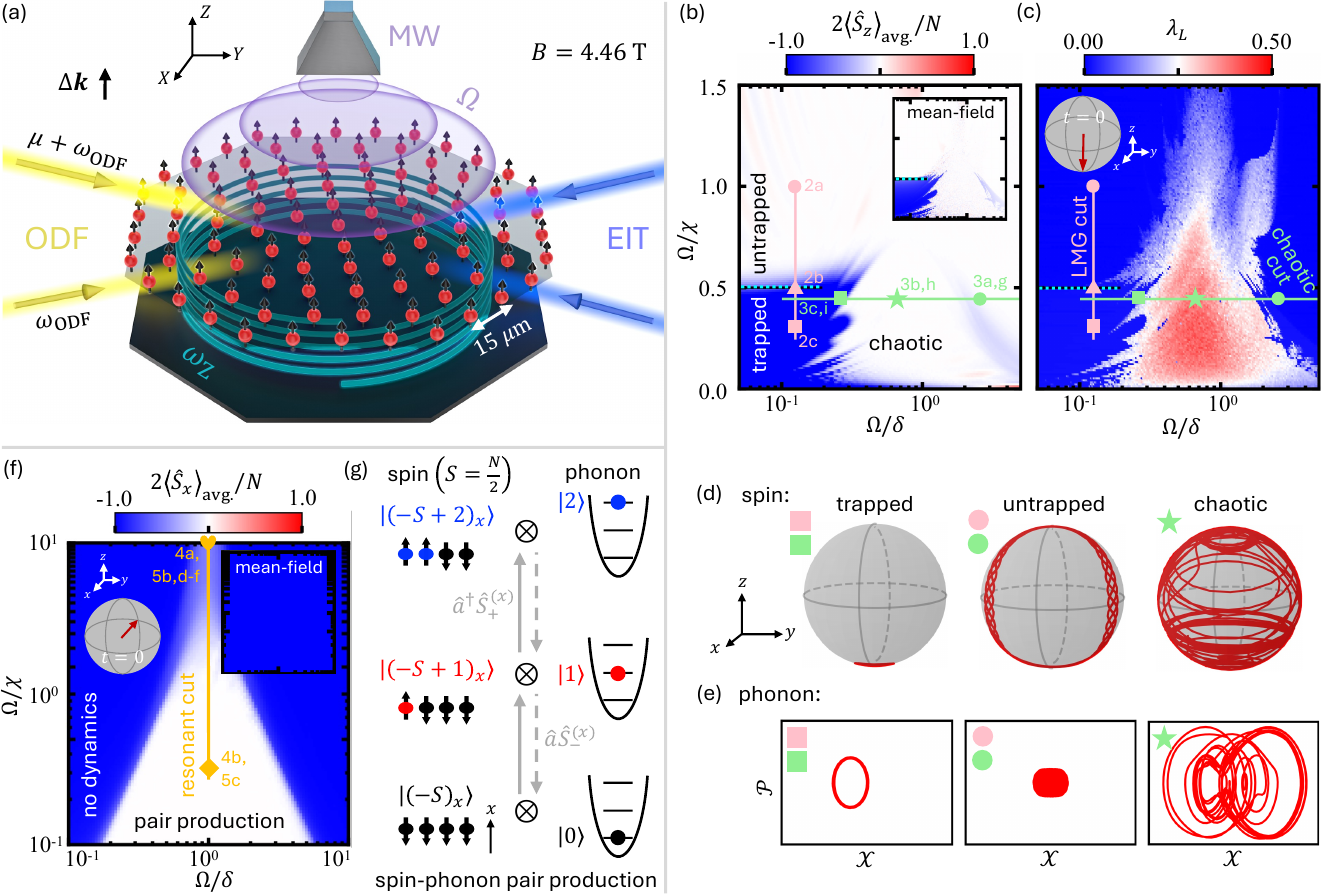}
\caption{Experimental setup and dynamical regimes. (a) $N\sim 10^2$ beryllium ions (red dots) are confined in a Penning trap, with an axial magnetic field $B=4.46$ T. The Doppler-cooled ions form a 2D crystal, whose axial center-of-mass (COM) motion realizes a high-Q mechanical oscillator with frequency $\omega_Z$ (teal spring), and which may be further cooled by crossed electromagnetically induced transparency (EIT) beams (blue). Phonon excitations of this COM motion collectively couple to the ions' valence electronic spin at rate $g$, induced by a spin-dependent optical dipole force (ODF) from crossed optical beams (yellow) with relative wavevector $\Delta \bs{k}$ and frequency difference $\mu \equiv \delta + \omega_Z$. The spins can also be globally addressed by microwaves (MW, purple) with Rabi rate $\Omega$. (b) Time-averaged magnetization computed including quantum fluctuations and (c) classical Lyapunov exponent, $\lambda_L$, for spins initially polarized along $-z$, as a function of $\Omega/\delta$ and $\Omega/\chi$, where $\chi\equiv 4g^2/\delta$; inset to (b) shows corresponding MF results. We consider two different experimental parameter sweeps: a { Lipkin-Meshkov-Glick} ``LMG cut'' (pink line)---dominated by an effective, integrable LMG spin-spin interaction---which traverses a dynamical phase transition {(dotted cyan line)} between spin-trapped and spin-untrapped behaviors; and a ``chaotic cut'' (green line), where the phonons play a more active role and can enable non-integrable chaotic behavior. Colored symbols label parameters considered later in the text, and we also show {representative} examples of corresponding classical dynamical trajectories of (d) the spins on the collective Bloch sphere and (e) the COM motion in canonical phase space ($\mathcal{X},\mathcal{P}$). (f) Time-averaged transverse magnetization including quantum fluctuations and experimental sweep (``resonant cut'', orange line and symbols) for spins initially polarized along $-x$, which exhibits non-trivial dynamics
close to resonance ($\Omega\sim \delta$), driven by quantum fluctuations for an initial bosonic vacuum state. Corresponding MF results in inset exhibit no dynamics. For $\Omega\gg \chi$, the build-up of entanglement manifests in the form of pair production of correlated spin-phonon excitations illustrated in (g). The spin excitations correspond to fully symmetrized spin flips along $x$, which occur together with a phonon excitation. This paired generation leads to exponential growth of mode populations from vacuum or thermal fluctuations and produces nonclassical correlations, including reduced fluctuations in collective quadratures characteristic of two-mode squeezing.}
\label{fig:Fig1}
\end{figure*}

Here we realize the Dicke model in a two-dimensional crystal of approximately 100 trapped $^{9}$Be$^{+}$ ions, whose internal states form a collective spin coupled to the center-of-mass vibrational mode (Fig.~\ref{fig:Fig1}a). The long coherence time of this mode enables controlled, nearly unitary spin–phonon evolution across a broad range of dynamical regimes.

When the bosonic mode can be adiabatically eliminated, we observe a dynamical phase transition between ferromagnetic (trapped) and paramagnetic (untrapped) dynamics (Fig.~\ref{fig:Fig1}b), distinguished by whether the average magnetization remains confined below the equator of the Bloch sphere~\cite{marino_dynamical_2022,smerzi_quantum_1997,albiez_direct_2005,abbarchi_macroscopic_2013,reinhard_self-trapping_2013,zhang_observation_2017,borish_transverse-field_2020,muniz_exploring_2020,li_improving_2023}. Retaining the coupled spin–boson dynamics takes the system beyond this integrable limit into a non-integrable regime predicted to be chaotic at the mean-field level~\cite{emary_chaos_2003,emary_quantum_2003}, as indicated by a calculated positive Lyapunov exponent (Fig.~\ref{fig:Fig1}c).

Preparing the system near an unstable fixed point allows us to probe a regime in which ideal classical dynamics predict no evolution. We observe exponential excitation growth consistent with correlated pair creation seeded by quantum fluctuations and accompanied by numerically inferred spin--phonon entanglement. The measured local observables approach values consistent with quantum thermalization, with an accompanying increase in the one-body R{\'e}nyi entropy; separating entanglement from decoherence requires additional measurements. In the near-integrable regime, simulations incorporating experimental imperfections further identify two-mode spin--phonon squeezing reaching 2.6~dB below the standard quantum limit (4.6~dB relative to the initial thermal state). At longer times, we observe generalized vacuum-Rabi collapses and revivals, demonstrating coherence of the many-body dynamics.

\section{Experimental implementation}
Our simulator comprises a single-layer Coulomb crystal of approximately $N=100$ $^{9}$Be$^{+}$ ions in a Penning trap~\cite{bollinger_simulating_2013,sawyer_spin_2014,bohnet_quantum_2016}. Each ion encodes a spin-$1/2$ in the ground-state electron-spin levels $\ket{\uparrow}$ and $\ket{\downarrow}$. Resonant microwaves, tuned to the 124 GHz Zeeman splitting induced by a strong external magnetic field of $B=4.46$ T, coherently rotate the spins and generate a transverse field with Rabi frequency $\Omega$.

Two off-resonant laser beams produce a spin-dependent optical dipole force (ODF). Their frequency difference $\mu$ is tuned near the axial COM mode at $\omega_Z/(2\pi)=1.59$ MHz, with detuning $\delta=\mu-\omega_Z$. The uniform COM motion couples all spins to the same bosonic mode; other drumhead modes remain far detuned and contribute small corrections discussed in Methods. Electromagnetically induced transparency (EIT) cooling reduces the initial COM occupation to $\bar n\simeq0.5$, compared with $\bar n\simeq5$ under Doppler cooling~\cite{jordan_near_2019}.

In the Lamb--Dicke regime, the combined optical dipole force and resonant microwave drive implement the Dicke Hamiltonian, $\hat H_{\rm Dicke}=\hat H_0+\hat H_{\rm s-p}$~\cite{safavi-naini_verification_2018}, with
\begin{equation}
\hat H_0/\hbar=-\delta\hat a^\dagger\hat a+\Omega\hat S_x,\qquad
\hat H_{\rm s-p}/\hbar=\frac{2g}{\sqrt N}(\hat a+\hat a^\dagger)\hat S_z.
\label{eq:H_Dicke}
\end{equation}
Here, $\hat a$ annihilates a COM phonon and $\hat S_\alpha=\sum_{j=1}^N\hat\sigma_{\alpha,j}/2$ is the collective spin. The coupling $g>0$ is set by the ODF amplitude. We characterize the Hamiltonian parameters, initial occupation and dominant decoherence channels independently, and actively track the collective spin and COM frequencies during data acquisition (Methods). These calibrations underpin the numerical comparisons below.

\section{ The Dicke Model Phase Diagram}
\noindent {\bf Equilibrium Phases:} At zero temperature the Dicke model undergoes a quantum phase transition (QPT) at $\Omega_c^{\rm (QPT)}=4g^2/|\delta|$ (for $\delta<0$ in our convention)~\cite{emary_chaos_2003,emary_quantum_2003} between a ``normal'' phase ($\Omega>\Omega_c^{\rm (QPT)}$; spins aligned along transverse field, unoccupied bosonic mode) and a ``superradiant'' phase ($\Omega<\Omega_c^{\rm (QPT)}$; ferromagnetic spin configuration with $\langle|\hat S_z|\rangle\sim N/2$ and macroscopic boson occupation), well captured by mean-field (MF) theory up to $1/N$ corrections~\cite{kirton_introduction_2019}.

\noindent {\bf Non-equilibrium phases:}
Out of equilibrium, the dynamics depend strongly on initial conditions and Hamiltonian parameters, even when restricted to the fully symmetric, collective manifold of total spin $S=N/2$, where $S(S+1)$ are the eigenvalues of the total spin operator $\bh{S}\cdot \bh{S}$. We focus on two initial conditions, with the bosonic mode ideally prepared in vacuum:

\noindent {\it All spins prepared along $-z$:}
For spins initially prepared in $\ket{(-N/2)_z}$---where $\ket{m_\alpha}$ are collective eigenstates of $\hat{S}_\alpha$ with eigenvalue $m_\alpha$--- the system features dynamical phase transitions (DPTs)~
\cite{schiro_time-dependent_2010,sciolla_quantum_2010,gambassi_quantum_2011,smacchia_exploring_2015}, where the long-time MF magnetization $\langle\hat S_z\rangle_{\rm avg.}=\lim_{T\to\infty}T^{-1}\int_0^T\langle\hat S_z(t)\rangle dt$ serves as a dynamical order parameter~\cite{marino_dynamical_2022}. In broad terms, these DPTs are well described by a classical MF analysis, which predicts rich dynamical behaviors ranging from simple, analytically tractable trapped or untrapped dynamics of the spins and bosons, to non-integrable regimes in which near-resonant spin-boson coupling gives rise to chaos~\cite{emary_chaos_2003,emary_quantum_2003,lewis-swan_characterizing_2021}.

When $|\delta| \gg|\Omega|, |g|$ the population of bosonic excitations is energetically suppressed, and to leading order their role reduces to virtual mediators of interactions between the spins (we take $g,\Omega,\delta>0$; see Supplementary Information for the role of signs). In this regime, the dynamics can be described by a pure spin model known as the Lipkin-Meshkov-Glick (LMG) model~\cite{lipkin_validity_1965,ribeiro_thermodynamical_2007,lerose_impact_2019}, with Hamiltonian
\begin{equation}
{ \hat{H}_{\rm LMG }/\hbar}= \frac{\chi}{N} \hat {S}_z^2 +\Omega \hat{S}_x \quad \chi\equiv 4g^2/\delta\label{eq:H_LMG}.
\end{equation}
In this regime, the spin-spin interaction strength $\chi=\Omega_c^{(QPT)}$ can be interpreted as setting the natural scale of competition between the two equilibrium phases. This model has been widely explored experimentally, from its MF dynamical phases in the infinite range interaction limit~\cite{muniz_exploring_2020,borish_transverse-field_2020} to the development of non-trivial quantum correlations in the presence of power-law decaying couplings~\cite{zhang_observation_2017}.

A critical drive $\Omega_c^{\rm (DPT)} = \chi/2$, separates a dynamical ferromagnetic phase (trapped), and a dynamical paramagnetic phase (untrapped),
as indicated by the {dotted cyan line} on the classical MF phase diagram shown in Fig.~\ref{fig:Fig1}b. In the former   the instantaneous magnetization $\langle \hat{S}_z \rangle$ oscillates about a non-zero time-averaged value, remaining trapped below the Bloch sphere equator (Fig.~\ref{fig:Fig1}d) with $\langle \hat{S}_z \rangle_{\rm avg.} \neq 0$. In the latter the dominant transverse field $\Omega$ leads to Rabi flopping with $\langle \hat{S}_z \rangle_{\rm avg.} = 0$.
The adiabatically eliminated bosons are bound to the motion of the collective spin, $\langle \hat{a}+ \hat{a}^\dagger \rangle \propto (2g/\delta \sqrt{N})\langle \hat{S}_z\rangle $, reflecting their dynamics in phase space (Fig.~\ref{fig:Fig1}e), with quadratures $\hat{\mathcal{X}}=(\hat{a}+\hat{a}^\dagger)/\sqrt{2}$ and $\hat{\mathcal{P}}=i(\hat{a}^\dagger -\hat{a})/\sqrt{2}$.

When $|\delta|\sim|\Omega|\sim|g|$, the phonons actively participate in the dynamics and cannot be adiabatically eliminated~\cite{wall_boson-mediated_2017}. MF trajectories can become irregular, featuring erratic oscillations with intermittent retrapping (Fig.~\ref{fig:Fig1}d,e). A positive calculated Lyapunov exponent $\lambda_L$ identifies the classically chaotic regime (Fig.~\ref{fig:Fig1}c)~\cite{emary_quantum_2003,lewis-swan_unifying_2019,chavez-carlos_classical_2016,lerose_chaotic_2018,lerose_impact_2019,altland_equilibration_2012}, and thus non-integrability \cite{strogatz_nonlinear_2014}. In contrast to the integrable LMG dynamics, here thermal and quantum fluctuations can dephase these trajectories and damp oscillations in averaged observables compared to MF theory~\cite{lewis-swan_characterizing_2021}. We test experimental signatures of this predicted regime below.

\noindent {\it All spins prepared along $-x$: }
Genuine quantum behaviors emerge in the Dicke model when the spins are initially in $\ket{(-N/2)_x}$. This state is a stationary point of the MF dynamics (see inset of Fig.~\ref{fig:Fig1}f,
showing  $\langle\hat S_x(t)\rangle/(N/2)=-1$ at all times), and non-trivial dynamics are only driven by thermal or quantum fluctuations away from the stationary point. Thus, for an initial bosonic vacuum, quantum noise is the sole driver of dynamics, requiring beyond MF methods to model the resulting behavior, such as the truncated Wigner-approximation (TWA) \cite{schachenmayer_many-body_2015}, (see Methods and Supplementary Information).
In Fig.~\ref{fig:Fig1}{f}, we show TWA dynamics of the time average transverse order parameter,
$\langle \hat{S}_x \rangle_{\rm avg.} $, which displays non trivial behavior in a narrow region about $\Omega/\delta \sim 1$.

In the interaction picture, the spin--phonon coupling separates into pair-production and excitation-exchange processes, oscillating at $\Omega-\delta$ and $\Omega+\delta$, respectively. The explicit rotating-frame Hamiltonian and the $x$-basis spin ladder operators are given in Methods.

At resonance, $\delta=\Omega\gg g$, the rapidly oscillating
$\hat H_{\rm osc}^{\rm rot}$ can be neglected, leaving
$\hat H_{\rm pair}^{\rm rot}$ to generate correlated spin-phonon
excitations. For an initial phonon vacuum and spins polarized along
$-x$, ideal Dicke evolution preserves
$\langle\hat a\rangle=\langle\hat S_y\rangle=\langle\hat S_z\rangle=0$,
while
\begin{equation}
\frac{d}{dt}\langle\hat S_x\rangle
=-\frac{2g}{\sqrt N}
\langle\hat S_y(\hat a+\hat a^\dagger)\rangle.
\end{equation}
The transverse magnetization therefore evolves entirely through a
connected spin-phonon correlator, which vanishes under MF
factorization. In this case, the ideal dynamics are fluctuation driven
(see Methods and Supplementary Information).

For low spin excitation, the Holstein-Primakoff (HP) approximation
\cite{holstein_field_1940} gives
$\hat S_+^{(x)}\simeq\sqrt N\,\hat b^\dagger$ and
$\hat S_x=-N/2+\hat b^\dagger\hat b$, reducing the interaction to
$\hat H_{\rm pair}^{\rm rot}/\hbar\simeq
-ig(\hat a^\dagger\hat b^\dagger-\hat a\hat b)$.
This two-mode squeezing Hamiltonian
exponentially generates entangled spin-phonon pairs from vacuum, with
$\langle\hat a^\dagger\hat a\rangle
=\langle\hat b^\dagger\hat b\rangle
\simeq\sinh^2(gt)$
(Fig.~\ref{fig:Fig1}f,g).
It squeezes the plus combinations of the hybrid quadratures
$\hat{\mathcal V}_{\pm}=\hat{\mathcal P}
\pm\hat S_z/\sqrt{N/2}$ and
$\hat{\mathcal W}_{\pm}=\hat{\mathcal X}
\pm\hat S_y/\sqrt{N/2}$,
while antisqueezing the minus combinations.
At larger excitation, finite-spin corrections beyond this approximation produce non-sinusoidal collapses and revivals---generalized vacuum-Rabi oscillations---with dephasing reflecting the Dicke spectrum.

When $\delta=\Omega$ is no longer large compared with $g$,
$\hat H_{\rm osc}^{\rm rot}$ must be retained, and the dynamics can
enter a chaotic, thermalizing regime. A local probe of quantum correlations is the single-spin  second-order R{\'e}nyi entropy.    For the initial
state considered here, parity gives $\langle\hat\sigma_{y,j}\rangle=\langle\hat\sigma_{z,j}\rangle=0$,
and permutation symmetry,
$\langle\hat\sigma_{x,j}\rangle=2\langle\hat S_x\rangle/N$, yielding
\begin{equation}
S_2(\hat\rho_j)
=-\log_2[{\rm Tr}(\hat\rho_j^2)]
=-\log_2\left[\frac12+
\frac{2\langle\hat S_x\rangle^2}{N^2}\right].
\label{Rey}
\end{equation}
For a globally pure state, this entropy quantifies entanglement
between spin $j$ and the remaining system~\cite{renyi_measures_1961};
for mixed states, it also includes contributions from initial
thermal mixedness and decoherence.

\begin{figure}[t]
\centering
\includegraphics[width = 0.48\textwidth]{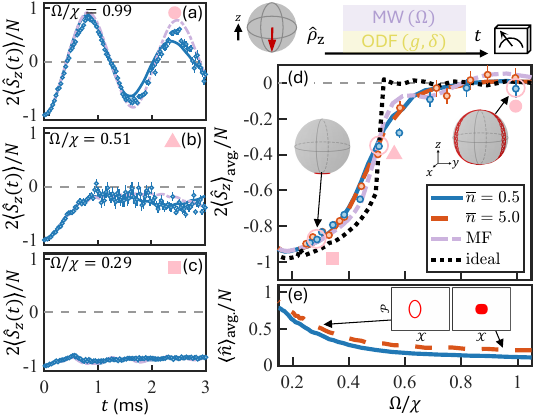}
\caption{Dynamical phase transition in the integrable, spin-dominated limit (see pink ``LMG cut'' in Fig.~1 and corresponding pink symbols). (a-c) Representative time traces of $\braket{\hat{S}_z(t)}$ for initial EIT-cooled phonons and $-z$ polarized spins, following a variable time pulse of the ODF beams and MW source to simulate an LMG model (see sketch, top right). We compare against semiclassical (blue-solid) and mean-field (MF) models (lavender-dashed-dot), including relevant experimental details. (d) Magnetization averaged over $3$ ms of evolution for initial EIT (blue, {$N\in [57,79]$}) and Doppler (orange, {$N\in[86,167$]}) cooling, compared to semiclassical (blue solid, $\overline{n} = 0.5$; orange-dashed, $\overline{n} = 5.0$) and MF (lavender dashed-dot) results with $N=100$; idealized MF results are also shown (black dotted), averaged over $0-20$ ms. (e) We also present corresponding semiclassical predictions for the time-averaged phonon occupancy, not directly accessible in the experiment. Bloch sphere/phase space schematics next to pink symbols correspond to the parameters and behaviors outlined in Fig.~\ref{fig:Fig1}b-e. $\Omega/\delta\in [0.11,0.13]$ and $g \in 2\pi \times [0.93,1.00]$ kHz throughout. Error bars denote standard-error of the mean.}
\label{fig:Fig2}
\end{figure}

\section{Probing the LMG dynamical phase diagram}
We first probe the integrable spin limit, $|\delta|\gg |g|,|\Omega|$, using spins polarized along $-z$ and a thermal phonon state prepared by either Doppler or EIT cooling. The interaction and transverse-field rates exceed the single-spin decoherence rate, $|\chi|,|\Omega|\gg\Gamma$. After evolution under the ODF and transverse microwave field for up to $T=3$ ms, fluorescence measurements determine the spin populations (Methods).

Figure~\ref{fig:Fig2}a--c shows magnetization traces at approximately fixed $\Omega/\delta\simeq0.1$ and $g\simeq2\pi\times0.9$ kHz, while varying $\Omega/\chi$ along the pink cut in Fig.~\ref{fig:Fig1}b. For $\Omega/\chi=0.29$, the spins remain trapped below the Bloch-sphere equator and the magnetization oscillates near its initial value. For $\Omega/\chi=0.99$, the transverse field instead produces untrapped Rabi oscillations. The magnetization averaged over the measured 3-ms window changes from a nonzero value to approximately zero near $\Omega_c^{\rm(DPT)}=\chi/2$ (Fig.~\ref{fig:Fig2}d).

Both mean-field (MF) theory and the truncated Wigner approximation (TWA) reproduce the measurements, confirming the predominantly MF character of this transition. The TWA includes independently characterized light scattering, magnetic-field noise and initial thermal occupation, whereas the MF model omits initial quantum and thermal fluctuations and noise-induced correlations (Methods and Supplementary Information). The similar dynamics for the two cooling schemes support the passive role of phonons in this limit. The simulated time-averaged phonon occupation increases as the spins become more strongly trapped and remains offset by the initial thermal occupation (Fig.~\ref{fig:Fig2}e); this quantity is not directly measured.

The transition is rounded by the finite averaging window. Ideal MF results averaged over 20 ms show a sharper non-analytic feature, whereas the experimentally accessed regimes remain robust over the different system sizes: $N=57$--79 for EIT cooling and $N=86$--167 for Doppler cooling. Finite-$N$ effects can affect individual traces, but the comparisons attribute the observed rounding primarily to the observation time.

\begin{figure}[t]
\centering
\includegraphics[width = 0.48\textwidth]{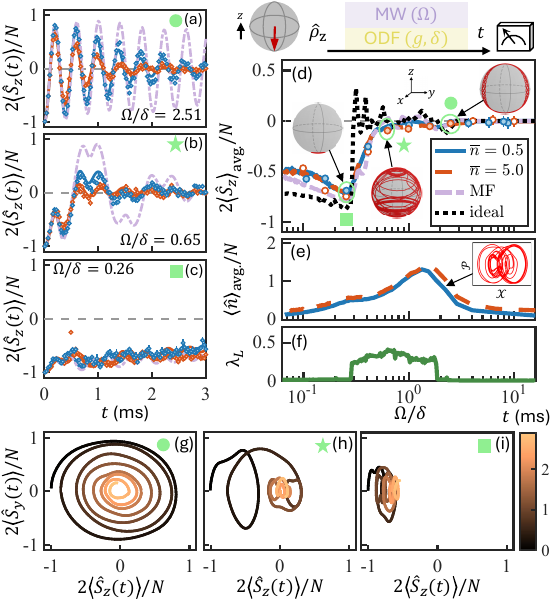}
\caption{Non-integrable dynamics and chaos in the Dicke model (see green cut and corresponding symbols in Fig.\ref{fig:Fig1}).
(a-c) Representative spin dynamics for initial EIT (blue) and Doppler (orange) cooling and $-z$ polarized spins, following an analogous sequence to Fig.~\ref{fig:Fig2} to simulate the Dicke model (see also sketch, top right), compared against semiclassical (blue/orange solid) and MF models (lavender dashed-dot), including relevant experimental details. (d) Magnetization averaged over $3$ ms of evolution, with filled (unfilled) symbols corresponding to results with $N\in [51,108]$ ($N\in [159,178]$). We compare to semiclassical (blue solid, $\overline{n} = 0.5$; orange dashed, $\overline{n} = 5.0$) and MF (lavender dashed-dot) results with $N=100$, {as well as ideal MF results (black dotted) averaged over $0-20$ ms. (e) We} also show the semiclassically predicted time-averaged phonon occupancy{, (f) as well as the corresponding classically computed Lyapunov exponent $\lambda_L$}. Bloch sphere and phase space diagrams associated with green symbols display typical classical behaviors and corresponding parameters outline in Fig.~\ref{fig:Fig1}b-e. {(g-i)} Experimental phase space dynamics corresponding to (a-c) of the mean magnetization (with EIT-cooled initial phonons), with color denoting the elapsed time. $\braket{\hat{S}_y(t)}$ is computed from the time-derivative of smoothed dynamics, see Methods and Supplementary Information for details. $\Omega/\chi \in [0.41,0.47]$ and $g\in 2\pi\times [1.07,1.15]$ kHz throughout (for unfilled symbols, $g \approx 2\pi \times 0.88$ kHz). Error bars denote standard-error of the mean.}
\label{fig:Fig3}
\end{figure}

\section{Exploring chaotic Dicke dynamics}\label{sec:chaos}
To activate both spin and phonon dynamics, we vary $\Omega/\delta$ at approximately fixed $\Omega/\chi\simeq0.4$ and $g\simeq2\pi\times1.1$ kHz, using the same $-z$-polarized initial state (green cut in Fig.~\ref{fig:Fig1}b). Regular motion occurs at small $\Omega/\delta$, where the system approaches the spin-trapped LMG limit, and at large $\Omega/\delta$, where rapid spin precession is bound to the phonon motion and produces oscillations about zero mean magnetization (Fig.~\ref{fig:Fig3}a,c).

At intermediate ratios, neither degree of freedom can be eliminated. The classical equations predict irregular trajectories with intermittent retrapping and a positive Lyapunov exponent (Fig.~\ref{fig:Fig3}b,f). Experimental phase portraits likewise exhibit irregular loops rather than smooth orbits (Fig.~\ref{fig:Fig3}g--i). Their vertical coordinate, $\langle\hat S_y\rangle$, is reconstructed from the time derivative of the measured $\langle\hat S_z\rangle$, with dissipative corrections and smoothing assessed against simulations (Methods). Measured power spectra broaden in this same region, and their noise-corrected spectral entropy increases where the calculated Lyapunov exponent is positive (Extended Data Fig.~\ref{fig:chaotic_spectra}).

In and near this regime, the measured magnetization is more strongly damped than MF predictions, even when technical noise is included. TWA simulations account for initial spin and phonon fluctuations and reproduce this damping. Doppler-cooled traces are more strongly damped than EIT-cooled traces, consistent with their larger thermal occupation. Nevertheless, simulations starting from the phonon vacuum differ little from those with the EIT occupation, while both differ markedly from MF dynamics (Supplementary Fig.~\ref{fig:noise_comparison_chaotic}; Supplementary Information). Damping alone is not unique to chaos; Even though the experiment does not directly measure a Lyapunov exponent, the coexistence of damping  with irregular phase portraits, spectral broadening and enhanced spectral entropy provides a consistent set of experimental signatures of the predicted chaotic regime.

Despite their different transient damping, the two cooling schemes yield similar magnetization averaged over 3 ms (Fig.~\ref{fig:Fig3}d). Ideal MF dynamics averaged over 20 ms retain strong fluctuations and a non-analytic feature at the boundary of the trapped regime. Simulations also predict a pronounced peak in phonon production in the chaotic region (Fig.~\ref{fig:Fig3}e), indicating access to a larger portion of the spin--boson Hilbert space. This is consistent with proposed connections between chaos, information delocalization, and scrambling~\cite{lewis-swan_unifying_2019,Cheng2020,vikram2026bidirectional}.

\section{Resonant dynamics: exponential growth, pair production and entanglement }

\begin{figure}[t]
\centering
\includegraphics[width = 0.48\textwidth]{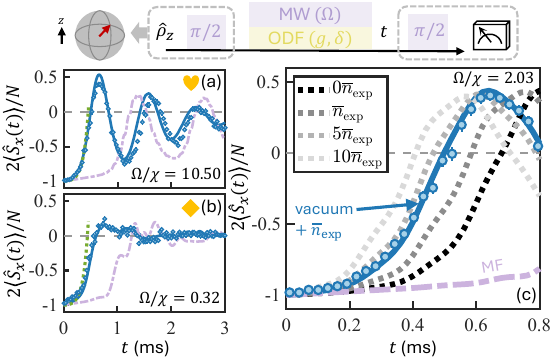}
\caption{Fluctuation-driven, unstable fixed-point dynamics in the resonant spin-phonon regime (see orange trace in Fig. \ref{fig:Fig1}f, and corresponding orange symbols). (a-b) Representative transverse spin dynamics for initial EIT-cooled phonons and $-x$ polarized spins, using a modification of the pulse sequence in Fig.~\ref{fig:Fig2} to prepare and access the transverse magnetization (see sketch, top). We compare against our semiclassical (blue-solid) and MF models (lavender dashed-dot), including relevant experimental details, as well as analytic results from a large-$N$ expansion (green dotted). (c) We also compare to a classical model accounting for classical correlations from different noise sources, with different levels of added initial thermal noise in the phonons, quantified as multiples of measured occupancy $\overline{n}_{\rm exp}\sim 0.5(2)$. $\Omega/\delta\in [0.89,1.02]$ and $g\in 2\pi\times [0.87,0.91]$ kHz throughout, and $N\in [89,112]$. Error bars denote standard-error of the mean.}
\label{fig:Fig4}
\end{figure}

To further illuminate the onset of genuine quantum effects and non-integrable dynamics in the Dicke model, we explore the dynamics when the initial spin state is instead polarized along the $x$ direction. We plot time traces of the transverse magnetization $\braket{\hat{S}_x(t)}$---accessed by an additional $\pi/2$-pulse before fluorescence measurements of the spin population---in Fig.~\ref{fig:Fig4}, taken through the orange cut shown in Fig.\ref{fig:Fig1}. For these measurements, we consider only initial states prepared with EIT cooling (blue circles), and our results are again validated by the numerical TWA dynamics (blue line).

\begin{figure}[t]
\centering
\includegraphics[width = 0.48\textwidth]{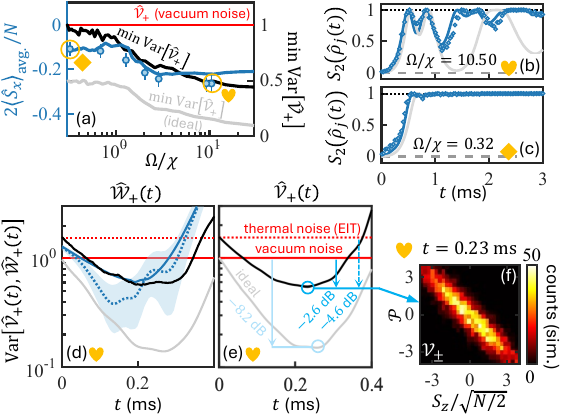}
\caption{Pair production and entanglement in the resonant spin-phonon regime. (a) Time-averaged transverse magnetization from results in Fig.~\ref{fig:Fig4}, compared to our semiclassical model (blue solid). {Magnetization is averaged over 0--3 ms}. Error bars denote standard-error of the mean. (b-c) Representative dynamics of the single-qubit R{\'e}nyi entropy, compared to our semiclassical model (blue solid{; ideal results in gray}). Maximum entropy indicated by black dotted line. {(d-e) Variance dynamics for composite $\mathcal{W}_{+}$ and $\mathcal{V}_{+}$ quadratures, computed in our semiclassical model (black solid; ideal results in gray), showing noise reduction relative to the vacuum state and also the initial EIT-cooled thermal state. For $\mathcal{W}_{+}$, we additionally plot estimates for this quantity extracted entirely from the dynamics of the transverse magnetization, both in our semiclassical model (solid blue) and with our measurement data (dotted blue, with shaded blue error region). We} plot the corresponding minimum achieved variance $\mathcal{V}_+$ versus $\Omega/\chi$ in panel (a); reference value for an uncorrelated vacuum phonon state displayed in red. {(f)} Semiclassical prediction for joint spin-phonon quadrature {histogram for $\mathcal{V}_+$}, simulating $5\times 10^3$ measurements. {Shaded error region in (d) corresponds to bootstrapped 68\% confidence intervals, see Methods and Supplementary Information for details.}}
\label{fig:Fig5}
\end{figure}

All time traces shown in Fig.~\ref{fig:Fig4}a-c display short-time transverse magnetization growth consistent with quantum-fluctuation-driven correlated spin--boson excitation production, which is fully consistent with the ideal exponential growth shown by the green dotted line. Nevertheless, at longer times there is a distinct change in behavior. In the small $\Omega/\chi=\delta^2/(4g^2)<1$ limit, the $x$ magnetization rapidly relaxes towards zero, consistent with thermalization in the non-integrable regime as spin--phonon excitations build up{, Fig.~\ref{fig:Fig4}b ($\Omega/\chi=0.32,\Omega\sim2\pi \times 1$ kHz)}. On the contrary, in the regime $\Omega/\chi=\delta^2/(4g^2)>1$, we clearly see the expected collapses and revivals at longer times, emerging from the more constrained growth of correlated spin-boson excitations{, Fig.~\ref{fig:Fig4}a ($\Omega/\chi=10.5,\Omega\sim2\pi \times 5.66$ kHz)}. These collapses and revivals arise from the resonant spin--phonon pair-creation dynamics and constitute generalized vacuum-Rabi oscillations~\cite{gietka_vacuum_2024-1}.

While the ideal MF solution exhibits no dynamics (see inset of Fig.~\ref{fig:Fig1}f), relevant sources of decoherence and thermal fluctuations---even at the coldest temperatures accessible in the experiment---can drive dynamics away from the classically unstable fixed point. Nevertheless, we still observe marked differences between our MF solution---which includes technical noise at MF level---and the full quantum dynamics. This is demonstrated in the time traces of the $x$ magnetization shown in Fig.~\ref{fig:Fig4}c{ ($\Omega/\chi=2.03,\Omega\sim2\pi \times 2.5$ kHz)}. Furthermore, properly accounting for classical correlations from thermal noise and stochastic noise sources in our MF model (gray dotted lines) reproduces the observed dynamics \emph{only} if the initial thermal population is artificially increased by a factor of five relative to experimental conditions. We therefore interpret the faster-than-classical evolution in the experiment as being seeded by intrinsically quantum fluctuations, namely phonon vacuum noise together with spin projection noise, which accelerate the escape from the unstable fixed point.

The time-averaged magnetization trends closer to zero with decreasing $\Omega/\chi$, {as shown in Fig. \ref{fig:Fig5}a,} consistent with entanglement growth in the chaotic regime even when unavoidable non-unitary decoherence sources are included in the simulations. In this regime, the collective magnetization determines the single-spin R{\'e}nyi entropy, which quantifies local mixedness (see Eq.~\eqref{Rey}). This remarkable robustness allows us to interpret the experimentally measured decay of the collective $x$-magnetization to zero,
when the single-spin R{\'e}nyi entropy in Fig.~\ref{fig:Fig5}c reaches its maximum value, as consistent with entanglement buildup and quantum thermalization in the chaotic regime \cite{lerose_bridging_2020}.
However,  explicitly separating  this
entanglement-induced mixedness from decoherence will require more elaborate measurements.

For the semi-integrable regime, such a conclusion can only be drawn at short times (Fig. \ref{fig:Fig5}b), when the collective magnetization is almost identical to the ideal result. At longer times, the collapses and revivals exhibited by the purely unitary solution become noticeably damped by decoherence in the experiment. In this regime, the inferred variance of the hybrid quadratures provides an alternative way to quantify entanglement, at least over short timescales. In Fig.~\ref{fig:Fig5}a we show the theoretically expected minimal variance of the hybrid quadrature $\hat{\mathcal V}_+$, which decreases with increasing $\Omega/\chi=\delta^2/(2g)^2$ (for $\Omega=\delta$). At $\Omega/\chi = 10.50$, as shown in {Fig.~\ref{fig:Fig5}e,d}, the variance {of the $\hat{\mathcal{V}}_+$ ($\hat{\mathcal{W}}_+$) quadrature in the ideal simulation, initialized in the phonon vacuum, evolves to a minimum of approximately $1/7$ of its initial value, corresponding to $\sim 8.2$~dB ($\sim 8.4$~dB) below the standard quantum limit (SQL). When decoherence is accounted for, the reduction remains significant at about $\sim 2.6$~dB ($2.4$~dB) below the SQL, corresponding to a net decrease of $\sim 4.6$~dB ($\sim 4.4$~dB) relative to the initial thermal noise.
Fig.~\ref{fig:Fig5}f displays simulated histograms of the hybrid spin-boson phase-space quadrature $\hat{\mathcal V}_\pm$, at the optimal squeezing time for $\Omega/\chi = 10.50$.}

Although we cannot directly measure spin-phonon correlations in the current setting, the relevant correlators entering the variance of the spin-phonon quadrature $\hat{\mathcal{W}}_+$ can nevertheless, be reconstructed from experimentally measured observables and their time derivatives (see Methods and Supplementary Information), provided the dynamics remains well described by the Dicke model, augmented with dissipative processes accounting for the dominant experimentally characterized decoherence channels. Despite the additional uncertainty associated with extracting time derivatives from experimental data, the reconstructed variance exhibits short-time behavior consistent with the buildup of nonclassical spin-phonon correlations associated with two-mode squeezing (Fig.~\ref{fig:Fig5}d). While this procedure does not constitute a direct experimental measurement of entanglement, several independent observations consistently support the quantum character of the observed dynamics.

\section{Discussion}
Our realization of the Dicke model accesses integrable dynamical phase transitions, fluctuation-driven excitation growth, and signatures of chaotic dynamics in a large ion crystal. Magnetization measurements reveal  long-time collapses and revivals, while numerical comparisons and reconstruction identify signatures of two-mode squeezing, entanglement built up and quantum thermalization. Direct higher-order spin--spin and spin--phonon measurements would establish these quantum correlations more conclusively; demonstrating classical intractability remains a future goal.

The rapidly generated spin--phonon correlations near the Dicke instability offer a resource for quantum-enhanced sensing~\cite{barberena_fast_2024,zhang_harnessing_2025}, including weak oscillatory electric fields and potential ultralight-dark-matter signals~\cite{gilmore_quantum-enhanced_2021,gilmore_amplitude_2017}. They also motivate exploration of EPR correlations and steering. Correlated pair production connects to thermofield double states, holography, teleportation and Hayden--Preskill information recovery~\cite{vikram2026bidirectional}. These directions extend the platform's relevance from collective dynamics to quantum sensing and information.

\acknowledgements{We thank Diego Fallas Padilla and Kurt A. Thompson for excellent feedback on the manuscript, and Diego Barberena and Justin Bohnet for useful discussions. This work was supported by ARO W911NF24-1-0128, AFOSR FA9550-25-1-0080, the NSF JILA-PFC PHY-2317149, The Heising-Simons Foundation and the U.S. Department of Energy, Office of Science, National Quantum Information Science Research Centers, Quantum Systems Accelerator, and NIST. S.R.M. is supported by the NSF QLCI award OMA-2120757. A.S.N. is supported by the Dutch Research Council (NWO/OCW) as a part of the Quantum Software Consortium (project number 024.003.037). J.F.L. acknowledges support from an Intelligence Community Postdoctoral Research Fellowship.}

\section*{Author contributions}
B.B., J.F.L., J.J.B. and A.L.C. collected and analysed the
experimental data. S.R.M., Y.Z., A.S.-N., R.J.L.-S. and A.M.R.
developed the theoretical model. All authors discussed the results
and contributed to the preparation of the manuscript.

\section*{Corresponding author}
Correspondence to Sean R. Muleady
(\href{mailto:sean.muleady@gmail.com}{sean.muleady@gmail.com})
or Ana Maria Rey
(\href{mailto:arey@jila.colorado.edu}{arey@jila.colorado.edu}).

\section*{Competing interests}
The authors declare no competing interests.

\appendix

\clearpage
\section*{Methods}
\setcounter{section}{0}
\renewcommand{\thesection}{M\arabic{section}}
\renewcommand{\theHsection}{M\arabic{section}}
\section{Experimental details}
Here we provide a more detailed description of the experimental ion trap set-up used to realize the Dicke model along with a discussion of the experimental protocols employed for controlling and measuring the relevant model parameters.

Crystals of $^9$Be$^+$ ions are confined in a room-temperature Penning ion trap and formed through Doppler laser cooling. The trap consists of cylindrical electrodes {with an inside diameter of 2.0 cm} that are used to generate a deep, harmonic confining potential along the axis of the cylinders, denoted by the $Z$-axis of the trap. The trap resides in the bore of a 4.46 T superconducting magnet with the $Z$-axis of the trap aligned with the magnetic field direction to better than a few hundredths of a degree~\cite{Bullock_thesis_2026,Gilmore_thesis_2020}. Radial confinement of the ions is obtained from the Lorentz force due to $\bs{E}\times\bs{B}$ induced rotation of the ion crystal at a frequency $\omega_r$. The ion crystal rotation frequency $\omega_r$ is precisely controlled with a rotating electric potential, dubbed the rotating wall potential \cite{Huang_rotwall_1998}. In a frame rotating at $\omega_r$, the effective confining potential of the trap, ignoring the weak rotating wall potential, is well approximated by
\begin{equation}
q\phi_\mathrm{trap}(\rho,Z) = \frac{1}{2}M {\omega_Z}^2 (Z^2+\beta \rho^2)\,,
\label{eq:trap_confine}
\end{equation}
where $q$ is the ion charge, $M$ is the ion mass, and $Z$ and $\rho$ are the respective axial and radial distance from trap center. $\omega_Z/(2\pi) \approx $ 1.59 MHz characterizes the ion confinement along the axial or magnetic field direction, and $\beta = \omega_r(${$\omega_c$}$-\omega_r)/\omega_Z^2 - 1/2$ is the relative strength of the radial confinement for rotation frequency $\omega_r$ and cyclotron frequency {$\omega_c$}. For the work discussed here, {$\omega_c$}$/2\pi = 7.6$ MHz and the rotation frequency was set to $\omega_r/2\pi \approx 180$ kHz, resulting in $\beta \approx 0.026$ and single plane crystals for ion number $N \lesssim 500$ ions.

In the strong 4.46 T magnetic field of the Penning trap, the $^9$Be$^+$ atomic eigenstates are approximately described by product states $\ket{m_I}\ket{m_J}$ of the $I=3/2$ nuclear spin and valence electron angular momentum $J$ quantum numbers. The nuclear spin is optically pumped to the $\ket{m_I = 3/2}$ level, where it remains throughout the experiment, simplifying the relevant energy level structure shown in Extended Data Fig.~\ref{fig:Be_levels}. The spin-1/2 or qubit degree of freedom in each ion consists of the ground valence electron spin states $\ket{\uparrow} \equiv \ket{2S_{1/2}\, m_j=1/2}$ and $\ket{\downarrow} \equiv \ket{2S_{1/2}\, m_j=-1/2}$. These levels are separated by approximately 124 GHz by the magnetic field.

A low phase noise microwave source~\cite{britton_vibration-induced_2016} at 124 GHz enables coherent global rotations of the ion qubits and implementation of the transverse field interaction of the Dicke model with Rabi rates as large as $\Omega/2\pi \approx 5.7$ kHz. A voltage controlled attenuator allows tuning $\Omega$ arbitrarily below this maximum. {The large trap electrodes enable the application of microwaves with a waist that is large compared to the ion crystal diameter ($\sim 300 \,\mu$m for $N\approx 200$), resulting
in very uniform microwave intensity across the crystal. A random benchmarking experiment performed in a previous NIST Penning trap set-up measured an error for a $\pi$-pulse rotation of the composite Bloch vector of less than $10^{-3}$~\cite{biercuk_quantum_contrl_Penning_2009}. In the current set-up, the microwave propagation k-vector makes a 20-degree angle with respect to the trap magnetic field axis. This will produce a radial inhomogeneity in the microwave $\pi$-pulse duration due to the Doppler shift from the in-plane component of the microwave k-vector. For the largest crystals ($N\approx 200$) employed in the Dicke model simulations we estimate a decrease in the $\pi$-pulse duration for ions on the radial edge of the crystal of less than $0.5 \%$. }

{\raggedright  Doppler cooling and state-selective readout is performed on the cycling transition $\ket{\uparrow} \rightarrow \ket{^2P_{3/2}\, m_J=+3/2}$. During cooling, laser beams directed parallel and perpendicular to the magnetic field are tuned $\approx$ 10-20 MHz below the $\ket{\uparrow} \rightarrow \ket{^2P_{3/2}\, m_J=+3/2}$ cycling transition frequency. For spin state detection, photons are collected on a photomultiplier tube (PMT) from the global fluorescence of ions in the $\ket{\uparrow}$ state using only the parallel beam detuned $\approx$ 1 MHz below the cycling transition resonance. The PMT counts are integrated over roughly 0.5 ms to give roughly 2 photons per bright ion, chosen based on the trade-off between reducing photon shot noise and limiting in-plane heating. Qubit state initialization is performed with optical pumping into the $\ket{\uparrow}$ state using the Doppler cooling lasers and a repump laser tuned to the $\ket{\downarrow} \rightarrow \ket{^2P_{3/2}\, m_J=+1/2}$ transition frequency. We also observe ion fluorescence using a camera that records both the timing and position information of photon arrivals. This information can be used in postprocessing to obtain an image in the rotating frame, as shown in Extended Data Fig.~\ref{fig:ions}. These images allow us to count the number of ions $N$ directly. We can also use these images to obtain a more accurate estimate of the average photons per ion recorded in 1 ms to improve the calibration of $N$ without images.\par}

Ion motion parallel to the magnetic field can be described in terms of a set of normal modes frequently called the drumhead modes \cite{shankar_broadening_2020}. The highest frequency drumhead mode is the axial center-of-mass (COM) mode, consisting of an in-phase, uniform oscillation of all the ions in the crystal at the trap frequency $\omega_Z$. The $Q$ factor of the COM mode has been measured to be $Q > 10^6$ through a ring down measurement~\cite{Gilmore_thesis_2020}. The drumhead modes are efficiently cooled by the Doppler laser cooling beams. Measurements of the ion temperature for motion parallel to the magnetic field are consistent with the 0.4 mK Doppler laser cooling limit for $^9$Be$^+$~\cite{jordan_near_2019}. For the $\omega_Z/2\pi = 1.59$ MHz axial center-of-mass (COM) mode, this corresponds to a mean phonon occupation number of $\Bar{n} \approx 4.6$. For $N \lesssim 150$, the axial COM mode is separated from the next highest frequency drumhead modes (the so-called tilt modes) by greater than 20 kHz.

The temperature of the drumhead modes can be reduced further through electromagnetically induced transparency (EIT) cooling \cite{jordan_near_2019,shankar_modeling_2019}. We implement EIT cooling by coherently coupling the two qubit levels to the same $\ket{{^2\mathrm{P}_{3/2}}\: m_J=1/2}$ excited state through two different laser beams as shown in Extended Data Fig.~\ref{fig:Be_levels}. The beams are tuned to a virtual level approximately 400 MHz to the high frequency side of the $\ket{{^2\mathrm{P}_{3/2}}\: m_J=1/2}$ level. The two EIT laser cooling beams intersect the ion crystal at $\pm 10$ degree angles with respect to the crystal plane, with a wave vector difference $\Delta\bs{k}$ that is aligned with the magnetic field axis ($Z$-axis) of the trap (see Fig.~\ref{fig:Fig1}a). Fine tuning of the EIT laser cooling parameters generates enhanced scattering on the motion subtracting versus the motion adding sidebands, while a dark state completely suppresses carrier transitions, resulting in lower motional temperatures. EIT cooling reduces the temperature of the entire spectrum of drumhead modes with a measured temperature of the axial COM mode of $\Bar{n} \lesssim 0.5$. {Immediately after EIT cooling, the occupation of the COM mode is measured to be $\bar{n} = 0.3(0.2)$ quanta~\cite{jordan_near_2019}. Heating rate measurements of the COM mode conducted during the data taking for the Dicke model simulation were less than 100 quanta/s, consistent with earlier measurements; see supplementary material of Refs. \cite{jordan_near_2019,gilmore_quantum-enhanced_2021}. A 100 quanta/s heating rate produces a small increase of $\sim 0.3$ quanta at the end of the 3 ms maximum duration of the simulation. This increase is comparable to the uncertainty with which we can measure $\bar{n}$. When comparing with measurements taken with EIT cooled ion crystals, the numerical calculations assumed a constant thermal occupation of $\bar{n} \sim 0.5$.} We note that, for the work discussed here, ion motion in the plane of the ion crystal was only Doppler cooled, resulting in higher temperatures than for motion parallel to the magnetic field \cite{shankar_broadening_2020,torrisi_perpendicular_2016}.

The axial COM mode is coupled to the spin degree of freedom through the application of a spin-dependent optical dipole force (ODF). Two laser beams, tuned within the ${{^2}P_{3/2}}$ manifold but off resonance with any optical transitions by approximately 12 GHz (see Extended Data Fig.~\ref{fig:Be_levels}), intersect the ion crystal at $\pm10$ degree angles with respect to the crystal plane (see Fig.~\ref{fig:Fig1}), forming a 1-dimensional (1D) optical lattice with a wavelength of approximately $0.9~\mu\mathrm{m}$. The wavefronts of the 1D lattice are aligned to be parallel with the ion crystal plane. In addition, the frequency $\omega_\mathrm{ODF}$ and polarizations of the ODF laser beams are tuned to values that null the AC Stark shift (ACSS) from each individual beam but generate a spin-dependent force from the interference of the beams with the force on $\ket{\uparrow}$ equal in magnitude but opposite in sign to the force on $\ket{\downarrow}$~\cite{britton_engineered_2012,bollinger_simulating_2013,Bullock_thesis_2026}. In the Lamb Dicke approximation, the ODF interaction can be written as
\begin{equation}
\hat{H}_{\mathrm{ODF}} = F_0 \cos{(\mu t)} \sum_{j=1}^{N} \hat{Z}_j \hat{\sigma}^z_j \,,
\label{eq:H_ODF}
\end{equation}
where $F_0$ is the magnitude of the spin-dependent force, $\mu$ is the frequency difference between the two ODF laser beams, and $\hat{Z}_j$ is the axial position operator for the $j^\mathrm{th}$ ion,
\begin{equation}
 \hat{Z}_j = \sum_{m=1}^{N} b_{j,m} \sqrt{\frac{\hbar}{2M\omega_m}} \left( \hat{a}_m e^{-i\omega_mt} +\hat{a}^{\dagger}_m e^{i\omega_mt} \right) \,.
\end{equation}
Here we use an interaction picture with respect to the free drumhead mode frequencies $\omega_m$. The sum is over all drumhead modes $m$ with frequency $\omega_m$, raising and lowering operators $\hat{a}^{\dagger}_m$ and $\hat{a}_m$, and participation $b_{j,m}$ of ion $j$. With the assumption that $\delta \equiv \mu-\omega_Z \ll \mu - \omega_m$ for any other drumhead mode $m$, $H_\mathrm{ODF}$ is approximately given by
\begin{equation}
 \hat{H}_\mathrm{ODF}\approx \frac{F_0 Z_0}{\sqrt{N}} \left( \hat{a} e^{i(\mu-\omega_Z)t} +\hat{a}^{\dagger} e^{-i(\mu-\omega_Z)t} \right) \sum_{j=1}^N\frac{\hat{\sigma}_j^{z}}{2}
\end{equation}
where $Z_0 \equiv \sqrt{\hbar/(2M\omega_Z)}$ is the axial ground state wave function size for a single ion and $\hat{a}$ and $\hat{a}^\dagger$ now refer to the axial COM mode raising and lowering operators. After a unitary transformation we obtain
\begin{equation}
 \hat{H}_\mathrm{ODF}\approx \hbar\frac{2g}{\sqrt{N}} \left( \hat{a} +\hat{a}^{\dagger} \right)\hat{S}_z -\hbar\delta \hat{a}^{\dagger} \hat{a}
\end{equation}
where $g\equiv F_0 Z_0/2\hbar$ and $\delta = \mu - \omega_Z$. Calibration of the strength $g$ is done measuring a MF spin precession as described in \cite{britton_engineered_2012,Gilmore_thesis_2020,Bullock_thesis_2026}.

{The previous paragraph neglects the weak coupling produced by the spin-dependent force of the spin degrees of freedom with other drumhead modes. For the simulations of this manuscript, the spin-dependent force frequency $\mu$ was tuned to a frequency higher that the COM mode frequency $\omega_Z$ with values of $\delta \equiv \mu-\omega_Z >0$ ranging from $0.5$ kHz $< \delta < 5.5$ kHz. The closest modes to the axial COM mode are the so-called tilt modes, which have frequencies more than 20 kHz below the axial COM mode, giving rise to a detuning $\delta_\text{tilt} \equiv \mu-\omega_\text{tilt}$ that is an order of magnitude larger than $\delta$. The large detunings mean the tilt modes, as well as other drumhead modes, will only be virtually excited, producing small one-axis-twisting type corrections to the spin degrees of freedom of the Hamiltonian. In the supplementary material to Ref.~\cite{bohnet_quantum_2016}, which measured squeezing of the spin degree of freedom at the decoupling time of the COM mode, a theoretical analysis showed that including coupling to other modes produced a few percent correction to the composite Bloch vector contrast and variance. The Dicke model simulations employed slightly larger detunings than Ref.~\cite{bohnet_quantum_2016} (5.5 kHz vs 3 kHz), but we anticipate corrections to the Bloch vector contrast and variance to be less than 10\%.}

For a single trapped ion, the ground state wavefunction size is $Z_0 \approx 19 \, \textrm{nm}$, producing a Lamb-Dicke parameter $\eta \equiv \delta k Z_0 = 0.13$ for the optical dipole force interaction with the axial COM mode. Here $\delta k \equiv 2\pi/(900 \, \textrm{nm})$, where 900 nm is the wavelength of the 1D optical lattice. The Lamb Dicke approximation requires
\begin{equation}
 \delta k\, Z_{rms,j} << 1
\end{equation}
for all ions $j$ in the array, where $Z_{rms,j} \equiv \sqrt{\left\langle Z_{j}^2 \right\rangle}$. For Doppler laser cooling to the 0.5 mK limit, $\delta k\, Z_{rms,j} \approx 0.52$ for typical ion arrays used in this work. With EIT cooling to $\bar{n}_\textrm{exp} = 0.5$, $\delta k\, Z_{rms,j} \approx 0.22$. The strength of the spin dependent force $F_0$ is reduced with increasing $\delta k\, Z_{rms,j}$ by the Debye-Waller factor~\cite{Wineland_NIST_1998}. The MF spin precession calibration of the Dicke model parameter $g$ measures the effective spin-dependent force, which includes the Debye-Waller factor.

Before carrying out a set of experiments, calibrations of all relevant experimental parameters are taken. These include the Hamiltonian parameters $g$ \cite{britton_engineered_2012} and $\Omega$. We also measure $\bar{n}$ \cite{sawyer_spectroscopy_2012}, the quasi-static COM mode frequency fluctuations \cite{gilmore_quantum-enhanced_2021}, the ODF laser beam polarizations that null each beam's AC Stark shift, the spin decoherence rate from magnetic field noise, and the spin decoherence rate from the application of the ODF beams, which is dominantly from off-resonant light scatter~\cite{Bullock_thesis_2026,uys_decoherence_2010}. The calibrations and measurements of the experimental noise and decoherence are used with TWA simulations to model non-ideal behavior of the experimental system (see the next section).

In addition to these calibrations, active tracking of the qubit (or spin-flip) frequency and the axial COM frequency are interleaved during a scan~\cite{Bullock_thesis_2026}. To track the qubit frequency, a Ramsey measurement with a 1 ms free precession period is carried out for three different microwave frequencies. The center frequency results in a dip in the bright fraction near resonance, while imbalances in the measured population at the points of steepest slope on the sides of the central Ramsey fringe, called the left and right points, measure drifts in the qubit frequency. The population imbalance at the two side fringes is scaled by the fringe width and a user-adjustable gain to actively track the qubit frequency and apply appropriate shifts to the applied microwave frequency. Similarly, a resonance feature is used to track the axial COM mode and stabilize the parameter $\delta$ by adjusting the detuning between the ODF beams $\mu$. Here the resonance feature was generated by driving with an oscillating axial electric field for 1 ms to excite COM motion, resulting in a decrease in fluorescence from Doppler shifts when heated.

Extended Data Fig.~\ref{fig:tracking} shows an example of interleaved tracking of the axial COM frequency and qubit frequency during a simulation of the Dicke model. For a given simulation, the dynamics are evolved for 100 different time steps from 0 to 3 ms, with 30 repeats at each time step and 5 passes over all time steps, giving a total of 15,000 measurements of the dynamics. Approximately every 150 measurements of the dynamics, the COM and qubit frequency were measured using $\sim$30 repetitions of each side fringe and $\sim$10 repetitions of the center frequency. From the statistical uncertainties of the measured populations the tracked frequencies are estimated to a precision of $\sim\pm$1 and $\sim\pm$4 Hz for the COM and qubit frequencies respectively. Drifts over the course of a scan would have resulted in errors of $\sim$ 100 Hz for most simulations without active compensation.

\section{Phase space plots and Chaos Characterization}

 As discussed in the main text, our identification of chaotic dynamics is based on a positive Lyapunov exponent $\lambda_L$ and the associated phase-space behavior. For two initially nearby trajectories in phase space, separated by a distance $|\Delta \mathbf{x}(0)|$, classical chaos leads to an exponential divergence,
$|\Delta \mathbf{x}(t)| \approx e^{\lambda_L t} |\Delta \mathbf{x}(0)|$. A positive Lyapunov exponent therefore signals chaotic behavior.

To visualize   the exponential divergence, we first plot the ideal  phase-space  mean-field dynamics for two initially nearby trajectories (red and green curves) in Extended Data Fig.~\ref{fig:chaotic_MFphasespace}. The parameters of the Dicke Hamiltonian used in this plot are the same as those in main text Fig.~\ref{fig:Fig3} (see the panels marked by the corresponding green symbols). For the regular dynamics (left and right panels), the two trajectories remain nearly indistinguishable throughout the evolution. In contrast, for the chaotic dynamics (center panels), the trajectories exhibit a clear divergence in phase space.

 Similar dynamics remains visible in the experiment even including quantum and thermal fluctuations, as can be seen from   the dynamical phase space plots in Fig.~\ref{fig:Fig3}. In these figures we plot the trajectory of the mean spin dynamics in terms of $\braket{\hat{S}_z(t)}$ and $\braket{\hat{S}_y(t)}$.  From the equations of motion for unitary evolution under $\hat{H}_{\rm Dicke}$, we have that $d\braket{\hat{S}_z}/dt = \Omega\braket{\hat{S}_y}$, so that this forms a canonical phase space for the $z$-magnetization. Therefore, even though we do not directly measure  $\braket{\hat{S}_y(t)}$, it  is  possible  to extract both the $z$ and $y$ collective spin components from the dynamics of $\braket{\hat{S}_z(t)}$ alone.

To  extract  the time  derivative, we first apply a smoothing process to the data for $\braket{\hat{S}_z(t)}$, utilizing MATLAB's smoothing spline with a smoothing parameter of $p = 0.98$, and utilizing the reciprocal variance of each timeseries point as a weighting function. In order to avoid issues arising from the smoothing process at the endpoints of the timeseries, we first mirror the data about $t=0$ before smoothing, so that we have an effective timeseries for $t \in [-3.0, 3.0]$ ms --- this effectively enforces a local extremum at $t=0$, so that the resulting derivative vanishes at this point, resulting in the expected condition $d\braket{\hat{S}_z(t)}/dt|_{t=0} \propto \braket{\hat{S}_y(0)} = 0$. For the opposite endpoint at the maximum evolution time,
we simply truncate the displayed data to a maximum time of $t=2.9$ ms.

We find that this process is suitable for removing local fluctuations of the time derivative owing to small variations in the mean value of neighboring time points, while not significantly altering trends in the global dataset. In Extended Data Fig.~\ref{fig:phase_space_comparison}, we directly compare our phase space plots using the smoothed experimental data to direct theoretical calculations of the $y$-magnetization via our semiclassical model, finding that our smoothing process and numerical derivative provide a good representation for the dynamics of $\braket{\hat{S}_y(t)}$.

In the presence
of light scattering, the unitary identity
$d\langle\hat S_z\rangle/dt=\Omega\langle\hat S_y\rangle$ acquires small
dissipative corrections. Nevertheless, the reconstructed and directly
calculated portraits show the same large-scale structure, validating the
reconstruction over the experimental time window.

We present additional examples of experimental and semiclassical spin phase-space dynamics in the chaotic regime in Extended Data Fig.~\ref{fig:chaotic_spinphasespace}, corresponding to parameter regimes with nonzero $\lambda_L$. At early times, before the decay envelope contracts the trajectories towards the phase space origin, the chaotic dynamics generically wander erratically over a large portion of the spin phase space, executing various-sized loops at irregular intervals---suggestive of the ergodic exploration expected of classically chaotic dynamics---whereas the untrapped dynamics remain confined to smooth rings near the edge of phase space for all parameters considered. The irregular and erratic phase-space trajectories resemble the example shown in Fig.~\ref{fig:Fig3}h of the main text.

{

To quantify the onset of chaotic behavior, Extended Data Fig.~\ref{fig:chaotic_spectra}c,d, shows the power spectra for the measured time dynamics.
For each measured time trace, sampled at times
$t_j= j\Delta t$ for $0\leq j \leq N_t-1$ for $N_t$ the number of sampled times, we calculate the discrete Fourier amplitude
\[
\widetilde S_z(\omega_n)
=
\sum_{j=0}^{N_t-1}
w_j\langle\hat S_z(t_j)\rangle e^{-i\omega_n t_j},
\qquad
\omega_n=2\pi n/N_t\Delta t,
\]
which is convolved with the Hann window function $w_j = \sin^2(\pi j/(N_t-1))$ to reduce the effect of spectral leakage. We define the normalized power spectrum as
\[
P(\omega_n)=
|\widetilde S_z(\omega_n)|^2
/\sum_m|\widetilde S_z(\omega_m)|^2,
\qquad
\sum_nP(\omega_n)=1.
\]
As our measured signal is real, we have $\widetilde S_z(\omega_{N_t-n})=\widetilde S_z^{*}(\omega_n)$, and we thus work with the one-sided power spectrum, with sums restricted to the $M_\omega$ non-negative frequencies up to the Nyquist frequency $\pi/\Delta t$. The power of each $\omega\neq 0$ frequency bin is doubled to account for the equal contribution of the redundant negative frequency components. The continuous curves in Extended Data Fig.~\ref{fig:chaotic_spectra}c,d are evaluated on an eightfold zero-padded frequency grid, in order to provide a smooth interpolation for visualization purposes.

Since each time point is obtained from an independent set of experimental runs, the measurement noise contributes an additional power $F = 2\sum_j w_j^2\sigma_j^2$ for statistical uncertainty $\sigma_j$ at time $t_j$, and the factor of $2$ reflecting our one-sided spectral convention.  This projection-noise floor is indicated in Extended Data Fig.~\ref{fig:chaotic_spectra}c,d for each data set by the faded, dotted horizontal lines. The regular trajectories feature spectra sharply peaked at the transverse drive frequency, whereas in the  chaotic regime the  spectra typically display no reproducible peaks, and instead exhibit significant support above the noise floor across a broad range of frequencies.

This distinction is summarized in Extended Data Fig.~\ref{fig:chaotic_spectra}e by the second order  R{\'e}nyi spectral entropy
\begin{align}
    H_2 = -\frac{\log\left(\sum_n p_n^2\right)}{\log M_\omega},
\end{align}
computed from the noise-corrected spectrum $p_n \propto \max\left(0,|\widetilde S_z(\omega_n)|^2-F\right)$, normalized such that $\sum_n p_n = 1$. We perform this subtraction of the expected noise floor to ensure that measurement noise does not contribute to the extracted entropy, so that $H_2$ characterizes the dynamics rather than the underlying noise level in each dataset. We also note that in computing the corresponding bootstrapped error bounds, resampling based on the statistical uncertainty results in an \emph{additional} power contribution $F$ to the resampled data sets, which must be fully subtracted off to avoid bias in the resampled statistics.

$H_2$ increases within the same
parameter window in which the classical calculation yields
$\lambda_L>0$, reproduced in Extended Data Fig.~\ref{fig:chaotic_spectra}f. Importantly,
this enhancement is observed for both atom-number ranges and for both EIT-
and Doppler-cooled initial states. Its location is therefore robust against
the experimentally explored variations in system size and initial thermal
phonon occupation.

Taken together, these results establish a consistent correspondence between
three independent features: exponential sensitivity to initial conditions
in the classical Dicke equations, irregular and broadly exploring
experimental spin phase portraits, and enhanced spectral entropy in the
measured dynamics. Even though  the experiment does not directly extract a Lyapunov
exponent from pairs of infinitesimally separated trajectories,  it does observe robust dynamical signatures that appear in---and track
the boundaries of---the parameter regime independently predicted to be
chaotic by the positive classical Lyapunov exponent.}

\section{Resonant interaction picture}
To elucidate the underlying physics, consider the joint rotating frame of the transverse drive and bosonic field, $\hat{H}_0$, where the interaction frame Hamiltonian can be written as $\hat{H}_{\rm s-p}^{\rm rot}=\hat{H}_{\rm{pair}}^{\rm rot}+ \hat{H}_{\rm{osc}}^{\rm rot}$ for
\begin{gather}
\hat{H}_{\rm{pair}}^{\rm rot}/\hbar\equiv \frac{ -i g}{\sqrt{N}} \left( \hat{a}^\dagger { \hat{S}^{(x)}_+} e^{i (\Omega-\delta) t} - \hat{a}\hat{S}_-^{(x)}e^{-i (\Omega-\delta) t}\right), \\ \hat{H}_{\rm{osc}}^{\rm rot}/\hbar\equiv \frac{-ig}{\sqrt{N}}\left(\hat{a} \hat{S}_+^{(x)} e^{i (\Omega+\delta) t} - \hat{a}^{\dagger}\hat{S}_-^{(x)} e^{-i (\Omega+\delta) t}\right), \label{Eq:res}
\end{gather}
with $x$-basis spin ladder operators $\hat{S}_y\equiv (\hat{ S}_+^{(x)} +\hat{S}_-^{(x)})/2$ and $\hat{S}_z\equiv (\hat{S}_+^{(x)} -\hat{S}_-^{(x)})/(2i)$.

\section{Extracting spin-phonon correlations}
\subsection{Ideal case}
We aim to extract a proxy for the spin-phonon squeezing for the quadrature $\hat{\mathcal{W}}_+ = \hat{\mathcal{X}} + \hat{S}_y/\sqrt{N/2}$ from the dynamics of $\hat{S}_x$ alone. We start by examining the equations of motion under the rotating wave approximation (RWA), assuming exact resonance conditions on our spin and boson degrees of freedom. { We have the resonant pair-creation Hamiltonian (anti-Tavis--Cummings in the stated ladder convention)}
\begin{align}
 { \hat{H}_{\rm pair}^{\rm rot}} = \hbar\frac{-ig}{\sqrt{N}}\left(\hat{a}^\dagger\hat{S}_+^{(x)} - \hat{a}\hat{S}_-^{(x)}\right)
\end{align}
for rotated spin ladder operators $\hat{S}_{\pm}^{(x)} = \hat{S}_y \pm i\hat{S}_z$. Under this Hamiltonian, we have relevant Heisenberg equations of motion
\begin{equation}
 \frac{d\hat{S}_x}{dt} =\frac{d\hat{n}}{dt} = \frac{-g}{\sqrt{N}}\left(\hat{a}^\dagger\hat{S}_+^{(x)} + \hat{a}\hat{S}_-^{(x)}\right),
\end{equation}
To compute the variance of our quadrature, we have (in the RWA)
\begin{align}
 \hat{\mathcal{W}}_+^2 &= \hat{\mathcal{X}}^2 + \frac{2}{N}\hat{S}_y^2 + 2\sqrt{\frac{2}{N}}\hat{\mathcal{X}}\hat{S}_y\\
 \begin{split}
 &\rightarrow \hat{n} + \frac{1}{2} + \frac{1}{2N}\left({ \hat{S}_+^{(x)}\hat{S}_-^{(x)} + \hat{S}_-^{(x)}\hat{S}_+^{(x)}}\right) \\
 &\qquad + \frac{1}{\sqrt{N}}\left(\hat{a}^\dagger\hat{S}_+^{(x)} + \hat{a}\hat{S}_-^{(x)}\right),
 \end{split}
\end{align}
where $\hat{n} = \hat{a}^\dagger\hat{a}$ is the bosonic number operator.

We first note that from our equations of motion, we have
\begin{gather}
 \braket{\hat{n}(t)} = \braket{\hat{n}(0)} - \braket{\hat{S}_x(0)} + \braket{\hat{S}_x(t)},\label{eq:n_intermsof_s}\\
 \frac{1}{\sqrt{N}}\Big\langle\left(\hat{a}^\dagger\hat{S}_+^{(x)} + \hat{a}\hat{S}_-^{(x)}\right)(t)\Big\rangle = -\frac{1}{g}\frac{d\braket{\hat{S}_x(t)}}{dt}.\label{eq:spinphononcorr_intermsof_dsdt}
\end{gather}
Eq.~\eqref{eq:n_intermsof_s} is consistent with our earlier Holstein-Primakoff approximation, where the number of total excitations in the phonon and spin degrees of freedom are inherently linked, and can be determined from one another given the initial occupation for each. Also, from Eq.~\eqref{eq:spinphononcorr_intermsof_dsdt}, we observe that the derivative of the transverse spin dynamics offers a direct way to extract the spin-phonon correlations, assuming we properly account for technical and fundamental noise sources (see next section; also Ref.~\cite{safavi-naini_verification_2018}).

Second, we have
\begin{align}
 \frac{1}{2N}\Big\langle\left({ \hat{S}_+^{(x)}\hat{S}_-^{(x)} + \hat{S}_-^{(x)}\hat{S}_+^{(x)}}\right)(t)\Big\rangle = \frac{1}{N}\left(\braket{\hat{S}^2(t)} - \braket{\hat{S}_x^2(t)}\right),
\end{align}
and { because the initial state lies in the fully symmetric $S=N/2$ sector and collective unitary evolution preserves this sector, we have} $\braket{\hat{S}^2(t)} = S(S+1)$ for total spin $S=N/2$. Thus, with these assumptions,
\begin{align}
\begin{split}
 \braket{\hat{\mathcal{W}}_+^2} &\rightarrow \braket{\hat{n}(0)} - \braket{\hat{S}_x(0)} + \braket{\hat{S}_x(t)} + \frac{1}{2} \\
 &\qquad + \frac{1}{N}\left(S(S+1) - \braket{\hat{S}_x^2(t)}\right) - \frac{1}{g}\frac{d\braket{\hat{S}_x(t)}}{dt}.
 \end{split}
\end{align}

\subsection{Accounting for noise}
We have relevant jump operators $\sqrt{\Gamma_z/4}\,\hat{\sigma}_{z,i}$, $\sqrt{\Gamma_{\uparrow\downarrow}}\,\hat{\sigma}_{-,i}$, and $\sqrt{\Gamma_{\downarrow\uparrow}}\,\hat{\sigma}_{+,i}$ for each qubit, as well as a collective jump operator $\sqrt{\Gamma_B}\hat{S}_z$. Going to the rotating frame of the applied transverse field, we instead have relevant jump operators (under the RWA) $\sqrt{(\Gamma_{\uparrow\downarrow} + \Gamma_{\downarrow\uparrow})/4}\,\hat{\sigma}_{x,i}$ and $\sqrt{(\Gamma_{z} + \Gamma_{\uparrow\downarrow} + \Gamma_{\downarrow\uparrow})/4}\,\hat{\sigma}_{\pm,i}^{(x)}$ for each qubit, as well as collective jump operators $\sqrt{\Gamma_B/4}\,\hat{S}_{\pm}^{(x)}$. Static, longitudinal noise is rotated out, and thus neglected. We also neglect the role of COM fluctuations, which were shown in Supplementary Fig.~\ref{fig:noise_sources_boson} to have a negligible effect on the resonant cut dynamics over relevant timescales.

Under these noise processes, we instead have the corresponding equations of motion
\begin{align}
\frac{d\braket{\hat{n}}}{dt} &= \frac{d\braket{\hat{S}_x}}{dt} + \left(\frac{\Gamma_z + \Gamma_{\uparrow\downarrow} + \Gamma_{\downarrow\uparrow} + \Gamma_B}{2}\right)\braket{\hat{S}_x(t)}\\
&=\frac{-g}{\sqrt{N}}\Big\langle\left(\hat{a}^\dagger\hat{S}_+^{(x)} + \hat{a}\hat{S}_-^{(x)}\right)(t)\Big\rangle,
\end{align}
from which we have the following replacement for our inferred spin-phonon correlations:
\begin{widetext}
\begin{align}
 \frac{1}{\sqrt{N}}\Big\langle\left(\hat{a}^\dagger\hat{S}_+^{(x)} + \hat{a}\hat{S}_-^{(x)}\right)(t)\Big\rangle = -\frac{1}{g}\left[\frac{d\braket{\hat{S}_x(t)}}{dt} + \left(\frac{\Gamma_z + \Gamma_{\uparrow\downarrow} + \Gamma_{\downarrow\uparrow} + \Gamma_B}{2}\right)\braket{\hat{S}_x(t)}\right].
\end{align}
\end{widetext}

Now, let us define $\tilde{S}_x(t) = e^{(\Gamma_z + \Gamma_{\uparrow\downarrow} + \Gamma_{\downarrow\uparrow} + \Gamma_B)t/2}\braket{\hat{S}_x(t)}$, so that we have
\begin{align}
 \frac{d\tilde{S}_x}{dt} = e^{(\Gamma_z + \Gamma_{\uparrow\downarrow} + \Gamma_{\downarrow\uparrow} + \Gamma_B)t/2} \times \frac{d\braket{\hat{n}(t)}}{dt},
\end{align}
and so we have the corrected boson number replacement
\begin{align}
\begin{split}
 \braket{\hat{n}(t)} &= \braket{\hat{n}(0)} - \braket{\hat{S}_x(0)} + \braket{\hat{S}_x(t)} + \\
 &\qquad \left(\frac{\Gamma_z
 + \Gamma_{\uparrow\downarrow} + \Gamma_{\downarrow\uparrow} + \Gamma_B}{2}\right) \int_0^t dt'\, \braket{\hat{S}_x(t')}.
 \end{split}
\end{align}
{ Finally, collective noise processes such as those proportional to $\Gamma_B$ do not change the total spin. For the local noise channels, define $R=\Gamma_{\uparrow\downarrow}+\Gamma_{\downarrow\uparrow}$, $A=(\Gamma_z+3R)/2$, $C=(\Gamma_z-R)/2$ and $K=N(\Gamma_z+2R)/2$. The total-spin equation is
\begin{align}
\frac{d\braket{\hat S^2(t)}}{dt}=-A\braket{\hat S^2(t)}-C\braket{\hat S_x^2(t)}+K.
\end{align}
Defining $\tilde S^2(t)=e^{At}\braket{\hat S^2(t)}$, we obtain
\begin{align}
\tilde S^2(t)=\tilde S^2(0)-C\int_0^t dt'\,e^{At'}\braket{\hat S_x^2(t')}+\frac KA(e^{At}-1).
\end{align}
Consequently,
\begin{align}
\braket{\hat S^2(t)}=e^{-At}\left[\braket{\hat S^2(0)}-C\int_0^t dt'\,e^{At'}\braket{\hat S_x^2(t')}\right]+\frac KA(1-e^{-At}).
\end{align}
The expressions containing $K/A$ are understood by continuity when $A=0$.
}
Thus, in total, we have
\begin{widetext}
\begin{align}
\begin{split}
 \braket{\hat{\mathcal{W}}_+^2(t)} &= \braket{\hat{n}(0)} - \braket{\hat{S}_x(0)} + \braket{\hat{S}_x(t)} + \left(\frac{\Gamma_z + \Gamma_{\uparrow\downarrow} + \Gamma_{\downarrow\uparrow} + \Gamma_B}{2}\right) \int_0^t dt'\, \braket{\hat{S}_x(t')}\\ &\qquad + \frac{1}{2}\\
 &\qquad + \frac{e^{-(\Gamma_z + 3(\Gamma_{\uparrow\downarrow} + \Gamma_{\downarrow\uparrow}))t/2}}{N}\Big[\braket{\hat{S}^2(0)} - \\ &\qquad \left(\frac{\Gamma_z - (\Gamma_{\uparrow\downarrow} + \Gamma_{\downarrow\uparrow})}{2}\right)\int_0^t dt'\, e^{(\Gamma_z + 3(\Gamma_{\uparrow\downarrow} + \Gamma_{\downarrow\uparrow}))t'/2} \braket{\hat{S}_x^2(t')}\Big]\\ &\qquad { +\frac{K}{NA}(1-e^{-At})} - \frac{1}{N}\braket{\hat{S}_x^2(t)}\\
 &\qquad -\frac{1}{g}\left[\frac{d\braket{\hat{S}_x(t)}}{dt} + \left(\frac{\Gamma_z + \Gamma_{\uparrow\downarrow} + \Gamma_{\downarrow\uparrow} + \Gamma_B}{2}\right)\braket{\hat{S}_x(t)}\right].\label{eq:Wsqueezing_proxy}
\end{split}
\end{align}
\end{widetext}
 Every term can be extracted from the dynamics of the distribution for $\hat{S}_x$, as well as characterization of the initial state and relevant noise processes.

{
\subsection{Extracting spin variance dynamics}
The dynamics of $\braket{\hat{S}_x^2(t)}$ can be directly extracted from our measurement data for $\braket{\hat{S}_x(t)}$. However, the computed raw moment must be corrected to account for the effects of photon shot-noise.

For each experimental measurement, the measured photon count $n$ is converted to a transverse spin projection measurement via
\begin{align}
 S_{x,{\rm meas}} = N\left(\frac{n-d}{A} - \frac{1}{2}\right),\quad A = b-d
\end{align}
where $b$ and $d$ are the average counts measured with all ions bright or dark, respectively. In order to obtain an estimate for the second spin moment, the photon shot-noise adds, on average, an additional contribution given by
\begin{align}
 V_{\rm photon} = \frac{dN^2}{A^2} + \frac{N}{A}\left(S_{x,{\rm meas}} + \frac{N}{2}\right).
\end{align}
Thus, we can obtain the corrected second moment by averaging the quantity
\begin{align}
 S_{x,{\rm corr}}^{(2)} = S_{x,{\rm meas}}^2 - V_{\rm photon}.
\end{align}
We plot the corresponding estimate for the variance of the transverse magnetization in the left panel of Extended Data Fig.~\ref{fig:spinboson_correlation_proxy} alongside corresponding estimates from our semiclassical model.}

{
\subsection{Extracting spin-phonon correlation dynamics}
To reconstruct the spin-phonon correlator indirectly from experimentally accessible observables, we must estimate the derivative of the transverse magnetization dynamics. To reduce sensitivity to noise compared to direct computation via finite differences, we first apply a gaussian filter to the data with a window size of $\sigma = 0.1$ ms, and then compute an interpolating spline through the resulting data points. From this, we numerically extract the derivative at arbitrary times within the measured time window.

As for our computation of the phase space plots, the initial boundary near $t=0$ requires special care, where the absence of data at $t<0$ can bias the filter. To mitigate this artifact, we extend our datasets with auxiliary data points to negative times, using the short-time behavior implied by $\langle\hat{S}_x(t)\rangle = t (d\langle\hat{S}_x(t)\rangle/dt)_{t=0} + \langle\hat{S}_x(0)\rangle$. These anchor points are fixed by the initial state and characterized noise rates via $\langle\hat{S}_x(0)\rangle = -N/2$ and $(d\langle\hat{S}_x(t)\rangle/dt)_{t=0} = (N/2)\times(\Gamma_z + \Gamma_{\uparrow\downarrow} + \Gamma_{\downarrow\uparrow} + \Gamma_B)/2$. We choose $N$ self-consistently so that the smoothed/splined curves satisfies $\langle\hat{S}_x(0)\rangle = -N/2$; this avoids a discontinuity between the physical initial condition and the smoothed experimental trace that would otherwise lead to a spurious derivative at early times. The resulting $N$ is found to be within $\pm 1$ of the value independently obtained from experiment for each set of parameters.

We show the resulting derivative in the center panel of Extended Data Fig.~\ref{fig:spinboson_correlation_proxy}, which includes a final gaussian filtration with a window of $0.36$ ms to suppress residual fast oscillatory components associated with counter-rotating terms neglected in our model. To compute bootstrapped error bounds, we resample from the measurement data and re-perform our entire data smoothing and numerical derivative procedure on these data sets. In addition to our comparison with the actual spin-phonon correlation dynamics computed from our semiclassical model, we also perform the same numerical derivative computation using the semiclassical dynamics for $\langle\hat{S}_x(t)\rangle$, utilizing the same filtering and splining procedures. We find that the proxy generally provides a decent approximation to the underlying correlations (See Extended Data Fig.~\ref{fig:spinboson_correlation_proxy}; our proxy also decently agrees with that computed from our measurement data, with deviations outside the computed error bounds expected to be the result of deviations already present in the underlying transverse magnetization dynamics.)

\subsection{Extracting spin-phonon squeezing proxy dynamics}

To compute our proxy for the spin-phonon squeezing, we combine the (unsmoothed) estimated spin-phonon correlation extracted from the derivative of $d \langle \hat{S}_x(t)\rangle/dt$, as described above, and $\langle\hat{S}_x(t)\rangle$, $\langle\hat{S}_x^2(t)\rangle$ from the measurement data. We also use this measurement data to compute the necessary integrals in Eq.~\ref{eq:Wsqueezing_proxy} to account for the effects of decoherence. We further apply a final gaussian filtration with a window of $0.36$ ms. For error bounds, we again resample from the measurement data, and perform the entire post-process analysis on each resampled data set. We then bootstrap our confidence intervals from the resulting ensemble of measurement estimates.

The resulting quantity is presented on the right side of Extended Data Fig.~\ref{fig:spinboson_correlation_proxy} as well as in Fig.~\ref{fig:Fig5} of the main text. This demonstrates good agreement with both the proxy computed from our semiclassical dynamics, as well as actual squeezing dynamics from our semiclassical model.

}

\section{Platform context}
The Dicke model and related self-organization phenomena have been explored
with thermal gases~\cite{black_observation_2003,domokos_mechanical_2003}
and Bose--Einstein condensates
\cite{baumann_dicke_2010,nagy_dicke-model_2010,brennecke_real-time_2013,
klinder_dynamical_2015,leonard_supersolid_2017,
kollar_supermode-density-wave-polariton_2017,landini_formation_2018,
morales_two-mode_2019}. In the latter experiments, two distinguishable
motional modes form an effective pseudospin coupled to the cavity field,
providing a direct realization of the Dicke model with generalized
implementations also incorporating atomic hyperfine states. Beyond atomic
gases, the model has been realized in circuit-QED
\cite{yoshihara_superconducting_2017,forn-diaz_ultrastrong_2019} and
solid-state microcavities~\cite{todorov_ultrastrong_2010}. More recently, the model has also been explored using optical cavities
containing individually trapped atoms
\cite{yan_superradiant_2023,ho_optomechanical_2025} and trapped ions
\cite{safavi-naini_verification_2018,aedo_analog_2018,
sutherland_analog_2019,bohnet_quantum_2016,
lewis-swan_characterizing_2021}.

\section{Representative parameter settings and dynamical context}
The representative LMG traces in Fig.~\ref{fig:Fig2} include $\Omega/\chi=0.29$ with $\Omega=2\pi\times0.35$ kHz, $\Omega/\chi=0.51$ with $\Omega=2\pi\times0.476$ kHz, and $\Omega/\chi=0.99$ with $\Omega=2\pi\times0.665$ kHz. Along the chaotic cut in Fig.~\ref{fig:Fig3}, the representative ratios and drives are $\Omega/\delta=0.26$ with $\Omega\simeq2\pi\times(0.6\text{--}0.7)$ kHz, $\Omega/\delta=0.65$ with $\Omega\simeq2\pi\times1.2$ kHz, and $\Omega/\delta=2.51$ with $\Omega\simeq2\pi\times2.3$ kHz.

Trapped and untrapped collective spin dynamics connect to self-trapping and dynamical transitions studied in other many-body platforms~\cite{marino_dynamical_2022,smerzi_quantum_1997,albiez_direct_2005,abbarchi_macroscopic_2013,reinhard_self-trapping_2013,zhang_observation_2017,borish_transverse-field_2020,muniz_exploring_2020,li_improving_2023}. Further context for Dicke-model dynamics and excited-state structure is given in Refs.~\cite{lewis-swan_exploiting_2024,brandes_excited-state_2013}, and phase-coherent control of the Penning-trap motion is described in Ref.~\cite{affolter_phase-coherent_2020}.

\section{Further dynamical and information-theoretic context}
Earlier work on dynamical transitions provides further context for the non-equilibrium phase structure~\cite{eckstein_thermalization_2009}. The forward-looking discussion of EPR correlations and steering is grounded in Refs.~\cite{reid_colloquium_2009,uola_quantum_2020}. The connections between correlated pair production, thermofield double states and holography are discussed in Refs.~\cite{takahashi_higher_1975,israel_thermo-field_1976,maldacena_eternal_2003}, with teleportation and information recovery addressed in Refs.~\cite{maldacena_diving_2017,hayden_black_2007}.

\renewcommand{\bibsection}{\section*{Methods references}}

\clearpage
\section*{Extended Data}
\suppressfloats[t]
\setcounter{figure}{0}
\renewcommand{\figurename}{Extended Data Fig.}
\renewcommand{\figseries}{ED}

\begin{figure}[!ht]
\centering
\includegraphics[width = 0.3\textwidth]{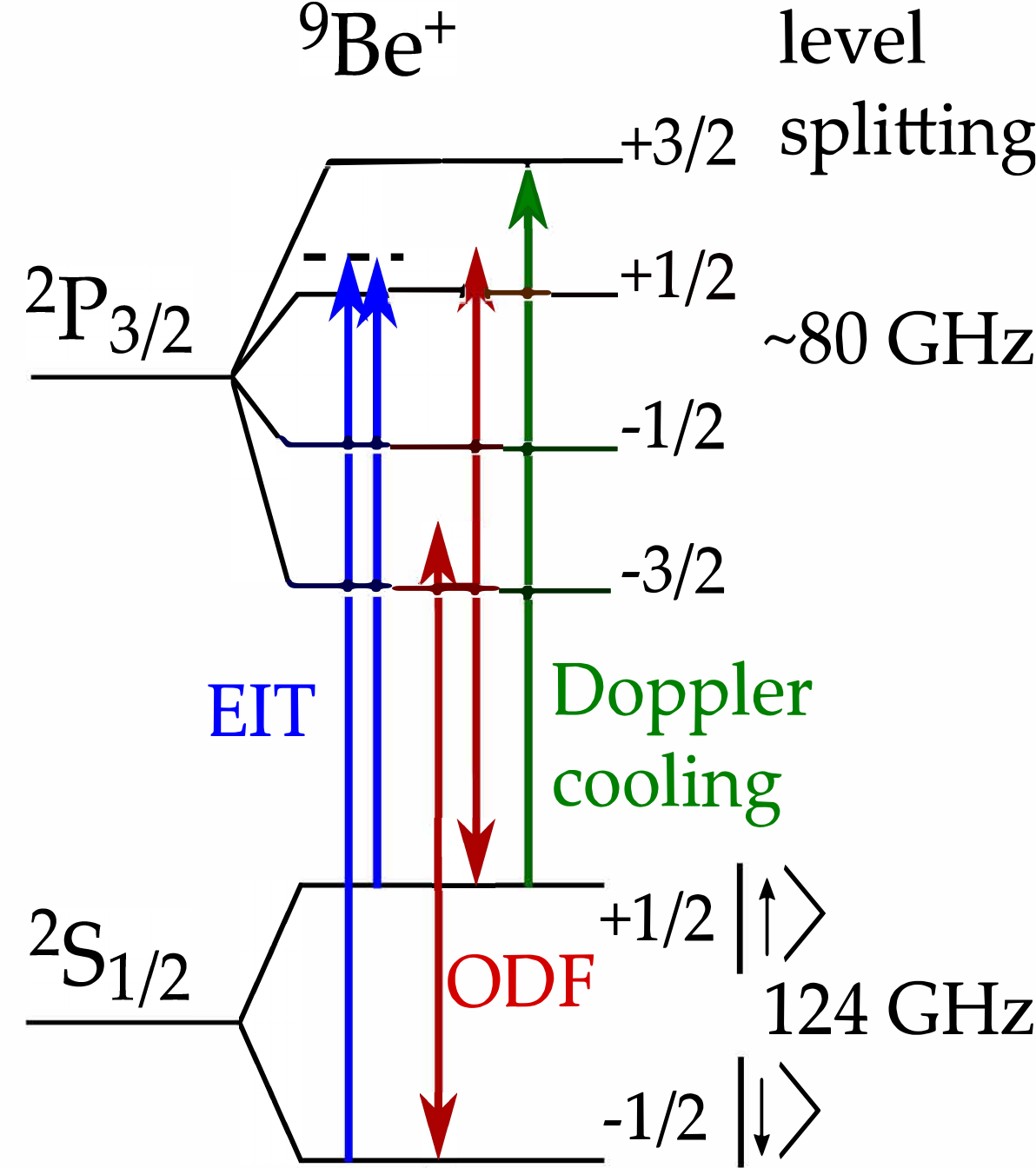}
\caption{$^9$Be$^+$ energy level diagram with corresponding laser frequencies.}
\label{fig:Be_levels}
\end{figure}
\clearpage
\begin{figure}[!ht]
\centering
\includegraphics[width=.4\textwidth]{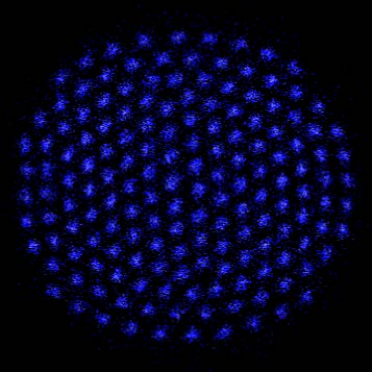}
\caption{Image of an ion crystal with $N=161$ in the rotating frame.}
\label{fig:ions}
\end{figure}
\clearpage
\begin{figure*}[!ht]
\centering
\includegraphics[width = 1\textwidth]{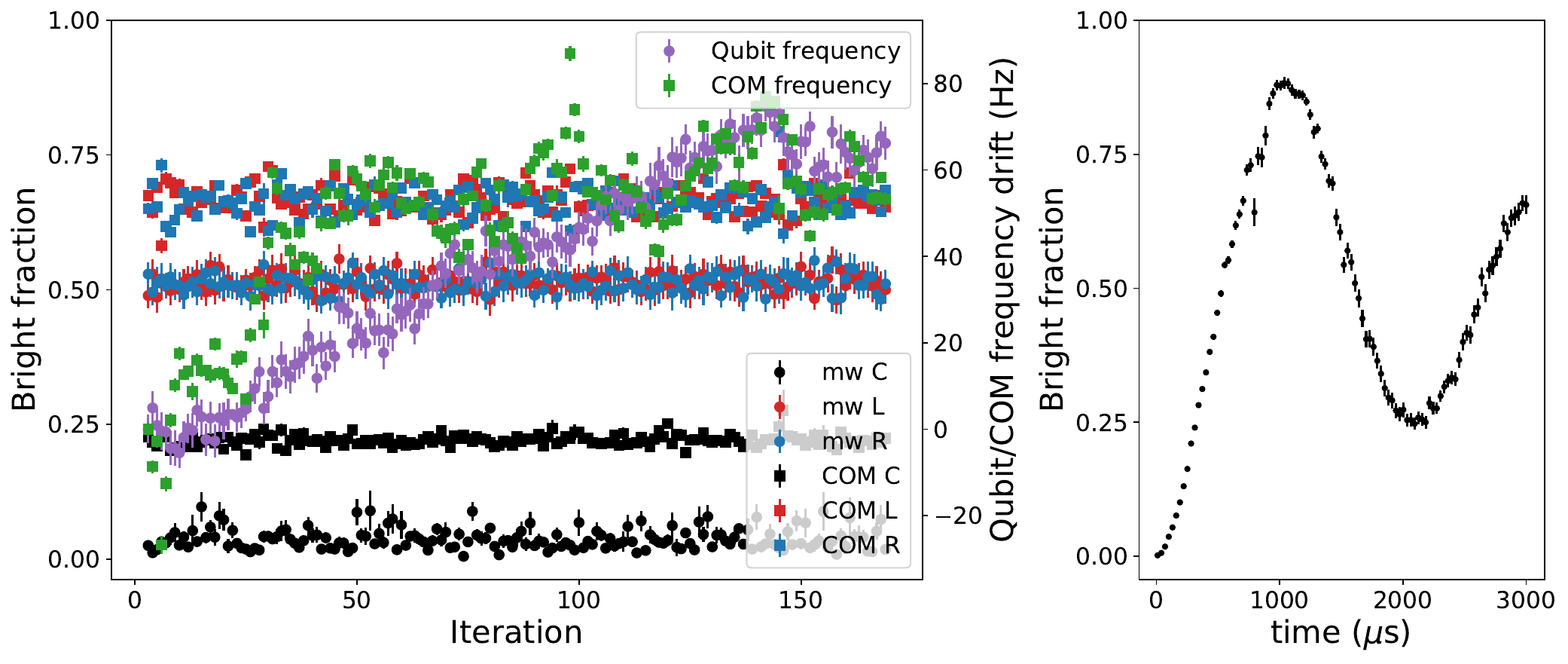}
\caption{Example Dicke model simulation with interleaved tracking. The right plot shows the simulated Dicke model dynamics with $\Omega/\delta=0.125$ and $\Omega/\chi=0.91$. Approximately every $\sim$150 measurements a COM and qubit frequency tracker measurement was carried out as shown on the left plot. On the left vertical axis is the bright fraction corresponding to the left (red), right (blue), and center (black) measurements for the qubit (circles) and COM (squares) frequency trackers. The right vertical axis shows the tracked center microwave qubit (purple circles) and COM (green squares) frequencies, offset by their starting values of 124.022438607 GHz and 1.591254 MHz respectively. Estimated uncertainties suggest the center frequencies are tracked to $\sim\pm$1 and $\sim\pm$4 Hz for the COM and qubit frequencies respectively. Drifts of the center frequencies over the course of the scan suggest without active feedback errors of order 100 Hz would have occurred.}
\label{fig:tracking}
\end{figure*}

\clearpage
\begin{figure}[!ht]
\centering
\includegraphics[width = 0.45\textwidth]{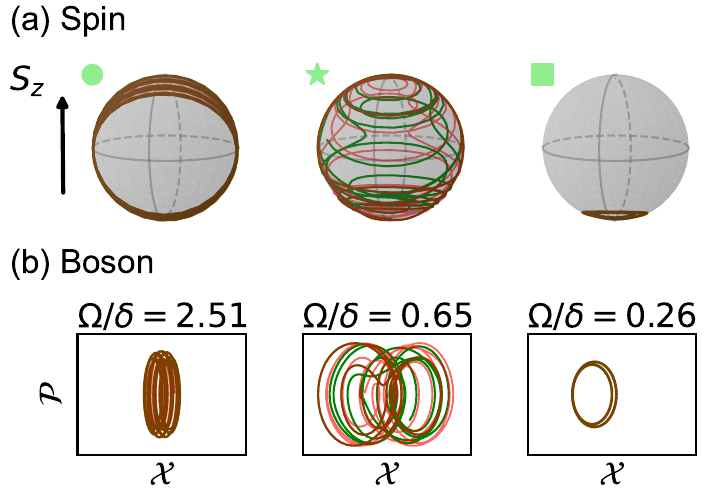}
\caption{Classical dynamical trajectories of (a) the spins on the collective Bloch sphere and (b) COM motion in the canonical phase space ($\cal{X}, \cal{P}$) along the ``chaotic cut". We plot two initially nearby trajectories in phase space (red and green curves) using the corresponding parameters in Fig.~\ref{fig:Fig3}a-c of the main text.}
\label{fig:chaotic_MFphasespace}
\end{figure}
\clearpage
\begin{figure}[!ht]
\centering
\includegraphics[width = 0.48\textwidth]{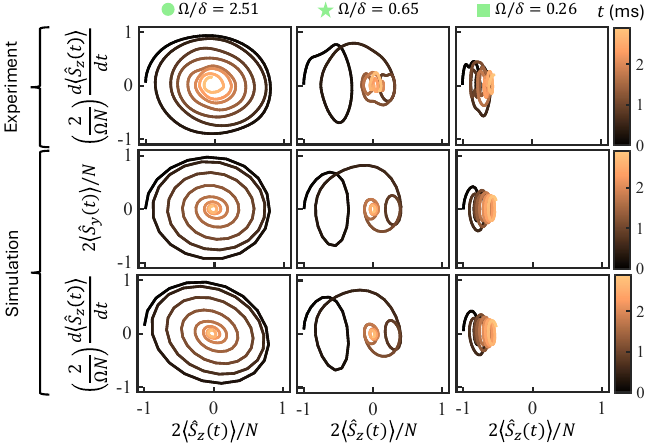}
\caption{Comparison of phase space plots. (Top) We plot the phase space plots from Fig.~\ref{fig:Fig3}, computed from the experimental data, where the y-axis is obtained from smoothing of the experimental $\braket{\hat{S}_z(t)}$ and subsequently taking the numerical derivative. (Middle) We compare to the corresponding semiclassical model, incorporating technical noise sources, where the y-axis is instead obtained directly from simulation results for $\braket{\hat{S}_y(t)}$. (Bottom) We also compare to the semiclassical model, where the y-axis is instead computed from the numerical derivative for $\braket{\hat{S}_z(t)}$; while expected to be exact for ideal unitary evolution under $\hat{H}_{\rm Dicke}$, up to errors arising from the finite timestep and statistical error from the number of trajectories, the presence of light scattering from the ODF beams invalidates any direct correspondence. Nonetheless, we observe fair agreement with our simulations to validate our approach.}
\label{fig:phase_space_comparison}
\end{figure}
\clearpage
\begin{figure*}[!ht]
\centering
\includegraphics[width = 0.9\textwidth]{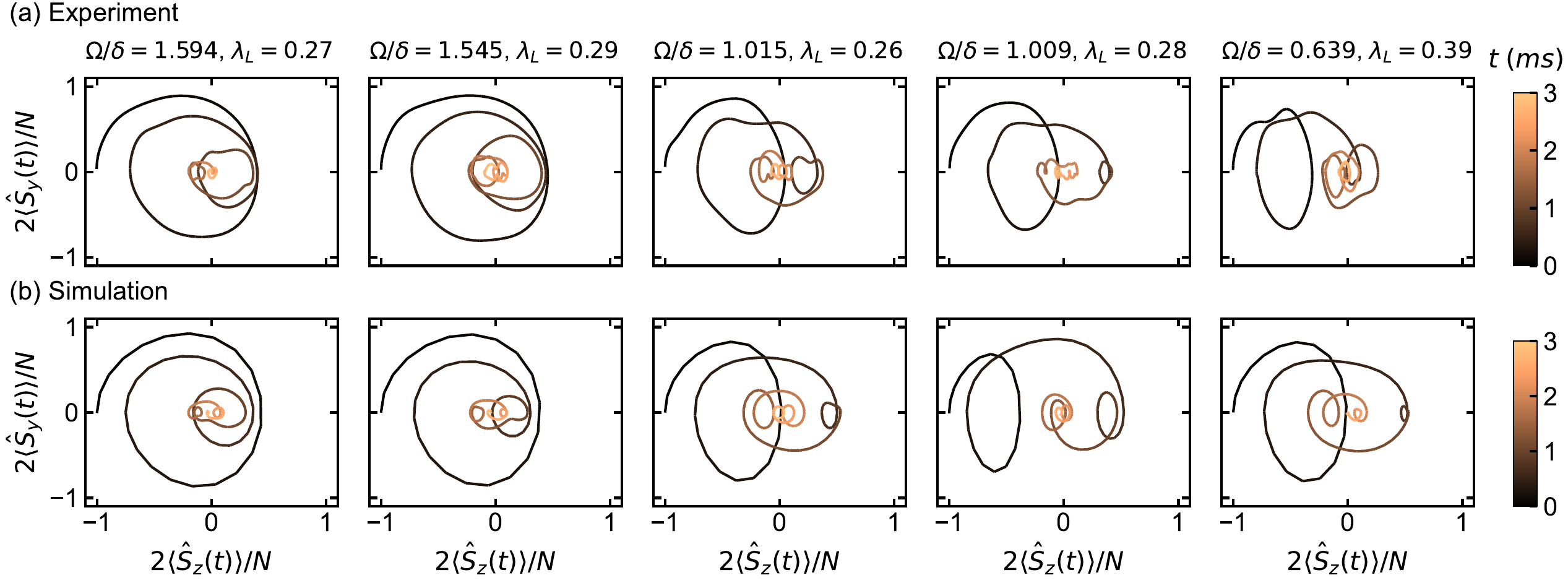}
\caption{  Additional phase space dynamics in the chaotic regime of the Dicke model. We plot (a) experimental and (b) semiclassical results for the case of initial EIT cooling and $-z$–polarized spins along the “chaotic cut” shown in Fig.~\ref{fig:Fig3}d of the main text. The corresponding classical Lyapunov exponent $\lambda_L$ is indicated in the figure. }
\label{fig:chaotic_spinphasespace}
\end{figure*}

\clearpage
\begin{figure*}[!ht]
\centering
\includegraphics[width = 0.9\textwidth]{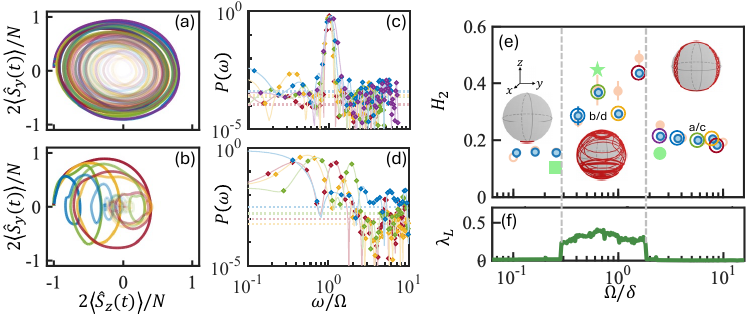}
\caption{
To highlight  the  correlation between the extracted experimental phase space curves,   the normalized power spectra, $P(\omega)$, the R{\'e}nyi spectral entropy, $H_2$, and the theoretically computed Lyapunov exponent  (see Methods)   we show all of them in the same figure.  Panels (a) and (b)  display overlapped experimental phase space curves  in the regular  and chaotic regimes, respectively,  along the ``chaotic cut'' of Fig.~\ref{fig:Fig3} in the main text,  with  curve transparency increasing with evolved time. Panels  (c) and (d) show the corresponding normalized power spectra $P(\omega)$ of  these curves versus  frequency scaled  by the transverse drive frequency. Only in the regular regime does $P(\omega)$ exhibit a single clear peak.  This behavior is quantified by the noise-floor-corrected spectral entropy $H_2$  shown  in  panel  (e) for all parameters shown in Fig.~\ref{fig:Fig3}d, with filled (unfilled) symbols corresponding to $N\in [51,108]$ ($N\in [159,178]$), and blue (faded orange) corresponding to initial EIT (Doppler) cooling. Circled data points and their colors correspond to the data plotted in a-d. $H_2$  is enhanced in the  regime where  the classically computed Lyapunov exponent $\lambda_L$, shown  in (f) (also shown in Fig.~\ref{fig:Fig3}f) is nonzero.  The grey dashed lines and Bloch sphere diagrams denote various classically predicted regimes for the  cut in Fig.~\ref{fig:Fig1}b-c. Error bars denote bootstrapped 68\% confidence intervals. Faded lines in c-d denote respective normalized noise floors for each power spectra.}
\label{fig:chaotic_spectra}
\end{figure*}
\clearpage
\begin{figure}[!ht]
\centering
\includegraphics[width = 0.4\textwidth]{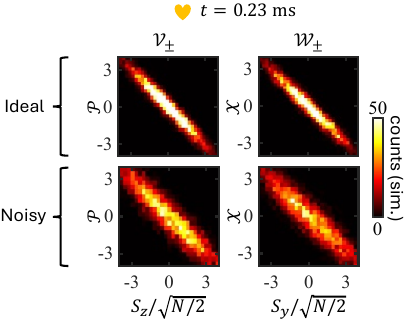}
\caption{Comparison of joint histograms for the squeezed/anti-squeezed mixed quadratures. (Top) We plot the histograms for the $\mathcal{V}_\pm$ (left) and $\mathcal{W}_\pm$ (right) quadratures obtained from ideal semiclassical evolution under $\hat{H}_{\rm Dicke}$, analogous to the result from Fig.~\ref{fig:Fig5}f computed in the presence of relevant technical noise sources, reiterated on the bottom left for comparison.}
\label{fig:squeezing_comparison}
\end{figure}
\clearpage
\begin{figure*}[!ht]
\centering
\includegraphics[width = 0.90\textwidth]{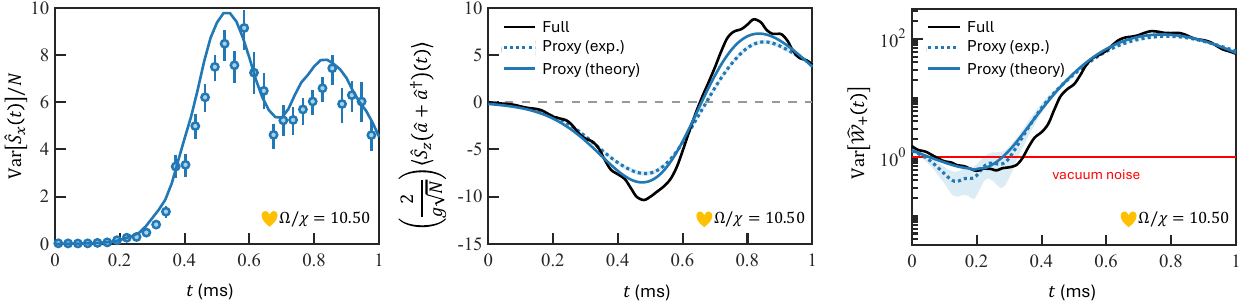}
\caption{{Reconstructed spin-phonon correlations. (Left) We compare the measured spin variance with the corresponding semiclassical estimate (blue solid) when all technical noise sources are included. Error bars denote 68\% confidence intervals. (Center) We show spin-phonon correlator dynamics from our semiclassical model (black solid), taking into account all technical noise sources. We also show the corresponding proxy estimate, extracted from the transverse magnetization dynamics in our semiclassical model (blue solid) as well as from the experimental results (blue dotted with blue shaded error region). (Right) Corresponding semiclassical result and corresponding semiclassical and experimental proxy estimates for the variance of the mixed spin-phonon $\hat{\mathcal{W}}_+$ quadrature, shown in Fig.~\ref{fig:Fig5}, and reiterated here to demonstrate the behavior of our proxy at later times. Blue shaded region denotes the bootstrapped 68\% confidence intervals for the experimental proxy, accounting for all post-processing used in our extraction procedure.}}
\label{fig:spinboson_correlation_proxy}
\end{figure*}

\clearpage

\clearpage

\clearpage

\clearpage

\clearpage
\setcounter{section}{0}
\setcounter{figure}{0}
\setcounter{equation}{0}
\renewcommand{\theHsection}{S\arabic{section}}
\renewcommand{\figseries}{S}
\renewcommand{\eqseries}{S}
\renewcommand{\figurename}{Supplementary Fig.}
\section*{Supplementary Information}
\renewcommand{\thesection}{S\arabic{section}}
\renewcommand{\theequation}{S\arabic{equation}}
Quantum simulation of the Dicke model in a two-dimensional ion crystal

\section{Theoretical model}

Here we describe our theoretical model, which captures relevant experimental details and noise sources. First, we account for the effects of spontaneous emission of the ions, which primarily affects our microwave qubit transition through its off-resonant coupling to the finite lifetime electronic states via the applied crossed optical beams used to engineer the spin-dependent ODF, see e.g.~\cite{uys_decoherence_2010,carter_comparison_2023}. For our spin states, this effectively manifests as two different types of scattering processes that we may account for via appropriate jump operators within a Lindblad master equation approach~\cite{uys_decoherence_2010,carter_comparison_2023}: an elastic Rayleigh scattering process, which can be accounted for by the collection of independent jump operators $\{\sqrt{\Gamma_{\rm z}/4}\hat{\sigma}_{z,j}\}$, and an inelastic Raman scattering process, which can be accounted for by the collection of independent jump operators $\{\sqrt{\Gamma_{\uparrow\downarrow}}\hat{\sigma}_{-,j}\}$ and $\{\sqrt{\Gamma_{\downarrow\uparrow}}\hat{\sigma}_{+,j}\}$. The relevant branching ratios of these different processes may be calculated from the relevant ODF and atomic parameters~\cite{uys_decoherence_2010,carter_comparison_2023}, and we take $\Gamma_{\uparrow\downarrow}/\Gamma_{\rm tot} = 0.12$, $\Gamma_{\downarrow\uparrow}/\Gamma_{\rm tot} = 0.08$, $\Gamma_{z}/\Gamma_{\rm tot} = 0.80$, where $\Gamma_{\rm tot} = \Gamma_z + \Gamma_{\uparrow\downarrow} + \Gamma_{\downarrow\uparrow} \sim 100-140$ s$^{-1}$ describes the decay of the transverse magnetization arising from the ODF beams in a Ramsey-type experiment: $\braket{\hat{S}_x(t)} = \braket{\hat{S}_x(0)} e^{-\Gamma_{\rm tot} t/2}$.

We also account for the role of low-frequency magnetic field fluctuations in the experiment. In principle, this can be modeled by including a global $z$-field, whose amplitude dynamically fluctuates with an appropriate power spectral density (PSD) matching that characterized in the experiment. Here, however, we opt for a simplified phenomenological model, where we split this noise into a quasi-static component, whose amplitude experiences shot-to-shot random variations described by a normal distribution $\mathcal{N}(0,\sigma_B^2)$, and a pure white-noise dynamical contribution, described by a jump operator $\sqrt{\Gamma_B} \hat{S}_z$ within a Lindblad master equation approach. The quasi-static/white-noise terms separately account for contributions to the magnetic field noise that fluctuate on timescales slow/fast compared to the relevant experimental timescale for each shot of the experiment. In order to choose the appropriate parameters for our model, we set $\Gamma_B$ and $\sigma_B$ to reproduce the expected Gaussian decay rates from a Ramsey-type experiment in the absence of the ODF beams. For an initial spin state $\ket{(N/2)_x}$, we fit the spin contrast decay as $2\braket{\hat{S}_x}/N = e^{-(\gamma_{\rm exp} t)^2/2}$, and find that using $\Gamma_B = 150$ s$^{-1}$ and $\sigma_B = 2\pi\times 45$ Hz in our model well reproduces the observed fitted gaussian decay rates{; that is, $\gamma_{\rm exp} = 100$ s$^{-1}$ ($350$ s$^{-1}$) with (without) an applied spin-echo during the decay.}

Finally, we account for the effects of previously studied quasi-static variations in the COM frequency~\cite{gilmore_quantum-enhanced_2021}, which we model as an effective shot-to-shot variation in the COM detuning, where we let $\delta \rightarrow \delta + \tilde{\delta}$ where $\tilde{\delta}$ is a random variable drawn from a normal distribution $\mathcal{N}(0,\sigma_{\delta}^2)$ for fixed $\sigma_{\delta} = 2\pi \times 40$ Hz.

In all, our dynamical evolution is described by the Lindblad master equation
\begin{widetext}
\begin{align}
 \frac{d}{dt} \hat{\rho}(t) = \frac{-i}{\hbar} \left[\hat{H}_{\rm tot},\hat{\rho}(t)\right] + \frac{\Gamma_z}{4} \sum_{j} \mathcal{D}[\hat{\sigma}_{z,j}]\hat{\rho}(t) + \Gamma_{\uparrow\downarrow} \sum_{j} \mathcal{D}[\hat{\sigma}_{-,j}]\hat{\rho}(t) + \Gamma_{\downarrow\uparrow} \sum_{j} \mathcal{D}[\hat{\sigma}_{+,j}]\hat{\rho}(t) + \Gamma_B \mathcal{D}[\hat{S}_z]\hat{\rho}(t)\label{eq:master_equation}
\end{align}
\end{widetext}
for Lindblad superoperators $\mathcal{D}[\hat{O}]\hat{\rho} \equiv \hat{O}\hat{\rho}\hat{O}^\dagger - \{\hat{O}^\dagger\hat{O},\hat{\rho}\}/2$ and anti-commutator $\{\cdot,\cdot\}$, and where the total Hamiltonian is defined by $\hat{H}_{\rm tot} = \hat{H}_{\rm Dicke} + \hat{H}_{\rm rand}$ with
\begin{align}
 \hat{H}_{\rm rand}/\hbar &= - \tilde{\delta}\hat{a}^\dagger\hat{a} + B\hat{S}_z,
\end{align}
where $\tilde{\delta}$ and $B$ are random variables drawn from the appropriate distributions described above. We then compute expectation values of observable operators $\hat{O}$ via $O(t)\equiv \mathbb{E}[{ \braket{\hat{O}(t)}}]$, where $\braket{\hat{O}(t)} = \Tr[\hat{O}\hat{\rho}(t)]$ and $\mathbb{E}[\cdot]$ denotes expectations with respect to the distributions for $\tilde{\delta}$ and $B$.
{
The resulting dynamics of relevant observables becomes:
\begin{widetext}
\begin{eqnarray}
 \frac{d}{dt}\langle \hat{S}_z\rangle &=&
 \Omega \langle \hat{S}_y \rangle
 {\color{Purple}\,-\, \left(\Gamma_{\uparrow\downarrow} + \Gamma_{\downarrow\uparrow}\right)\langle \hat{S}_z \rangle}
 {\color{Green}\,- \,N \left(\frac{\Gamma_{\uparrow\downarrow} - \Gamma_{\downarrow\uparrow}}{2}\right)},\label{Eqs} \\
\frac{d}{dt}\langle \hat{S}_y\rangle &=&
-\Omega \langle \hat{S}_z \rangle
+\frac{2g}{\sqrt{N}} \langle \hat{S}_x(\hat{a} + \hat{a}^\dagger)\rangle
{\color{Purple} \,-\, \left(\frac{\Gamma_z + \Gamma_{\uparrow\downarrow} + \Gamma_{\downarrow\uparrow} + \Gamma_B}{2} \right)\langle \hat{S}_y \rangle}
{\color{Green}\,+\, B \langle\hat{S}_x\rangle}, \nonumber \\
\frac{d}{dt}\langle \hat{S}_x \rangle &=&
-\frac{2g}{\sqrt{N}}\langle \hat{S}_y(\hat{a} + \hat{a}^\dagger)\rangle
{\color{Purple}\,-\, \left(\frac{\Gamma_z + \Gamma_{\uparrow\downarrow} + \Gamma_{\downarrow\uparrow} + \Gamma_B}{2} \right)\langle \hat{S}_x\rangle}
{\color{Green} \,-\, B \langle\hat{S}_y\rangle}, \\
\frac{d}{dt}\langle \hat{a}\rangle &=&
i(\delta + \tilde{\delta})\langle \hat{a}\rangle
-\frac{2ig}{\sqrt{N}}\langle \hat{S}_z\rangle.\label{eq:eom_end}
\end{eqnarray}
In the above equations, the dominant single-particle decoherence channels included in our model are shown in purple, whereas sub-leading single-particle decoherence mechanisms are highlighted in green.
\end{widetext}
}

For our initial state, as described in Fig.~\ref{fig:Fig1}(a) of the main text, we let $\hat{\rho}(0) = \hat{\rho}_\nu \equiv \hat{\rho}_{\rm th} \otimes \ket{(-N/2)_\nu}\bra{(-N/2)_\nu}$, where $\nu = z,x$ depending on the relevant spin state we consider, and $\hat{\rho}_{\rm th}$ is a thermal state of the phonons, given in a Fock state basis via
\begin{align}
 \hat{\rho}_{\rm th} \equiv \frac{1}{\mathcal{Z}}\sum_{n=0}^\infty e^{-\beta \hbar \omega_Z n} \ket{n}\bra{n},\quad \mathcal{Z} = \sum_{n=0}^\infty e^{-\beta \hbar \omega_Z n}
\end{align}
for inverse temperature $\beta \equiv 1/k_BT$, where $k_B$ is the Boltzmann factor and $T$ is the temperature. In terms of the initial thermal occupation $\overline{n} \equiv \Tr[\hat{\rho}(0)\hat{a}^\dagger\hat{a}]$, we have $\beta\hbar\omega_Z = \ln(1 + 1/\overline{n})$.

We make two comments regarding this model. First, note that since
\begin{align}
 \Big(\mathcal{D}[\hat{\sigma}_{-,j}] + \mathcal{D}[\hat{\sigma}_{+,j}]\Big)\hat{\rho}(t) &= \frac{1}{2}\Big(\mathcal{D}[\hat{\sigma}_{x,j}] + \mathcal{D}[\hat{\sigma}_{y,j}]\Big)\hat{\rho}(t),
\end{align}
then with $\Gamma_{\uparrow\downarrow} \geq \Gamma_{\downarrow\uparrow}$, we have
\begin{widetext}
\begin{align}
 \Big(\Gamma_{\uparrow\downarrow}\mathcal{D}[\hat{\sigma}_{-,j}] + \Gamma_{\downarrow\uparrow}\mathcal{D}[\hat{\sigma}_{+,j}]\Big)\hat{\rho}(t) &= \frac{\Gamma_{\downarrow\uparrow}}{2}\Big(\mathcal{D}[\hat{\sigma}_{x,j}] + \mathcal{D}[\hat{\sigma}_{y,j}]\Big)\hat{\rho}(t) + \left(\Gamma_{\uparrow\downarrow} - \Gamma_{\downarrow\uparrow}\right)\mathcal{D}[\hat{\sigma}_{-,j}]\hat{\rho}(t).\label{eq:jump_op_rewriting}
\end{align}
\end{widetext}
We use this formulation in our semiclassical model, outlined in Supplementary Information, section~\ref{app:ddtwa}.

Second, for the initial state $\hat{\rho}_z$, explored in Figs.~\ref{fig:Fig2},~\ref{fig:Fig3}, the dynamics of $\mathbb{E}[\braket{\hat{S}_z}]$ are invariant with respect to the sign of $\delta$. Likewise, for the initial state $\hat{\rho}_x$ utilized in Figs.~\ref{fig:Fig4} and \ref{fig:Fig5}, the dynamics of $\mathbb{E}[\braket{\hat{S}_x}]$ are also invariant with respect to inverting the sign of $\delta$ \emph{if} we also invert the sign of $\Omega$.

{
\subsection*{ Understanding the role of the sign of the detuning in the quantum dynamics}
{ For ideal Hamiltonian evolution, with the noise terms omitted,} at the level of the equations of motion, the effect of changing the sign of $\delta$ can be understood with a simple dynamical mapping. Defining
\begin{equation}
\hat {\mathcal{X}}={ (\hat a+\hat a^\dagger)/\sqrt{2}},\qquad
\hat {\mathcal{P}}={ i(\hat a^\dagger-\hat a)/\sqrt{2}},
\end{equation}
the Heisenberg equations of motion include
\begin{equation}
\dot{\hat {\mathcal{X}}}=-\delta \hat {\mathcal{P}},\qquad
\dot{\hat {\mathcal{P}}}=\delta \hat {\mathcal{X}}-{ \frac{2\sqrt{2}g}{\sqrt{N}}}\hat S_z,
\end{equation}
\begin{equation}
\dot{\hat S}_x=-{ \frac{2\sqrt{2}g}{\sqrt{N}}}\hat{\mathcal{X}}\hat S_y,\quad
\dot{\hat S}_y=-\Omega \hat S_z+{ \frac{2\sqrt{2}g}{\sqrt{N}}}\hat {\mathcal{X}}\hat S_x,\quad
\dot{\hat S}_z=\Omega \hat S_y .
\end{equation}
These equations are invariant under the transformation
\begin{equation}
\delta\rightarrow -\delta,\quad
\hat {\mathcal{P}}\rightarrow -\hat {\mathcal{P}},\quad
\hat S_\alpha\rightarrow -\hat S_\alpha (\alpha=x,y,z),\quad
\hat {\mathcal{X}}\rightarrow \hat {\mathcal{X}},
\end{equation}
with $\Omega$ and $g$ kept fixed.

Consequently, the dynamics at detuning $-\delta$ with the initial spin state inverted is mapped into the dynamics at detuning $+\delta$. For example, an initial state with all spins polarized along $+x$ at $+\delta$ is equivalent, under the above mapping, to an initial state with all spins polarized along $-x$ at $-\delta$. Similarly, an initial state polarized along $+z$ maps to one polarized along $-z$. The motional part of the initial state can be kept the same, provided the state is parity-symmetric in momentum, as is the case for the vacuum state or a thermal state with no coherent displacement.

Under this mapping, spin observables change sign,
\begin{equation}
\langle \hat S_\alpha(t)\rangle_{-\delta,-\mathbf{S}(0)}
=
-\langle \hat S_\alpha(t)\rangle_{+\delta,+\mathbf{S}(0)},
\end{equation}
and the momentum quadrature changes sign,
\begin{equation}
\langle \hat {\mathcal{P}}(t)\rangle_{-\delta,-\mathbf{S}(0)}
=
-\langle \hat {\mathcal{P}}(t)\rangle_{+\delta,+\mathbf{S}(0)}.
\end{equation}
In contrast, position-like and number-like observables are unchanged,
\begin{equation}
\langle \hat {\mathcal{X}}(t)\rangle_{-\delta,-\mathbf{S}(0)}
=
\langle \hat {\mathcal{X}}(t)\rangle_{+\delta,+\mathbf{S}(0)},
\end{equation}

\begin{equation}  \langle \hat a^\dagger \hat a(t)\rangle_{-\delta,-\mathbf{S}(0)}
=
\langle \hat a^\dagger \hat a(t)\rangle_{+\delta,+\mathbf{S}(0)}.
\end{equation}
{  The mapping is exact for ideal Hamiltonian evolution. Unequal Raman rates break this correspondence in the dissipative model; their effect is included explicitly in our calibrated simulations.}

This explains why changing the sign of $\delta$ does not lead to fundamentally new dynamics when accompanied by the corresponding inversion of the initial spin state; rather, it only changes the sign convention for spin and momentum-like observables.}

\section{Mean-field dynamics}
\label{app:MFT}
Let us define classical, mean-field (MF) variables $\tilde{\mathcal{S}}_\mu$ and $\tilde{\alpha}$, representing the expectation values of $\braket{\hat{S}_\mu}/N$ and $\braket{\hat{a}}/\sqrt{N}$, respectively. Then, for evolution described by Eq.~\eqref{eq:master_equation}, we have equations of motion within the MF approximation
\begin{widetext}
\begin{gather}
 \frac{d}{dt}\tilde{\mathcal{S}}_z = \Omega \tilde{\mathcal{S}}_y - \left(\Gamma_{\uparrow\downarrow} + \Gamma_{\downarrow\uparrow}\right)\tilde{\mathcal{S}}_z - \left(\frac{\Gamma_{\uparrow\downarrow} - \Gamma_{\downarrow\uparrow}}{2}\right),\label{eq:MFT_start}\\
{\frac{d}{dt}\tilde{\mathcal{S}}_y = -\Omega \tilde{\mathcal{S}}_z + 2g \tilde{\mathcal{S}}_x(\tilde{\alpha} + \tilde{\alpha}^*) + B \tilde{\mathcal{S}}_x - \left(\frac{\Gamma_z + \Gamma_{\uparrow\downarrow} + \Gamma_{\downarrow\uparrow} + \Gamma_B}{2} \right)\tilde{\mathcal{S}}_y} \\
 \frac{d}{dt}\tilde{\mathcal{S}}_x = -2g\tilde{\mathcal{S}}_y(\tilde{\alpha} + \tilde{\alpha}^*) - B \tilde{\mathcal{S}}_y - \left(\frac{\Gamma_z + \Gamma_{\uparrow\downarrow} + \Gamma_{\downarrow\uparrow} + \Gamma_B}{2} \right)\tilde{\mathcal{S}}_x,\\ \frac{d}{dt}\tilde{\alpha} = i(\delta+\tilde{\delta})\tilde{\alpha} - 2ig\tilde{\mathcal{S}}_z,\label{eq:MFT}
\end{gather}
\end{widetext}
for fixed $\tilde{\delta}$, $B$. At this level, the effects of collective and individual dephasing are indistinguishable from each other. The system size dependence also drops out from these equations in our normalization of the corresponding classical variables.

For the dynamics considered in the main text, $\braket{\hat{a}(0)}=\Tr[\hat{\rho}_{\rm th}\hat{a}] = 0$ and \begin{equation}\braket{\hat{S}_\mu(0)} = \braket{(-N/2)_\nu|\hat{S}_\mu|(-N/2)_\nu} = -N\delta_{\mu\nu}/2
\end{equation}for the initial state $\hat{\rho}_{\nu} = \hat{\rho}_{\rm th} \otimes \ket{(-N/2)_{\nu}} \bra{(-N/2)_\nu}$, with $\nu = z$ for Fig.~\ref{fig:Fig1}-\ref{fig:Fig3} or $\nu = x$ for Figs.~\ref{fig:Fig4} and \ref{fig:Fig5}. Thus, for our MF model of the dynamics, we evolve Eq.~\eqref{eq:MFT} with initial conditions $\tilde{\alpha} = 0$ and $\tilde{\mathcal{S}}_\mu = -\delta_{\mu\nu}/2$. To account for quasi-static noise in the evolution, we average all observables over an ensemble of $n_{\rm traj} = 10^3$ solutions---or trajectories---each computed for a different choice of $\tilde{\delta}$, $B$ drawn from their respective distributions, so that our MF prediction for the spin dynamics is given by $\mathbb{E}[\braket{\hat{S}_\mu}] \rightarrow N\times \mathbb{E}[\tilde{\mathcal{S}}_\mu]$. Where relevant, we also have the MF prediction for the phonon dynamics $\mathbb{E}[\hat{\mathcal{X}}] \rightarrow \sqrt{2N}\times\mathbb{E}[\rm{Re}\{\tilde{\alpha}\}]$ and
$\mathbb{E}[\hat{\mathcal{P}}] \rightarrow \sqrt{2N}\times\mathbb{E}[\rm{Im}\{\tilde{\alpha}\}]$ for $\hat{\mathcal{X}} = (\hat{a} + \hat{a}^\dagger)/\sqrt{2}$, $\hat{\mathcal{P}} = i(\hat{a}^\dagger - \hat{a})/\sqrt{2}$. We note that in Fig.~\ref{fig:Fig1}, as well as for the ``ideal'' results shown in Fig.~\ref{fig:Fig2}d, we take $\sigma_\delta$, $\sigma_B = 0$, as well as $\Gamma_z = \Gamma_{\uparrow\downarrow} = \Gamma_{\downarrow\uparrow} = \Gamma_{B} = 0$.

\section{Semiclassical dynamics}
\label{app:ddtwa}
For our semiclassical model, which accounts for beyond MF effects arising from both quantum and thermal fluctuations, we employ a discrete, dissipative version of the truncated Wigner approximation (TWA)~\cite{polkovnikov_phase_2010,barberena_fast_2024,huber_realistic_2022,schachenmayer_many-body_2015} that enables a treatment of both collective as well as individual dynamical noise sources in the spins. In essence, this approximation consists of evolving the MF equations of motion for $\hat{H}_{\rm tot}$ for $n_{\rm traj}$ random choices of our initial conditions, and averaging relevant observables over the ensemble of resulting dynamical trajectories. This sampling reproduces the effect of both thermal and quantum fluctuations in our initial state, and can provide an accurate approximation of these fluctuations in the dynamics.

Similar to Supplementary Information, section~\ref{app:MFT}, quasi-static noise may be accounted for by different choices of $B$ and $\tilde{\delta}$ for each trajectory. We also supplement the MF evolution for our Hamiltonian with stochastic terms, which properly account for the effects of dynamical noise modeled by our jump operators in the master equation. While Hermitian jump operators---such as those corresponding to $\Gamma_z$ and $\Gamma_B$---may be readily modeled by classical stochastic fields, spontaneous emission and absorption ($\Gamma_{\uparrow\downarrow},\Gamma_{\downarrow\uparrow}$) corresponding to an amplitude damping channel require a more elaborate approach. We first rewrite the relevant Lindblad terms as in Eq.~\eqref{eq:jump_op_rewriting}, where balanced contributions to decay and absorption may be equivalently modeled as independent $\hat{\sigma}_{x,j}$ and $\hat{\sigma}_{y,j}$ channels that are exactly modeled within our approach (up to statistical errors, and errors arising from the MF treatment of the unitary evolution). For the residual unbalanced evolution, proportional to $\Gamma_{\uparrow\downarrow} - \Gamma_{\downarrow\uparrow}$, we opt for an approximate approach outlined in Ref.~\cite{huber_realistic_2022}, which remains valid in the limit of small noise. In any case, as the main dynamical noise contributions arise from individual dephasing ($\Gamma_z/\Gamma_{\rm tot} = 0.80$), and the spontaneous decay/absorption rates are relatively balanced ($\Gamma_{\uparrow\downarrow}/\Gamma_{\rm tot} = 0.12$, $\Gamma_{\downarrow\uparrow}/\Gamma_{\rm tot} = 0.08$), we don't expect effects from this unbalanced decay, much less errors from our approximate treatment, to play any consequential role in the evolution.

Let $\mathcal{S}_{\mu,j}$ represent the expectation value of $\braket{\hat{\sigma}_{\mu,j}}/2$, and $\alpha$ represent the expectation value of $\braket{\hat{a}}$. Then, the relevant evolution equations for each trajectory are given by the Stratonovich stochastic differential equations
\begin{widetext}
\begin{gather}
\begin{split}
 d\mathcal{S}_{z,j} = \left[\Omega\mathcal{S}_{y,j} - \frac{3}{2}\left(\Gamma_{\uparrow\downarrow} - \Gamma_{\downarrow\uparrow}\right)\left(\mathcal{S}_{z,j} + 1/2\right)\right]dt + \sqrt{2\Gamma_{\downarrow\uparrow}}\left(\mathcal{S}_{y,j} dW_{x,j}(t) - \mathcal{S}_{x,j} dW_{y,j}(t)\right) \\
 + \sqrt{\Gamma_{\uparrow\downarrow} - \Gamma_{\downarrow\uparrow}}\left(\mathcal{S}_{z,j} + 1/2\right)dW_{d,j}(t)
 \end{split}\label{eq:ddtwa_sz}\\
 {\begin{split}d\mathcal{S}_{y,j} = \left[-\Omega\mathcal{S}_{z,j} + \frac{2g}{\sqrt{N}}\mathcal{S}_{x,j}\left(\alpha + \alpha^*\right) + B\mathcal{S}_{x,j}\right]dt - \sqrt{2\Gamma_{\downarrow\uparrow}}\mathcal{S}_{z,j} dW_{x,j}(t) + \sqrt{\Gamma_z}\mathcal{S}_{x,j} dW_{z,j}(t)\\ + \sqrt{\Gamma_B}\mathcal{S}_{x,j} dW_B(t) + \sqrt{\Gamma_{\uparrow\downarrow} - \Gamma_{\downarrow\uparrow}} \mathcal{S}_{x,j}dW_{d,j}(t)\end{split}}\label{eq:ddtwa_sy}\\
 \begin{split}d\mathcal{S}_{x,j} = \left[-\frac{2g}{\sqrt{N}}\mathcal{S}_{y,j}\left(\alpha + \alpha^*\right) - B\mathcal{S}_{y,j}\right]dt + \sqrt{2\Gamma_{\downarrow\uparrow}}\mathcal{S}_{z,j} dW_{y,j}(t) - \sqrt{\Gamma_z}\mathcal{S}_{y,j} dW_{z,j}(t)\\ - \sqrt{\Gamma_B}\mathcal{S}_{y,j} dW_B(t) - \sqrt{\Gamma_{\uparrow\downarrow} - \Gamma_{\downarrow\uparrow}}\mathcal{S}_{y,j}dW_{d,j}(t)\end{split}\label{eq:ddtwa_sx}\\
 d\alpha = i\left[\left(\delta + \tilde{\delta}\right)\alpha - \frac{2g}{\sqrt{N}}\sum_j\mathcal{S}_{z,j}\right]dt.\label{eq:ddtwa_alpha}
\end{gather}
\end{widetext}
Here, $dW_{\mu,j}(t)$ are independent, real Wiener increments for $\mu = x,y,z$ and $j = 1,...,N$, describing individual dephasing along each axis; $dW_B(t)$ is a real Wiener increment describing global dephasing; and $dW_{d,j}(t)$ for $j = 1,...,N$ are real, independent Wiener increments corresponding to residual imbalanced decay/absorption for each qubit. These stochastic Wiener increments are chosen independently for each trajectory and timestep such that $\mathbb{E}[dW(t)dW(t')] = dt \delta_{t,t'}$, where $t$, $t'$ denote timestep labels and $\mathbb{E}[\cdot]$ denotes averaging over all trajectories (and distinct Wiener increments are always chosen independently). For our semi-impl{i}cit solver, we utilize a typical timestep $dt = 2$ $\mu s$.

For initializing each trajectory, we set $\mathcal{S}_{\mu,j} = \pm 1/2$ with equal probability, independently for each $\mu\neq \nu$ and $j$, and set $\mathcal{S}_{\nu,j} = -1/2$ for initial state $\hat{\rho}_{\nu}$. Meanwhile, we choose $\Re\{\alpha\}$ and $\Im\{\alpha\}$ independently from a distribution $\mathcal{N}(0,\overline{n}/2 + 1/4)$; even as $\overline{n}\rightarrow0$, there is residual noise reflecting vacuum fluctuations of the phonon mode. We thus have that $\mathbb{E}[\alpha] = 0$, $\mathbb{E}[\alpha^*\alpha] = \overline{n} + 1/2$, where the latter quantity corresponds to the Weyl-ordered observable $\braket{(\hat{a}^\dagger\hat{a} + \hat{a}\hat{a}^\dagger)/2} = \braket{\hat{a}^\dagger\hat{a}} + 1/2$.

For the calculation of observables within this semiclassical approximation, we have the correspondence $\mathbb{E}[\braket{\hat{S}_\mu}] \rightarrow \sum_j \mathbb{E}[\mathcal{S}_{\mu,j}]$, $\mathbb{E}[\braket{\hat{\mathcal{X}}}] \rightarrow \sqrt{2}\times \mathbb{E}[\rm{Re}\{\alpha\}]$, and $\mathbb{E}[\braket{\hat{\mathcal{P}}}] \rightarrow \sqrt{2}\times \mathbb{E}[\rm{Im}\{\alpha\}]$. We can also obtain estimates for Weyl-ordered correlators of different observables, e.g. via the correspondence $\mathbb{E}[(\hat{S}_\mu\hat{S}_\nu + \hat{S}_\nu\hat{S}_\mu)/2] \rightarrow \sum_{j,j'}\mathbb{E}[\mathcal{S}_{\mu,j}\mathcal{S}_{\nu,j'}]$ and $\mathbb{E}[(\hat{a}^\dagger\hat{a} + \hat{a}\hat{a}^\dagger)/2] \rightarrow \mathbb{E}[\alpha^*\alpha]$, and with analogous equations for mixed spin-phonon observables and correlators. Here, $\mathbb{E}$ denotes averaging over all random trajectories, which incorporates averaging over all initial conditions and stochastic noise sources for our semiclassical model, as well as quasi-static noise terms in the Hamiltonian for both the semiclassical model and the full evolution.

In Supplementary Fig.~\ref{fig:noise_sources_spin},~\ref{fig:noise_sources_boson}, we utilize this semiclassical model to examine the effect of different noise sources in the evolution. We select parameters relevant to Figs.~\ref{fig:Fig3}-\ref{fig:Fig5} in the main text, for which we have closely examined the role of thermal noise. We note that for both sets of initial conditions, the main deviations of our results from the unitary dynamics are accounted for by magnetic field noise (Supplementary Fig.~\ref{fig:noise_sources_spin}, yellow dotted line). Light scattering from the ODF beams (Supplementary Fig.~\ref{fig:noise_sources_spin}, orange dotted line) and initial thermal occupation of the phonons (Supplementary Fig.~\ref{fig:noise_sources_boson}, green dotted line) play a noticeable but less prominent role, particularly for the $x$ polarized initial state when operating in the near-integrable regime. In all cases, COM noise (Supplementary Fig.~\ref{fig:noise_sources_boson}, pink dotted line) plays virtually no role in the observed spin dynamics, so far as we are independently considering the effects of different noise sources within our model. Magnetic field noise can be mitigated in future experiments by floating the optical table~\cite{britton_vibration-induced_2016,Gilmore_thesis_2020}, or by performing a spin-echo sequence compatible with the Dicke Hamiltonian. Furthermore, increased ODF power and stronger microwave drives can speed up the dynamics relative to the magnetic field decoherence rates.

\begin{figure}[H]
\centering
\includegraphics[width = 0.90\textwidth]{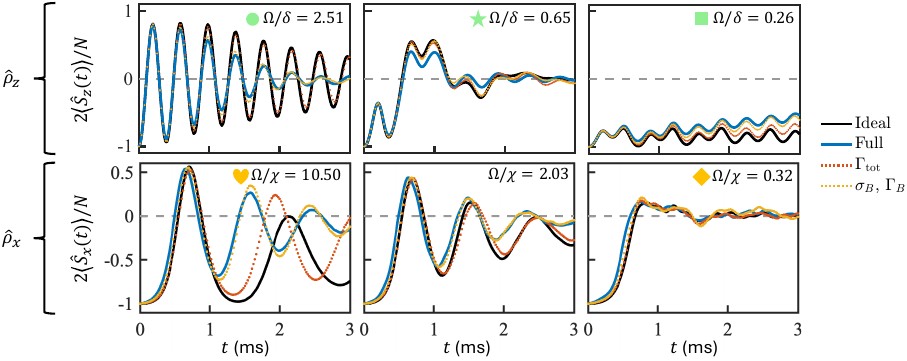}
\caption{Characterization of noise sources on spin degree of freedom in semiclassical model. (Top) We re-plot the results for the EIT cooled initial conditions from Fig.~\ref{fig:Fig3}, showing the results of our semiclassical model when different noise sources are included. We show our full results (blue solid), as well as the ideal semiclassical model (only including effects of $\hat{H}_{\rm Dicke}$, with an initial vacuum state for the phonons; black solid). Other dotted lines show results of ideal evolution in addition to either light scattering from the ODF beams characterized by the total rate $\Gamma_{\rm tot}$ (orange dotted), or quasi-static collective magnetic field noise, with field amplitude $B$ chosen from a normal distribution $\mathcal{N}(0,\sigma_B^2)$ and dynamical collective magnetic field noise (with a white-noise spectrum) with decay rate $\Gamma_B$ (yellow dotted). (Bottom) We also show analogous results corresponding to the results from Figs.~\ref{fig:Fig4} in the main text.}
\label{fig:noise_sources_spin}
\end{figure}
\begin{figure}[H]
\centering
\includegraphics[width = 0.90\textwidth]{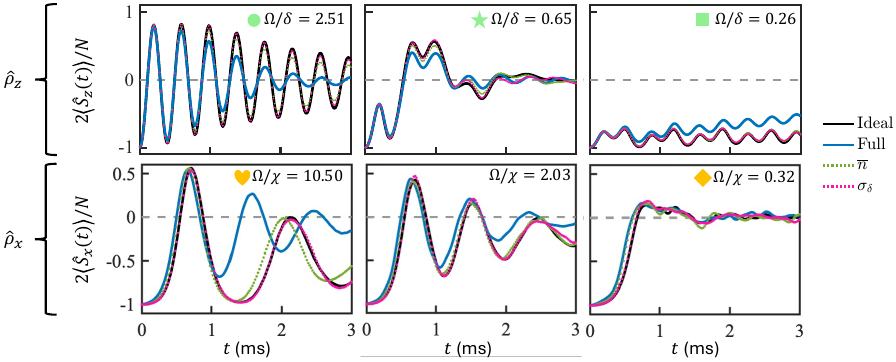}
\caption{Characterization of noise sources on phonon degree of freedom in semiclassical model. Analogous plot to Supplementary Fig.~\ref{fig:noise_sources_spin}, where we instead show results of ideal evolution in addition to either initial thermal noise in the phonons, with initial occupancy $n_0 \sim 0.5$ (green dotted), or quasi-static COM noise, with $\delta \rightarrow \delta + \tilde{\delta}$ for $\tilde{\delta}$ chosen from a normal distribution $\mathcal{N}(0,\sigma_{\delta}^2)$ (pink dotted). Full results with all noise sources (blue solid) as well as the ideal semiclassical model (black solid) are also plotted.}
\label{fig:noise_sources_boson}
\end{figure}

\begin{figure}[H]
\centering
\includegraphics[width = 0.90\textwidth]{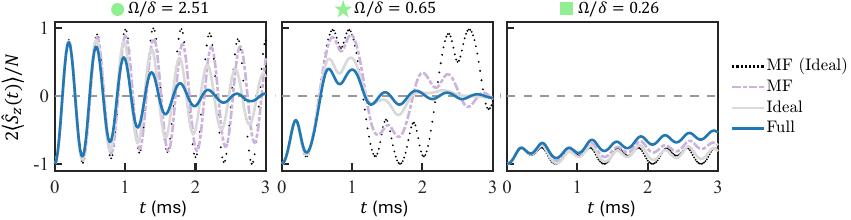}
\caption{Comparison of ``chaotic cut'' with and without technical noise sources. We re-plot the results for the EIT cooled initial conditions from Fig.~\ref{fig:Fig3}, showing the results of our semiclassical (full blue solid) and MF (lavender dashed-dot) models when all technical noise sources are included. We also show the corresponding ideal semiclassical (``Ideal'' gray solid) and MF dynamics (``Ideal'' black dotted).}
\label{fig:noise_comparison_chaotic}
\end{figure}

In Fig.~\ref{fig:Fig4}c of the main text, we closely examined the effects of thermal noise on the $x$ polarized initial state, suggesting that the observed dynamics are largely driven by quantum fluctuations in the initial state. However, we also note from Supplementary Fig.~\ref{fig:noise_comparison_chaotic} that for our $z$ polarized initial state in the chaotic (middle panel, denoted by the green star) and untrapped phases (left panel, denoted by the green circle), the ideal solutions, which neglect the effects of thermal noise appear to differ considerably from the MF solution shown in the main text (see Fig.~\ref{fig:Fig3}), suggesting that even here, quantum fluctuations play a dominant role in the observed damping compared to the classical solution.

For the analysis in Fig.~\ref{fig:Fig4}c, we showed results for a classical model where we take into account the effects of varying levels of thermal noise in the phonons. This model is essentially the semiclassical model outlined here, except that we initialize $\mathcal{S}_{\mu,j}$ at its MF values for each trajectory (i.e. $\bs{\mathcal{S}} = (-1/2,0,0)$), and draw $\Re\{\alpha\}$ and $\Im\{\alpha\}$ independently from a distribution $\mathcal{N}(0,\overline{n}/2)$. Thus, for $\overline{n}\rightarrow 0$, we recover the MF model up to our stochastic treatment of the individual decoherence processes. In this way, we ensure the \emph{only} differences between the TWA solution and our classical, thermal solutions are the absence of projection noise in the spins and the absence of vacuum noise in the phonons.

\section{Holstein-Primakoff {Approximation}}
For evolution under $\hat{H}_{\rm Dicke}$ in the resonant case with $\Omega = \delta$, let us define rotated spin ladder operators, $\hat{S}^{(x)}_{\pm} \equiv \hat{S}_y \pm i \hat{S}_z$, such that $\hat{S}_z = -i(\hat{S}^{(x)}_+ - \hat{S}^{(x)}_-)/2$, $\hat{S}_y = (\hat{S}^{(x)}_+ + \hat{S}^{(x)}_-)/2$, and which act on eigenstates of $\hat{S}_x$ as ${ \hat S_\pm^{(x)}} \ket{(m)_x} \propto \ket{(m\pm 1)_x}$. Thus, we may write
\begin{align}
 \hat{H}_{\rm Dicke}/\hbar = -\delta\hat{a}^\dagger\hat{a} + \delta \hat{S}_x + \frac{-ig}{\sqrt{N}}\left(\hat{a} + \hat{a}^\dagger\right)\left(\hat{S}_{+}^{(x)} - \hat{S}_{-}^{(x)}\right).
\end{align}
In the rotating frame defined by $\hat{H}_0/\hbar = -\delta\hat{a}^\dagger\hat{a} + \delta\hat{S}_x$, we obtain the Hamiltonian $\hat{H}^{\rm rot} = \hat{H}^{\rm rot}_{{\rm pair}} + \hat{H}^{\rm rot}_{\rm osc}$ where
\begin{widetext}
\begin{align}
 \hat{H}^{\rm rot}_{{\rm pair}} \equiv \hbar \frac{-ig}{\sqrt{N}}\left(\hat{a}^\dagger\hat{S}_{+}^{(x)} - \hat{a}\hat{S}_{-}^{(x)}\right),\quad \hat{H}^{\rm rot}_{{\rm osc}} \equiv \hbar \frac{-ig}{\sqrt{N}}\left(\hat{a}\hat{S}_{+}^{(x)}e^{2i\delta t} - \hat{a}^\dagger\hat{S}_{-}^{(x)}e^{-2i\delta t}\right).
\end{align}
\end{widetext}
As $\delta \rightarrow \infty$, we may neglect the effects of $\hat{H}_{\rm osc}^{\rm rot}$ in the rotating wave approximation, whose terms rapidly oscillate, averaging to $0$. For the remaining $\hat{H}_{\rm pair}^{\rm rot}$, we can then employ a Holstein-Primakoff (HP) approximation, valid for states with $(m)_x \sim -N/2$, where let ${ \hat S_+^{(x)}} \approx \sqrt{N}\hat{b}^\dagger$, ${ \hat S_-^{(x)}} \approx \sqrt{N}\hat{b}$ for bosonic creation/annihilation operators $\hat{b}^\dagger\hat{b}$ representing the collective spin mode. Then,
\begin{align}
 \hat{H}_{\rm pair}^{\rm rot} \approx -i\hbar g\left(\hat{a}^\dagger\hat{b}^\dagger - \hat{a}\hat{b}\right).
\end{align}
Now, let us define the mixed quadrature operators $\hat{c}_{\pm} = (\hat{a}\pm i\hat{b})/\sqrt{2}$, so that with $\hat{\mathcal{X}}_\pm = (\hat{c}_{\pm} + \hat{c}_{\pm}^\dagger)/\sqrt{2}$ and $\hat{\mathcal{P}}_\pm = i(\hat{c}_{\pm}^\dagger - \hat{c}_{\pm})/\sqrt{2}$ we have
\begin{align}
 \hat{H}_{\rm pair}^{\rm rot} \approx \frac{-\hbar g}{2}\left(\hat{\mathcal{P}}_+^2 - \hat{\mathcal{X}}_+^2 - \hat{\mathcal{P}}_-^2 + \hat{\mathcal{X}}_-^2\right).\label{eq:Hrot_invertedmass}
\end{align}
The $\hat{c}_{\pm}$ modes thus correspond to two independent, inverted mass oscillators. For the initial state $\hat{\rho}_x$ with $\overline{n} = 0$, this corresponds to the joint vacuum state $\ket{0}_{a}\otimes \ket{0}_b$ for our bosonic degrees of freedom within the Holstein-Primakoff approximation. Evolved under the dynamics of Eq.~\eqref{eq:Hrot_invertedmass}, this results in the squeezing of the mixed quadratures $\hat{\mathcal{P}}_\pm \pm \hat{\mathcal{X}}_\pm$, as well as the anti-squeezing of the canonically conjugate quadratures $\hat{\mathcal{P}}_\pm \mp \hat{\mathcal{X}}_\pm$. Thus, any linear combination of these two squeezed/anti-squeezed quadratures will likewise be squeezed/anti-squeezed.

In terms of our original phonon and collective spin operators, the simplest such combinations are
\begin{align*}
\hat{\mathcal V}_{\pm} &= \frac{(\hat{\mathcal P}_-\mp\hat{\mathcal X}_-)+(\hat{\mathcal P}_+\pm\hat{\mathcal X}_+)}{\sqrt2}
\approx \hat{\mathcal P}\pm\frac{\hat S_z}{\sqrt{N/2}},\\
{\hat{\mathcal W}_{\pm}} &{=\mp\frac{(\hat{\mathcal P}_-\pm\hat{\mathcal X}_-)-(\hat{\mathcal P}_+\mp\hat{\mathcal X}_+)}{\sqrt2}}
\approx\hat{\mathcal X}\pm\frac{\hat S_y}{\sqrt{N/2}}.
\end{align*}
The variances $\mathrm{Var}(\hat{\mathcal V}_+)$ and $\mathrm{Var}(\hat{\mathcal W}_+)$ decrease exponentially below their vacuum values of $1$.

In Extended Data Fig.~\ref{fig:squeezing_comparison}, we show joint histograms for the various joint spin-phonon quadratures, analogous to Fig.~\ref{fig:Fig5}f of the main text. These are computed in our semiclassical model outlined in Supplementary Information, section~\ref{app:ddtwa} by plotting the relevant histogram for our ensemble of dynamical trajectories. On the top, we show quadrature distributions without noise, for which the evolution at short times is relatively similar, up to correlation/anti-correlation of the relevant quadratures. The presence of technical noise generally leads to a reduction in the overall amount of squeezing for each quadrature. However, even in ideal case, we find that the $\hat{\mathcal{V}}_+$ quadrature (left) remains more squeezed than the $\hat{\mathcal{W}}_+$ quadrature (right)---an asymmetry arising from the curvature of the Bloch sphere and breakdown of the HP approximation.

\section{Fluctuation-driven vacuum Rabi dynamics}
\subsection{Vacuum Rabi oscillations}\label{sec:vacuum_rabi}
The distinction between the oscillatory dynamics observed in the collective magnetization $\langle \hat{S}_x (t)\rangle$ and conventional single-particle Rabi oscillations can be understood directly from the equations of motion shown in Eqs.~\eqref{Eqs}--\eqref{eq:eom_end}.

The terms proportional to $\Omega$ correspond to ordinary single-particle spin precession. In contrast, the terms proportional to $g$ couple the spin dynamics to spin-phonon correlations. For the initial state used in our experiment, in which the bosonic mode is---to great approximation---in vacuum and the spin is polarized along the $x$ direction, the Hamiltonian has a parity symmetry under
\begin{equation}
\hat{a}\rightarrow -\hat{a},\qquad
\hat{S}_y\rightarrow -\hat{S}_y,\qquad
\hat{S}_z\rightarrow -\hat{S}_z,
\qquad
\hat{S}_x\rightarrow \hat{S}_x .
\end{equation}
Consequently,
\begin{equation}
\langle \hat{a}\rangle=\langle \hat{S}_y\rangle=\langle \hat{S}_z\rangle=0
\end{equation}
remain zero during the ideal evolution. The dynamics of $\langle \hat{S}_x\rangle$ is therefore governed entirely by the connected spin-phonon correlator
\begin{equation}
\frac{d}{dt}\langle \hat{S}_x\rangle =
\frac{-2g}{\sqrt{N}}
\langle \hat{S}_y(\hat{a}+\hat{a}^\dagger)\rangle .
\end{equation}

At the strict mean-field level (see Eqs.~\eqref{eq:MFT_start}--\eqref{eq:MFT}) this correlator factorizes as
\begin{equation}
\langle \hat{S}_y(\hat{a}+\hat{a}^\dagger)\rangle
\rightarrow
\langle \hat{S}_y\rangle\langle \hat{a}+\hat{a}^\dagger\rangle=0 ,
\end{equation}
and the dynamics vanish identically. Thus, in the ideal vacuum case, the oscillatory dynamics of $\langle \hat{S}_x\rangle$ is fluctuation-driven and originates from the buildup of spin-phonon correlations seeded by vacuum fluctuations. This is the sense in which we use the term ``vacuum-Rabi oscillations.''

This interpretation must be qualified in the presence of decoherence and thermal occupation. The dominant single-particle decoherence channels included in our model, shown in purple in Eqs.~\eqref{Eqs}--\eqref{eq:eom_end}, modify decay rates and frequency shifts, but preserve the relevant parity symmetry. As a result, starting from the vacuum state, these dominant terms do not generate nonzero $\langle \hat{a}\rangle$, $\langle \hat{S}_y\rangle$, or $\langle \hat{S}_z\rangle$, and therefore do not convert the dynamics into ordinary one-body Rabi oscillations. They damp or renormalize the fluctuation-driven dynamics, but do not by themselves seed the short-time oscillations.

Thermal fluctuations play a different role. Even when $\langle \hat{a}\rangle=0$, a thermal state has enhanced quadrature variance. { For the initially factorized, $x$-polarized thermal state, the ideal Hamiltonian gives
\begin{equation}
\left.\frac{d}{dt}\langle\hat S_y(\hat a+\hat a^\dagger)\rangle\right|_{t=0}
=\frac{2g}{\sqrt N}\langle\hat S_x(0)\rangle(2\overline n+1).
\end{equation}
The remaining contributions to this derivative vanish in the initial state.
}
Thus, sufficiently large thermal fluctuations can generate spin-phonon correlations and drive oscillations in $\langle \hat{S}_x\rangle$, even though the one-body averages remain zero.

Sub-leading single-particle decoherence mechanisms, shown in green in Eqs.~\eqref{Eqs}--\eqref{eq:eom_end}, weakly break the ideal symmetry. These terms can generate small nonzero one-body fields, such as $\langle \hat{a}\rangle$, $\langle \hat{S}_y\rangle$, or $\langle \hat{S}_z\rangle$. When combined with thermal noise, they can seed oscillatory behavior at longer times; we thus do not claim that the long-time dynamics are purely vacuum driven.

The simulations included in the main text make this distinction explicit. For the calibrated decoherence rates, the mean-field dynamics in Fig.~4a remain nearly flat at short times, and the observed short-time oscillations are instead consistent with fluctuation-driven spin-phonon correlations. In contrast, reproducing comparable oscillations within a thermal-seeded model requires thermal fluctuations substantially larger than those independently characterized in the experiment.

This claim is further supported by our theoretical estimates for the spin dynamics along each axis, which shows that the resulting oscillatory behavior is not explained by Rabi oscillations of our spin degree of freedom, see Supplementary Fig.~\ref{fig:spin_comps_unstable}. We further show the corresponding variances in the two rightmost panels. In the mean-field trajectories (lavender dashed-dot lines), the dynamics of $\langle \hat{S}_x\rangle$ is activated by noise from the symmetry breaking single-particle decoherence mechanisms. This is clear since the magnetization growth at longer times is directly correlated with the growth of the $S_y$ and $S_z$ variances. The symmetry-broken noise sources are nevertheless subdominant at short times at which genuine quantum fluctuations dominate for the experimental parameters in consideration.
The fast oscillations at frequency $\propto 2\Omega$ in the variance in the MF theory are generated by the fast counter-rotating terms, and are heavily damped in the corresponding quantum system.

\begin{figure}[H]
\centering
\includegraphics[width = 0.9\textwidth]{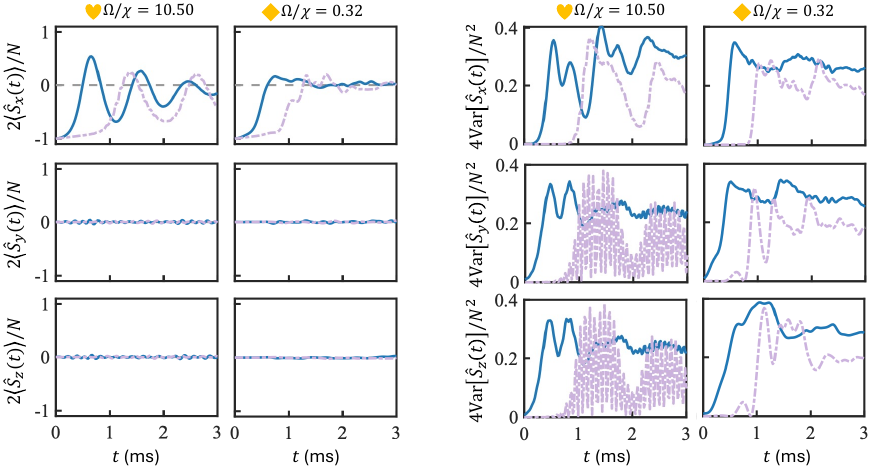}
\caption{{Additional spin components for the ``resonant cut''. Theoretical estimates for the spin dynamics for the initial conditions in Fig.~\ref{fig:Fig4}, showing the results of our semiclassical (blue solid) and MF (lavender dashed-dot) models when all technical noise sources are included. We include the mean spin (left), as well as the corresponding variance (right).}}
\label{fig:spin_comps_unstable}
\end{figure}

\renewcommand{\bibsection}{\section*{Supplementary references}}


\begin{thebibliography}{85}
\providecommand{\natexlab}[1]{#1}
\providecommand{\url}[1]{\texttt{#1}}
\expandafter\ifx\csname urlstyle\endcsname\relax
  \providecommand{\doi}[1]{doi: #1}\else
  \providecommand{\doi}{doi: \begingroup \urlstyle{rm}\Url}\fi

\bibitem[Marino et~al.(2022)Marino, Eckstein, Foster, and
  Rey]{marino_dynamical_2022}
Jamir Marino, Martin Eckstein, Matthew~S Foster, and Ana~Maria Rey.
\newblock Dynamical phase transitions in the collisionless pre-thermal states
  of isolated quantum systems: Theory and experiments.
\newblock \emph{Reports on Progress in Physics}, 85\penalty0 (11):\penalty0
  116001, October 2022.
\newblock ISSN 0034-4885, 1361-6633.
\newblock \doi{10.1088/1361-6633/ac906c}.

\bibitem[{Lewis-Swan} et~al.(2019){Lewis-Swan}, {Safavi-Naini}, Bollinger, and
  Rey]{lewis-swan_unifying_2019}
R.~J. {Lewis-Swan}, A.~{Safavi-Naini}, J.~J. Bollinger, and A.~M. Rey.
\newblock Unifying scrambling, thermalization and entanglement through
  measurement of fidelity out-of-time-order correlators in the {{Dicke}} model.
\newblock \emph{Nature Communications}, 10\penalty0 (1):\penalty0 1581, April
  2019.
\newblock ISSN 2041-1723.
\newblock \doi{10.1038/s41467-019-09436-y}.

\bibitem[Dicke(1954)]{dicke_coherence_1954}
R.~H. Dicke.
\newblock Coherence in {{Spontaneous Radiation Processes}}.
\newblock \emph{Physical Review}, 93\penalty0 (1):\penalty0 99--110, January
  1954.
\newblock \doi{10.1103/PhysRev.93.99}.

\bibitem[Hepp and Lieb(1973)]{hepp_superradiant_1973}
Klaus Hepp and Elliott~H Lieb.
\newblock On the superradiant phase transition for molecules in a quantized
  radiation field: The {Dicke} maser model.
\newblock \emph{Annals of Physics}, 76\penalty0 (2):\penalty0 360--404, April
  1973.
\newblock ISSN 00034916.
\newblock \doi{10.1016/0003-4916(73)90039-0}.

\bibitem[Barberena et~al.(2024)Barberena, Muleady, Bollinger, {Lewis-Swan}, and
  Rey]{barberena_fast_2024}
Diego Barberena, Sean~R. Muleady, John~J. Bollinger, Robert~J. {Lewis-Swan},
  and Ana~Maria Rey.
\newblock Fast generation of spin squeezing via resonant spin-boson coupling.
\newblock \emph{Quantum Science and Technology}, 9\penalty0 (2):\penalty0
  025013, February 2024.
\newblock ISSN 2058-9565.
\newblock \doi{10.1088/2058-9565/ad2186}.

\bibitem[Zhang et~al.(2025)Zhang, Zu{\~n}iga~Castro, and
  {Lewis-Swan}]{zhang_harnessing_2025}
Y.~Zhang, J.~C. Zu{\~n}iga~Castro, and R.~J. {Lewis-Swan}.
\newblock Harnessing quantum chaos in spin-boson models for all-purpose
  quantum-enhanced sensing.
\newblock \emph{Physical Review Research}, 7\penalty0 (1):\penalty0 013227,
  February 2025.
\newblock ISSN 2643-1564.
\newblock \doi{10.1103/PhysRevResearch.7.013227}.

\bibitem[{Lewis-Swan} et~al.(2024){Lewis-Swan}, Castro, Barberena, and
  Rey]{lewis-swan_exploiting_2024}
R.~J. {Lewis-Swan}, J.~C.~Zu{\~n}iga Castro, D.~Barberena, and A.~M. Rey.
\newblock Exploiting {{Nonclassical Motion}} of a {{Trapped Ion Crystal}} for
  {{Quantum-Enhanced Metrology}} of {{Global}} and {{Differential Spin
  Rotations}}.
\newblock \emph{Physical Review Letters}, 132\penalty0 (16):\penalty0 163601,
  April 2024.
\newblock ISSN 0031-9007, 1079-7114.
\newblock \doi{10.1103/PhysRevLett.132.163601}.

\bibitem[Emary and Brandes(2003{\natexlab{a}})]{emary_chaos_2003}
Clive Emary and Tobias Brandes.
\newblock Chaos and the quantum phase transition in the {{Dicke}} model.
\newblock \emph{Physical Review E}, 67\penalty0 (6):\penalty0 066203, June
  2003{\natexlab{a}}.
\newblock \doi{10.1103/PhysRevE.67.066203}.

\bibitem[Emary and Brandes(2003{\natexlab{b}})]{emary_quantum_2003}
Clive Emary and Tobias Brandes.
\newblock Quantum {{Chaos Triggered}} by {{Precursors}} of a {{Quantum Phase
  Transition}}: {{The Dicke Model}}.
\newblock \emph{Physical Review Letters}, 90\penalty0 (4):\penalty0 044101,
  January 2003{\natexlab{b}}.
\newblock \doi{10.1103/PhysRevLett.90.044101}.

\bibitem[Brandes(2013)]{brandes_excited-state_2013}
Tobias Brandes.
\newblock Excited-state quantum phase transitions in {{Dicke}} superradiance
  models.
\newblock \emph{Physical Review E}, 88\penalty0 (3):\penalty0 032133, September
  2013.
\newblock \doi{10.1103/PhysRevE.88.032133}.

\bibitem[Klinder et~al.(2015)Klinder, Ke{\ss}ler, Wolke, Mathey, and
  Hemmerich]{klinder_dynamical_2015}
Jens Klinder, Hans Ke{\ss}ler, Matthias Wolke, Ludwig Mathey, and Andreas
  Hemmerich.
\newblock Dynamical phase transition in the open {{Dicke}} model.
\newblock \emph{Proceedings of the National Academy of Sciences}, 112\penalty0
  (11):\penalty0 3290--3295, March 2015.
\newblock ISSN 0027-8424, 1091-6490.
\newblock \doi{10.1073/pnas.1417132112}.

\bibitem[Black et~al.(2003)Black, Chan, and
  Vuleti{\'c}]{black_observation_2003}
Adam~T. Black, Hilton~W. Chan, and Vladan Vuleti{\'c}.
\newblock Observation of {{Collective Friction Forces}} due to {{Spatial
  Self-Organization}} of {{Atoms}}: {{From Rayleigh}} to {{Bragg Scattering}}.
\newblock \emph{Physical Review Letters}, 91\penalty0 (20):\penalty0 203001,
  November 2003.
\newblock \doi{10.1103/PhysRevLett.91.203001}.

\bibitem[Domokos and Ritsch(2003)]{domokos_mechanical_2003}
Peter Domokos and Helmut Ritsch.
\newblock Mechanical effects of light in optical resonators.
\newblock \emph{Journal of the Optical Society of America B}, 20\penalty0
  (5):\penalty0 1098, May 2003.
\newblock ISSN 0740-3224, 1520-8540.
\newblock \doi{10.1364/JOSAB.20.001098}.

\bibitem[Baumann et~al.(2010)Baumann, Guerlin, Brennecke, and
  Esslinger]{baumann_dicke_2010}
Kristian Baumann, Christine Guerlin, Ferdinand Brennecke, and Tilman Esslinger.
\newblock Dicke quantum phase transition with a superfluid gas in an optical
  cavity.
\newblock \emph{Nature}, 464\penalty0 (7293):\penalty0 1301--1306, April 2010.
\newblock ISSN 1476-4687.
\newblock \doi{10.1038/nature09009}.

\bibitem[Nagy et~al.(2010)Nagy, K{\'o}nya, Szirmai, and
  Domokos]{nagy_dicke-model_2010}
D.~Nagy, G.~K{\'o}nya, G.~Szirmai, and P.~Domokos.
\newblock Dicke-{{Model Phase Transition}} in the {{Quantum Motion}} of a
  {{Bose-Einstein Condensate}} in an {{Optical Cavity}}.
\newblock \emph{Physical Review Letters}, 104\penalty0 (13):\penalty0 130401,
  April 2010.
\newblock \doi{10.1103/PhysRevLett.104.130401}.

\bibitem[Brennecke et~al.(2013)Brennecke, Mottl, Baumann, Landig, Donner, and
  Esslinger]{brennecke_real-time_2013}
Ferdinand Brennecke, Rafael Mottl, Kristian Baumann, Renate Landig, Tobias
  Donner, and Tilman Esslinger.
\newblock Real-time observation of fluctuations at the driven-dissipative
  {{Dicke}} phase transition.
\newblock \emph{Proceedings of the National Academy of Sciences}, 110\penalty0
  (29):\penalty0 11763--11767, July 2013.
\newblock \doi{10.1073/pnas.1306993110}.

\bibitem[L{\'e}onard et~al.(2017)L{\'e}onard, Morales, Zupancic, Esslinger, and
  Donner]{leonard_supersolid_2017}
Julian L{\'e}onard, Andrea Morales, Philip Zupancic, Tilman Esslinger, and
  Tobias Donner.
\newblock Supersolid formation in a quantum gas breaking a continuous
  translational symmetry.
\newblock \emph{Nature}, 543\penalty0 (7643):\penalty0 87--90, March 2017.
\newblock ISSN 1476-4687.
\newblock \doi{10.1038/nature21067}.

\bibitem[Koll{\'a}r et~al.(2017)Koll{\'a}r, Papageorge, Vaidya, Guo, Keeling,
  and Lev]{kollar_supermode-density-wave-polariton_2017}
Alicia~J. Koll{\'a}r, Alexander~T. Papageorge, Varun~D. Vaidya, Yudan Guo,
  Jonathan Keeling, and Benjamin~L. Lev.
\newblock Supermode-density-wave-polariton condensation with a
  {{Bose}}--{{Einstein}} condensate in a multimode cavity.
\newblock \emph{Nature Communications}, 8\penalty0 (1):\penalty0 14386,
  February 2017.
\newblock ISSN 2041-1723.
\newblock \doi{10.1038/ncomms14386}.

\bibitem[Landini et~al.(2018)Landini, Dogra, Kroeger, Hruby, Donner, and
  Esslinger]{landini_formation_2018}
M.~Landini, N.~Dogra, K.~Kroeger, L.~Hruby, T.~Donner, and T.~Esslinger.
\newblock Formation of a {{Spin Texture}} in a {{Quantum Gas Coupled}} to a
  {{Cavity}}.
\newblock \emph{Physical Review Letters}, 120\penalty0 (22):\penalty0 223602,
  May 2018.
\newblock \doi{10.1103/PhysRevLett.120.223602}.

\bibitem[Morales et~al.(2019)Morales, Dreon, Li, Baumg{\"a}rtner, Zupancic,
  Donner, and Esslinger]{morales_two-mode_2019}
Andrea Morales, Davide Dreon, Xiangliang Li, Alexander Baumg{\"a}rtner, Philip
  Zupancic, Tobias Donner, and Tilman Esslinger.
\newblock Two-mode {{Dicke}} model from nondegenerate polarization modes.
\newblock \emph{Physical Review A}, 100\penalty0 (1):\penalty0 013816, July
  2019.
\newblock \doi{10.1103/PhysRevA.100.013816}.

\bibitem[Yan et~al.(2023)Yan, Ho, Lu, Masson, {Asenjo-Garcia}, and
  {Stamper-Kurn}]{yan_superradiant_2023}
Zhenjie Yan, Jacquelyn Ho, Yue-Hui Lu, Stuart~J. Masson, Ana {Asenjo-Garcia},
  and Dan~M. {Stamper-Kurn}.
\newblock Superradiant and {{Subradiant Cavity Scattering}} by {{Atom Arrays}}.
\newblock \emph{Physical Review Letters}, 131\penalty0 (25):\penalty0 253603,
  December 2023.
\newblock \doi{10.1103/PhysRevLett.131.253603}.

\bibitem[Yoshihara et~al.(2017)Yoshihara, Fuse, Ashhab, Kakuyanagi, Saito, and
  Semba]{yoshihara_superconducting_2017}
Fumiki Yoshihara, Tomoko Fuse, Sahel Ashhab, Kosuke Kakuyanagi, Shiro Saito,
  and Kouichi Semba.
\newblock Superconducting qubit--oscillator circuit beyond the
  ultrastrong-coupling regime.
\newblock \emph{Nature Physics}, 13\penalty0 (1):\penalty0 44--47, January
  2017.
\newblock ISSN 1745-2481.
\newblock \doi{10.1038/nphys3906}.

\bibitem[{Forn-D{\'i}az} et~al.(2019){Forn-D{\'i}az}, Lamata, Rico, Kono, and
  Solano]{forn-diaz_ultrastrong_2019}
P.~{Forn-D{\'i}az}, L.~Lamata, E.~Rico, J.~Kono, and E.~Solano.
\newblock Ultrastrong coupling regimes of light-matter interaction.
\newblock \emph{Reviews of Modern Physics}, 91\penalty0 (2):\penalty0 025005,
  June 2019.
\newblock ISSN 0034-6861, 1539-0756.
\newblock \doi{10.1103/RevModPhys.91.025005}.

\bibitem[Todorov et~al.(2010)Todorov, Andrews, Colombelli, De~Liberato, Ciuti,
  Klang, Strasser, and Sirtori]{todorov_ultrastrong_2010}
Y.~Todorov, A.~M. Andrews, R.~Colombelli, S.~De~Liberato, C.~Ciuti, P.~Klang,
  G.~Strasser, and C.~Sirtori.
\newblock Ultrastrong {{Light-Matter Coupling Regime}} with {{Polariton Dots}}.
\newblock \emph{Physical Review Letters}, 105\penalty0 (19):\penalty0 196402,
  November 2010.
\newblock \doi{10.1103/PhysRevLett.105.196402}.

\bibitem[Smerzi et~al.(1997)Smerzi, Fantoni, Giovanazzi, and
  Shenoy]{smerzi_quantum_1997}
A.~Smerzi, S.~Fantoni, S.~Giovanazzi, and S.~R. Shenoy.
\newblock Quantum {{Coherent Atomic Tunneling}} between {{Two Trapped
  Bose-Einstein Condensates}}.
\newblock \emph{Physical Review Letters}, 79\penalty0 (25):\penalty0
  4950--4953, December 1997.
\newblock \doi{10.1103/PhysRevLett.79.4950}.

\bibitem[Albiez et~al.(2005)Albiez, Gati, F{\"o}lling, Hunsmann, Cristiani, and
  Oberthaler]{albiez_direct_2005}
Michael Albiez, Rudolf Gati, Jonas F{\"o}lling, Stefan Hunsmann, Matteo
  Cristiani, and Markus~K. Oberthaler.
\newblock Direct {{Observation}} of {{Tunneling}} and {{Nonlinear
  Self-Trapping}} in a {{Single Bosonic Josephson Junction}}.
\newblock \emph{Physical Review Letters}, 95\penalty0 (1):\penalty0 010402,
  June 2005.
\newblock \doi{10.1103/PhysRevLett.95.010402}.

\bibitem[Abbarchi et~al.(2013)Abbarchi, Amo, Sala, Solnyshkov, Flayac, Ferrier,
  Sagnes, Galopin, Lema{\^i}tre, Malpuech, and
  Bloch]{abbarchi_macroscopic_2013}
M.~Abbarchi, A.~Amo, V.~G. Sala, D.~D. Solnyshkov, H.~Flayac, L.~Ferrier,
  I.~Sagnes, E.~Galopin, A.~Lema{\^i}tre, G.~Malpuech, and J.~Bloch.
\newblock Macroscopic quantum self-trapping and {{Josephson}} oscillations of
  exciton polaritons.
\newblock \emph{Nature Physics}, 9\penalty0 (5):\penalty0 275--279, May 2013.
\newblock ISSN 1745-2481.
\newblock \doi{10.1038/nphys2609}.

\bibitem[Reinhard et~al.(2013)Reinhard, Riou, Zundel, Weiss, Li, Rey, and
  Hipolito]{reinhard_self-trapping_2013}
Aaron Reinhard, Jean-F{\'e}lix Riou, Laura~A. Zundel, David~S. Weiss, Shuming
  Li, Ana~Maria Rey, and Rafael Hipolito.
\newblock Self-{{Trapping}} in an {{Array}} of {{Coupled 1D Bose Gases}}.
\newblock \emph{Physical Review Letters}, 110\penalty0 (3):\penalty0 033001,
  January 2013.
\newblock \doi{10.1103/PhysRevLett.110.033001}.

\bibitem[Zhang et~al.(2017)Zhang, Pagano, Hess, Kyprianidis, Becker, Kaplan,
  Gorshkov, Gong, and Monroe]{zhang_observation_2017}
J.~Zhang, G.~Pagano, P.~W. Hess, A.~Kyprianidis, P.~Becker, H.~Kaplan, A.~V.
  Gorshkov, Z.-X. Gong, and C.~Monroe.
\newblock Observation of a many-body dynamical phase transition with a 53-qubit
  quantum simulator.
\newblock \emph{Nature}, 551\penalty0 (7682):\penalty0 601--604, November 2017.
\newblock ISSN 1476-4687.
\newblock \doi{10.1038/nature24654}.

\bibitem[Borish et~al.(2020)Borish, Markovi{\'c}, Hines, Rajagopal, and
  {Schleier-Smith}]{borish_transverse-field_2020}
V.~Borish, O.~Markovi{\'c}, J.~A. Hines, S.~V. Rajagopal, and
  M.~{Schleier-Smith}.
\newblock Transverse-{{Field Ising Dynamics}} in a {{Rydberg-Dressed Atomic
  Gas}}.
\newblock \emph{Physical Review Letters}, 124\penalty0 (6):\penalty0 063601,
  February 2020.
\newblock ISSN 0031-9007, 1079-7114.
\newblock \doi{10.1103/PhysRevLett.124.063601}.

\bibitem[Muniz et~al.(2020)Muniz, Barberena, {Lewis-Swan}, Young, Cline, Rey,
  and Thompson]{muniz_exploring_2020}
Juan~A. Muniz, Diego Barberena, Robert~J. {Lewis-Swan}, Dylan~J. Young, Julia
  R.~K. Cline, Ana~Maria Rey, and James~K. Thompson.
\newblock Exploring dynamical phase transitions with cold atoms in an optical
  cavity.
\newblock \emph{Nature}, 580\penalty0 (7805):\penalty0 602--607, April 2020.
\newblock ISSN 0028-0836, 1476-4687.
\newblock \doi{10.1038/s41586-020-2224-x}.

\bibitem[Li et~al.(2023)Li, Colombo, Shu, Velez, {Pilatowsky-Cameo}, Schmied,
  Choi, Lukin, {Pedrozo-Pe{\~n}afiel}, and Vuleti{\'c}]{li_improving_2023}
Zeyang Li, Simone Colombo, Chi Shu, Gustavo Velez, Sa{\'u}l {Pilatowsky-Cameo},
  Roman Schmied, Soonwon Choi, Mikhail Lukin, Edwin {Pedrozo-Pe{\~n}afiel}, and
  Vladan Vuleti{\'c}.
\newblock Improving metrology with quantum scrambling.
\newblock \emph{Science}, 380\penalty0 (6652):\penalty0 1381--1384, June 2023.
\newblock ISSN 0036-8075, 1095-9203.
\newblock \doi{10.1126/science.adg9500}.

\bibitem[Bollinger et~al.(2013)Bollinger, Britton, and
  Sawyer]{bollinger_simulating_2013}
J.~J. Bollinger, J.~W. Britton, and B.~C. Sawyer.
\newblock Simulating quantum magnetism with correlated non-neutral ion plasmas.
\newblock In \emph{{{NON-NEUTRAL PLASMA PHYSICS VIII}}: 10th {{International
  Workshop}} on {{Non-Neutral Plasmas}}}, pages 200--209, Greifswald, Germany,
  2013.
\newblock \doi{10.1063/1.4796076}.

\bibitem[Sawyer et~al.(2014)Sawyer, Britton, and Bollinger]{sawyer_spin_2014}
Brian~C. Sawyer, Joseph~W. Britton, and John~J. Bollinger.
\newblock Spin dephasing as a probe of mode temperature, motional state
  distributions, and heating rates in a two-dimensional ion crystal.
\newblock \emph{Physical Review A}, 89\penalty0 (3):\penalty0 033408, March
  2014.
\newblock ISSN 1050-2947, 1094-1622.
\newblock \doi{10.1103/PhysRevA.89.033408}.

\bibitem[Bohnet et~al.(2016)Bohnet, Sawyer, Britton, Wall, Rey, {Foss-Feig},
  and Bollinger]{bohnet_quantum_2016}
J.~G. Bohnet, B.~C. Sawyer, J.~W. Britton, M.~L. Wall, A.~M. Rey,
  M.~{Foss-Feig}, and J.~J. Bollinger.
\newblock Quantum spin dynamics and entanglement generation with hundreds of
  trapped ions.
\newblock \emph{Science}, 352\penalty0 (6291):\penalty0 1297--1301, June 2016.
\newblock ISSN 1095-9203 (Electronic) 0036-8075 (Linking).
\newblock \doi{10.1126/science.aad9958}.

\bibitem[Jordan et~al.(2019)Jordan, Gilmore, Shankar, {Safavi-Naini}, Bohnet,
  Holland, and Bollinger]{jordan_near_2019}
Elena Jordan, Kevin~A. Gilmore, Athreya Shankar, Arghavan {Safavi-Naini},
  Justin~G. Bohnet, Murray~J. Holland, and John~J. Bollinger.
\newblock Near {{Ground-State Cooling}} of {{Two-Dimensional Trapped-Ion
  Crystals}} with {{More}} than 100 {{Ions}}.
\newblock \emph{Physical Review Letters}, 122\penalty0 (5):\penalty0 053603,
  February 2019.
\newblock ISSN 0031-9007, 1079-7114.
\newblock \doi{10.1103/PhysRevLett.122.053603}.

\bibitem[{Safavi-Naini} et~al.(2018){Safavi-Naini}, {Lewis-Swan}, Bohnet,
  G{\"a}rttner, Gilmore, Jordan, Cohn, Freericks, Rey, and
  Bollinger]{safavi-naini_verification_2018}
A.~{Safavi-Naini}, R.~J. {Lewis-Swan}, J.~G. Bohnet, M.~G{\"a}rttner, K.~A.
  Gilmore, J.~E. Jordan, J.~Cohn, J.~K. Freericks, A.~M. Rey, and J.~J.
  Bollinger.
\newblock Verification of a {{Many-Ion Simulator}} of the {{Dicke Model Through
  Slow Quenches}} across a {{Phase Transition}}.
\newblock \emph{Physical Review Letters}, 121\penalty0 (4):\penalty0 040503,
  July 2018.
\newblock \doi{10.1103/PhysRevLett.121.040503}.

\bibitem[Kirton et~al.(2019)Kirton, Roses, Keeling, and
  Dalla~Torre]{kirton_introduction_2019}
Peter Kirton, Mor~M. Roses, Jonathan Keeling, and Emanuele~G. Dalla~Torre.
\newblock Introduction to the {{Dicke Model}}: {{From Equilibrium}} to
  {{Nonequilibrium}}, and {{{\emph{Vice Versa}}}}.
\newblock \emph{Advanced Quantum Technologies}, 2\penalty0 (1-2):\penalty0
  1800043, February 2019.
\newblock ISSN 25119044.
\newblock \doi{10.1002/qute.201800043}.

\bibitem[Schir{\'o} and Fabrizio(2010)]{schiro_time-dependent_2010}
Marco Schir{\'o} and Michele Fabrizio.
\newblock Time-{{Dependent Mean Field Theory}} for {{Quench Dynamics}} in
  {{Correlated Electron Systems}}.
\newblock \emph{Physical Review Letters}, 105\penalty0 (7):\penalty0 076401,
  August 2010.
\newblock \doi{10.1103/PhysRevLett.105.076401}.

\bibitem[Sciolla and Biroli(2010)]{sciolla_quantum_2010}
Bruno Sciolla and Giulio Biroli.
\newblock Quantum {{Quenches}} and {{Off-Equilibrium Dynamical Transition}} in
  the {{Infinite-Dimensional Bose-Hubbard Model}}.
\newblock \emph{Physical Review Letters}, 105\penalty0 (22):\penalty0 220401,
  November 2010.
\newblock \doi{10.1103/PhysRevLett.105.220401}.

\bibitem[Gambassi and Calabrese(2011)]{gambassi_quantum_2011}
A.~Gambassi and P.~Calabrese.
\newblock Quantum quenches as classical critical films.
\newblock \emph{Europhysics Letters}, 95\penalty0 (6):\penalty0 66007,
  September 2011.
\newblock ISSN 0295-5075.
\newblock \doi{10.1209/0295-5075/95/66007}.

\bibitem[Smacchia et~al.(2015)Smacchia, Knap, Demler, and
  Silva]{smacchia_exploring_2015}
Pietro Smacchia, Michael Knap, Eugene Demler, and Alessandro Silva.
\newblock Exploring dynamical phase transitions and prethermalization with
  quantum noise of excitations.
\newblock \emph{Physical Review B}, 91\penalty0 (20):\penalty0 205136, May
  2015.
\newblock \doi{10.1103/PhysRevB.91.205136}.

\bibitem[{Lewis-Swan} et~al.(2021){Lewis-Swan}, Muleady, Barberena, Bollinger,
  and Rey]{lewis-swan_characterizing_2021}
R.~J. {Lewis-Swan}, S.~R. Muleady, D.~Barberena, J.~J. Bollinger, and A.~M.
  Rey.
\newblock Characterizing the dynamical phase diagram of the {{Dicke}} model via
  classical and quantum probes.
\newblock \emph{Physical Review Research}, 3\penalty0 (2):\penalty0 L022020,
  June 2021.
\newblock ISSN 2643-1564.
\newblock \doi{10.1103/PhysRevResearch.3.L022020}.

\bibitem[Lipkin et~al.(1965)Lipkin, Meshkov, and Glick]{lipkin_validity_1965}
H.~J. Lipkin, N.~Meshkov, and A.~J. Glick.
\newblock Validity of many-body approximation methods for a solvable model:
  ({{I}}). {{Exact}} solutions and perturbation theory.
\newblock \emph{Nuclear Physics}, 62\penalty0 (2):\penalty0 188--198, February
  1965.
\newblock ISSN 0029-5582.
\newblock \doi{10.1016/0029-5582(65)90862-X}.

\bibitem[Ribeiro et~al.(2007)Ribeiro, Vidal, and
  Mosseri]{ribeiro_thermodynamical_2007}
Pedro Ribeiro, Julien Vidal, and R{\'e}my Mosseri.
\newblock Thermodynamical {{Limit}} of the {{Lipkin-Meshkov-Glick Model}}.
\newblock \emph{Physical Review Letters}, 99\penalty0 (5):\penalty0 050402,
  August 2007.
\newblock \doi{10.1103/PhysRevLett.99.050402}.

\bibitem[Lerose et~al.(2019)Lerose, {\v Z}unkovi{\v c}, Marino, Gambassi, and
  Silva]{lerose_impact_2019}
Alessio Lerose, Bojan {\v Z}unkovi{\v c}, Jamir Marino, Andrea Gambassi, and
  Alessandro Silva.
\newblock Impact of nonequilibrium fluctuations on prethermal dynamical phase
  transitions in long-range interacting spin chains.
\newblock \emph{Physical Review B}, 99\penalty0 (4):\penalty0 045128, January
  2019.
\newblock \doi{10.1103/PhysRevB.99.045128}.

\bibitem[Wall et~al.(2017)Wall, {Safavi-Naini}, and
  Rey]{wall_boson-mediated_2017}
Michael~L. Wall, Arghavan {Safavi-Naini}, and Ana~Maria Rey.
\newblock Boson-mediated quantum spin simulators in transverse fields:
  ${{XY}}$ model and spin-boson entanglement.
\newblock \emph{Physical Review A}, 95\penalty0 (1):\penalty0 013602, January
  2017.
\newblock \doi{10.1103/PhysRevA.95.013602}

\bibitem[{Ch{\'a}vez-Carlos} et~al.(2016){Ch{\'a}vez-Carlos},
  {Bastarrachea-Magnani}, {Lerma-Hern{\'a}ndez}, and
  Hirsch]{chavez-carlos_classical_2016}
J.~{Ch{\'a}vez-Carlos}, M.~A. {Bastarrachea-Magnani}, S.~{Lerma-Hern{\'a}ndez},
  and J.~G. Hirsch.
\newblock Classical chaos in atom-field systems.
\newblock \emph{Physical Review E}, 94\penalty0 (2):\penalty0 022209, August
  2016.
\newblock \doi{10.1103/PhysRevE.94.022209}.

\bibitem[Lerose et~al.(2018)Lerose, Marino, {\v Z}unkovi{\v c}, Gambassi, and
  Silva]{lerose_chaotic_2018}
Alessio Lerose, Jamir Marino, Bojan {\v Z}unkovi{\v c}, Andrea Gambassi, and
  Alessandro Silva.
\newblock Chaotic {{Dynamical Ferromagnetic Phase Induced}} by {{Nonequilibrium
  Quantum Fluctuations}}.
\newblock \emph{Physical Review Letters}, 120\penalty0 (13):\penalty0 130603,
  March 2018.
\newblock \doi{10.1103/PhysRevLett.120.130603}.

\bibitem[Altland and Haake(2012)]{altland_equilibration_2012}
Alexander Altland and Fritz Haake.
\newblock Equilibration and macroscopic quantum fluctuations in the {{Dicke}}
  model.
\newblock \emph{New Journal of Physics}, 14\penalty0 (7):\penalty0 073011, July
  2012.
\newblock ISSN 1367-2630.
\newblock \doi{10.1088/1367-2630/14/7/073011}.

\bibitem[Strogatz(2014)]{strogatz_nonlinear_2014}
S.H. Strogatz.
\newblock \emph{Nonlinear Dynamics and Chaos: With Applications to Physics,
  Biology, Chemistry, and Engineering}.
\newblock Studies in Nonlinearity. Avalon Publishing, 2014.
\newblock ISBN 9780813349114.
\newblock URL \url{https://books.google.com/books?id=JDQGAwAAQBAJ}.

\bibitem[Schachenmayer et~al.(2015)Schachenmayer, Pikovski, and
  Rey]{schachenmayer_many-body_2015}
J.~Schachenmayer, A.~Pikovski, and A.~M. Rey.
\newblock Many-{{Body Quantum Spin Dynamics}} with {{Monte Carlo Trajectories}}
  on a {{Discrete Phase Space}}.
\newblock \emph{Physical Review X}, 5\penalty0 (1):\penalty0 011022, 2015.
\newblock ISSN 2160-3308.
\newblock \doi{10.1103/PhysRevX.5.011022}.

\bibitem[Holstein and Primakoff(1940)]{holstein_field_1940}
T.~Holstein and H.~Primakoff.
\newblock Field {{Dependence}} of the {{Intrinsic Domain Magnetization}} of a
  {{Ferromagnet}}.
\newblock \emph{Physical Review}, 58\penalty0 (12):\penalty0 1098--1113,
  December 1940.
\newblock ISSN 0031-899X.
\newblock \doi{10.1103/PhysRev.58.1098}.


\bibitem[R{\'e}nyi(1961)]{renyi_measures_1961}
Alfr{\'e}d R{\'e}nyi.
\newblock On {{Measures}} of {{Entropy}} and {{Information}}.
\newblock \emph{Berkeley Symposium on Mathematical Statistics and Probability},
  4\penalty0 (1):\penalty0 547--561, 1961.

\bibitem[Cheng et~al.(2020)Cheng, Liu, Guo, Chen, Zhang, and Zhai]{Cheng2020}
Yanting Cheng, Chang Liu, Jinkang Guo, Yu~Chen, Pengfei Zhang, and Hui Zhai.
\newblock Realizing the {Hayden}-{Preskill} protocol with coupled {Dicke} models.
\newblock \emph{Phys. Rev. Res.}, 2\penalty0 (4):\penalty0 043024, Oct 2020.
\newblock \doi{10.1103/PhysRevResearch.2.043024}.
\newblock URL \url{https://link.aps.org/doi/10.1103/PhysRevResearch.2.043024}.

\bibitem[Vikram et~al.(2026)Vikram, Chaparro, Khan, Lucas, Akers, and
  Rey]{vikram2026bidirectional}
Amit Vikram, Edwin Chaparro, Muhammad~Miskeen Khan, Andrew Lucas, Chris Akers,
  and Ana~Maria Rey.
\newblock Bidirectional teleportation using scrambling dynamics: a practical
  protocol, 2026.
\newblock URL \url{https://arxiv.org/abs/2601.15536}.

\bibitem[Gietka(2024)]{gietka_vacuum_2024-1}
Karol Gietka.
\newblock Vacuum {{Rabi}} splitting as a manifestation of virtual two-mode
  squeezing: {{Extracting}} the squeezing parameters from frequency shifts.
\newblock \emph{Physical Review A}, 110\penalty0 (6):\penalty0 063703, December
  2024.
\newblock ISSN 2469-9926, 2469-9934.
\newblock \doi{10.1103/PhysRevA.110.063703}.


\bibitem[Lerose and Pappalardi(2020)]{lerose_bridging_2020}
Alessio Lerose and Silvia Pappalardi.
\newblock Bridging entanglement dynamics and chaos in semiclassical systems.
\newblock \emph{Physical Review A}, 102\penalty0 (3):\penalty0 032404, September 2020.
\newblock ISSN 2469-9926, 2469-9934.
\newblock \doi{10.1103/PhysRevA.102.032404}

\bibitem[Gilmore et~al.(2021)Gilmore, Affolter, {Lewis-Swan}, Barberena,
  Jordan, Rey, and Bollinger]{gilmore_quantum-enhanced_2021}
Kevin~A. Gilmore, Matthew Affolter, Robert~J. {Lewis-Swan}, Diego Barberena,
  Elena Jordan, Ana~Maria Rey, and John~J. Bollinger.
\newblock Quantum-enhanced sensing of displacements and electric fields with
  two-dimensional trapped-ion crystals.
\newblock \emph{Science}, 373\penalty0 (6555):\penalty0 673--678, August 2021.
\newblock ISSN 0036-8075, 1095-9203.
\newblock \doi{10.1126/science.abi5226}.

\bibitem[Gilmore et~al.(2017)Gilmore, Bohnet, Sawyer, Britton, and
  Bollinger]{gilmore_amplitude_2017}
K.~A. Gilmore, J.~G. Bohnet, B.~C. Sawyer, J.~W. Britton, and J.~J. Bollinger.
\newblock Amplitude {{Sensing}} below the {{Zero-Point Fluctuations}} with a
  {{Two-Dimensional Trapped-Ion Mechanical Oscillator}}.
\newblock \emph{Physical Review Letters}, 118\penalty0 (26):\penalty0 263602,
  June 2017.
\newblock \doi{10.1103/PhysRevLett.118.263602}.

\setcounter{firstbib}{\value{NAT@ctr}}
\end{thebibliography}

\begin{thebibliography}{85}
\providecommand{\natexlab}[1]{#1}
\providecommand{\url}[1]{\texttt{#1}}
\expandafter\ifx\csname urlstyle\endcsname\relax
  \providecommand{\doi}[1]{doi: #1}\else
  \providecommand{\doi}{doi: \begingroup \urlstyle{rm}\Url}\fi

\setcounter{enumiv}{\value{firstbib}}
\setcounter{NAT@ctr}{\value{firstbib}}
\bibitem[Bullock(2026)]{Bullock_thesis_2026}
B.~B. Bullock.
\newblock \emph{Improved quantum control of two-dimensional ion crystals in a
  Penning trap}.
\newblock PhD thesis, Boulder, CO, 2026.

\bibitem[Gilmore(2020)]{Gilmore_thesis_2020}
K.~A. Gilmore.
\newblock \emph{Quantum sensing with large two-dimensional crystals of ions in
  a Penning trap}.
\newblock PhD thesis, Boulder, CO, 2020.

\bibitem[Huang et~al.(1998)Huang, Bollinger, Mitchell, Itano, and
  Dubin]{Huang_rotwall_1998}
X.-P. Huang, J.~J. Bollinger, T.~B. Mitchell, W.~M. Itano, and D.~H.~E. Dubin.
\newblock Precise control of the global rotation of strongly coupled ion
  plasmas in a penning trap.
\newblock \emph{Physics of Plasmas}, 5\penalty0 (5):\penalty0 1656--1663, May
  1998.
\newblock ISSN 1070-664X.
\newblock \doi{10.1063/1.872834}.
\newblock URL \url{https://doi.org/10.1063/1.872834}.

\bibitem[Britton et~al.(2016)Britton, Bohnet, Sawyer, Uys, Biercuk, and
  Bollinger]{britton_vibration-induced_2016}
J.~W. Britton, J.~G. Bohnet, B.~C. Sawyer, H.~Uys, M.~J. Biercuk, and J.~J.
  Bollinger.
\newblock Vibration-induced field fluctuations in a superconducting magnet.
\newblock \emph{Physical Review A}, 93\penalty0 (6):\penalty0 062511, June
  2016.
\newblock ISSN 2469-9926, 2469-9934.
\newblock \doi{10.1103/PhysRevA.93.062511}.

\bibitem[Biercuk et~al.(2009)Biercuk, Uys, VanDevender, Shiga, Itano, and
  Bollinger]{biercuk_quantum_contrl_Penning_2009}
Michael~J. Biercuk, Hermann Uys, Aaron~P. VanDevender, Nobuyasu Shiga, Wayne~M.
  Itano, and John~J. Bollinger.
\newblock High-fidelity quantum control using ion crystals in a {Penning} trap.
\newblock \emph{Quantum Info. and Comp.}, 9:\penalty0 920--949, 2009.
\newblock \doi{10.48550/arXiv.0906.0398}.

\bibitem[Shankar et~al.(2020)Shankar, Tang, Affolter, Gilmore, Dubin, Parker,
  Holland, and Bollinger]{shankar_broadening_2020}
Athreya Shankar, Chen Tang, Matthew Affolter, Kevin Gilmore, Daniel H.~E.
  Dubin, Scott Parker, Murray~J. Holland, and John~J. Bollinger.
\newblock Broadening of the drumhead-mode spectrum due to in-plane thermal
  fluctuations of two-dimensional trapped ion crystals in a {{Penning}} trap.
\newblock \emph{Physical Review A}, 102\penalty0 (5):\penalty0 053106, November
  2020.
\newblock \doi{10.1103/PhysRevA.102.053106}.

\bibitem[Shankar et~al.(2019)Shankar, Jordan, Gilmore, {Safavi-Naini},
  Bollinger, and Holland]{shankar_modeling_2019}
Athreya Shankar, Elena Jordan, Kevin~A. Gilmore, Arghavan {Safavi-Naini},
  John~J. Bollinger, and Murray~J. Holland.
\newblock Modeling near ground-state cooling of two-dimensional ion crystals in
  a {{Penning}} trap using electromagnetically induced transparency.
\newblock \emph{Physical Review A}, 99\penalty0 (2):\penalty0 023409, February
  2019.
\newblock \doi{10.1103/PhysRevA.99.023409}.

\bibitem[Torrisi et~al.(2016)Torrisi, Britton, Bohnet, and
  Bollinger]{torrisi_perpendicular_2016}
Steven~B. Torrisi, Joseph~W. Britton, Justin~G. Bohnet, and John~J. Bollinger.
\newblock Perpendicular laser cooling with a rotating-wall potential in a
  {{Penning}} trap.
\newblock \emph{Physical Review A}, 93\penalty0 (4):\penalty0 043421, April
  2016.
\newblock ISSN 2469-9926, 2469-9934.
\newblock \doi{10.1103/PhysRevA.93.043421}.

\bibitem[Britton et~al.(2012)Britton, Sawyer, Keith, Wang, Freericks, Uys,
  Biercuk, and Bollinger]{britton_engineered_2012}
Joseph~W. Britton, Brian~C. Sawyer, Adam~C. Keith, C.-C.~Joseph Wang, James~K.
  Freericks, Hermann Uys, Michael~J. Biercuk, and John~J. Bollinger.
\newblock Engineered two-dimensional {{Ising}} interactions in a trapped-ion
  quantum simulator with hundreds of spins.
\newblock \emph{Nature}, 484\penalty0 (7395):\penalty0 489--492, April 2012.
\newblock ISSN 1476-4687.
\newblock \doi{10.1038/nature10981}.

\bibitem[Wineland et~al.(1998)Wineland, Monroe, Itano, Leibfried, King, and
  Meekhof]{Wineland_NIST_1998}
D.~J. Wineland, C.~Monroe, W.~M. Itano, D.~Leibfried, B.~E. King, and D.~M.
  Meekhof.
\newblock Experimental issues in coherent quantum-state manipulation of trapped
  atomic ions.
\newblock \emph{J. Res. Natl. Inst. Stand. Technol.}, 103:\penalty0 259, 1998.

\bibitem[Sawyer et~al.(2012)Sawyer, Britton, Keith, Wang, Freericks, Uys,
  Biercuk, and Bollinger]{sawyer_spectroscopy_2012}
Brian~C. Sawyer, Joseph~W. Britton, Adam~C. Keith, C.-C.~Joseph Wang, James~K.
  Freericks, Hermann Uys, Michael~J. Biercuk, and John~J. Bollinger.
\newblock Spectroscopy and {{Thermometry}} of {{Drumhead Modes}} in a
  {{Mesoscopic Trapped-Ion Crystal Using Entanglement}}.
\newblock \emph{Physical Review Letters}, 108\penalty0 (21):\penalty0 213003,
  May 2012.
\newblock ISSN 0031-9007, 1079-7114.
\newblock \doi{10.1103/PhysRevLett.108.213003}.

\bibitem[Uys et~al.(2010)Uys, Biercuk, VanDevender, Ospelkaus, Meiser, Ozeri,
  and Bollinger]{uys_decoherence_2010}
H.~Uys, M.~J. Biercuk, A.~P. VanDevender, C.~Ospelkaus, D.~Meiser, R.~Ozeri,
  and J.~J. Bollinger.
\newblock Decoherence due to {{Elastic Rayleigh Scattering}}.
\newblock \emph{Physical Review Letters}, 105\penalty0 (20):\penalty0 200401,
  November 2010.
\newblock ISSN 0031-9007, 1079-7114.
\newblock \doi{10.1103/PhysRevLett.105.200401}.

\bibitem[Carter et~al.(2023)Carter, Muleady, Shankar, Lilieholm, Bullock,
  Affolter, Rey, and Bollinger]{carter_comparison_2023}
Allison~L. Carter, Sean~R. Muleady, Athreya Shankar, Jennifer~F. Lilieholm,
  Bryce~B. Bullock, Matthew Affolter, Ana~Maria Rey, and John~J. Bollinger.
\newblock Comparison of spontaneous emission in trapped-ion multiqubit gates at
  high magnetic fields.
\newblock \emph{Physical Review A}, 107\penalty0 (4):\penalty0 042618, April
  2023.
\newblock ISSN 2469-9926, 2469-9934.
\newblock \doi{10.1103/PhysRevA.107.042618}.

\bibitem[Ho et~al.(2025)Ho, Lu, Xiang, Rusconi, Masson, {Asenjo-Garcia}, Yan,
  and {Stamper-Kurn}]{ho_optomechanical_2025}
Jacquelyn Ho, Yue-Hui Lu, Tai Xiang, Cosimo~C. Rusconi, Stuart~J. Masson, Ana
  {Asenjo-Garcia}, Zhenjie Yan, and Dan~M. {Stamper-Kurn}.
\newblock Optomechanical self-organization in a mesoscopic atom array.
\newblock \emph{Nature Physics}, 21\penalty0 (7):\penalty0 1071--1077, July
  2025.
\newblock ISSN 1745-2473, 1745-2481.
\newblock \doi{10.1038/s41567-025-02916-7}.

\bibitem[Aedo and Lamata(2018)]{aedo_analog_2018}
Ibai Aedo and Lucas Lamata.
\newblock Analog quantum simulation of generalized {{Dicke}} models in trapped
  ions.
\newblock \emph{Physical Review A}, 97\penalty0 (4):\penalty0 042317, April
  2018.
\newblock ISSN 2469-9926, 2469-9934.
\newblock \doi{10.1103/PhysRevA.97.042317}.

\bibitem[Sutherland(2019)]{sutherland_analog_2019}
R.~T. Sutherland.
\newblock Analog quantum simulation of superradiance and subradiance in trapped
  ions.
\newblock \emph{Physical Review A}, 100\penalty0 (6):\penalty0 061405, December
  2019.
\newblock \doi{10.1103/PhysRevA.100.061405}.

\bibitem[Affolter et~al.(2020)Affolter, Gilmore, Jordan, and
  Bollinger]{affolter_phase-coherent_2020}
M.~Affolter, K.~A. Gilmore, J.~E. Jordan, and J.~J. Bollinger.
\newblock Phase-coherent sensing of the center-of-mass motion of trapped-ion
  crystals.
\newblock \emph{Physical Review A}, 102\penalty0 (5):\penalty0 052609, November
  2020.
\newblock \doi{10.1103/PhysRevA.102.052609}.

\bibitem[Eckstein et~al.(2009)Eckstein, Kollar, and
  Werner]{eckstein_thermalization_2009}
Martin Eckstein, Marcus Kollar, and Philipp Werner.
\newblock Thermalization after an {{Interaction Quench}} in the {{Hubbard
  Model}}.
\newblock \emph{Physical Review Letters}, 103\penalty0 (5):\penalty0 056403,
  July 2009.
\newblock \doi{10.1103/PhysRevLett.103.056403}.

\bibitem[Reid et~al.(2009)Reid, Drummond, Bowen, Cavalcanti, Lam, Bachor,
  Andersen, and Leuchs]{reid_colloquium_2009}
M.~D. Reid, P.~D. Drummond, W.~P. Bowen, E.~G. Cavalcanti, P.~K. Lam, H.~A.
  Bachor, U.~L. Andersen, and G.~Leuchs.
\newblock {\emph{Colloquium}} : {{The Einstein-Podolsky-Rosen}} paradox:
  {{From}} concepts to applications.
\newblock \emph{Reviews of Modern Physics}, 81\penalty0 (4):\penalty0
  1727--1751, December 2009.
\newblock ISSN 0034-6861, 1539-0756.
\newblock \doi{10.1103/RevModPhys.81.1727}.

\bibitem[Uola et~al.(2020)Uola, Costa, Nguyen, and
  G{\"u}hne]{uola_quantum_2020}
Roope Uola, Ana C.~S. Costa, H.~Chau Nguyen, and Otfried G{\"u}hne.
\newblock Quantum steering.
\newblock \emph{Reviews of Modern Physics}, 92\penalty0 (1):\penalty0 015001,
  March 2020.
\newblock ISSN 0034-6861, 1539-0756.
\newblock \doi{10.1103/RevModPhys.92.015001}.

\bibitem[Takahashi and Umezawa(1975)]{takahashi_higher_1975}
Ya~Takahashi and H~Umezawa.
\newblock Higher order calculation in thermo field theory.
\newblock \emph{Collective phenomena}, 2:\penalty0 55, 1975.

\bibitem[Israel(1976)]{israel_thermo-field_1976}
W.~Israel.
\newblock Thermo-field dynamics of black holes.
\newblock \emph{Physics Letters A}, 57\penalty0 (2):\penalty0 107--110, May
  1976.
\newblock ISSN 03759601.
\newblock \doi{10.1016/0375-9601(76)90178-X}.

\bibitem[Maldacena(2003)]{maldacena_eternal_2003}
Juan Maldacena.
\newblock Eternal black holes in anti-de {{Sitter}}.
\newblock \emph{Journal of High Energy Physics}, 2003\penalty0 (04):\penalty0
  021--021, April 2003.
\newblock ISSN 1029-8479.
\newblock \doi{10.1088/1126-6708/2003/04/021}.

\bibitem[Maldacena et~al.(2017)Maldacena, Stanford, and
  Yang]{maldacena_diving_2017}
Juan Maldacena, Douglas Stanford, and Zhenbin Yang.
\newblock Diving into traversable wormholes.
\newblock \emph{Fortschritte der Physik}, 65\penalty0 (5):\penalty0 1700034,
  May 2017.
\newblock ISSN 0015-8208, 1521-3978.
\newblock \doi{10.1002/prop.201700034}.

\bibitem[Hayden and Preskill(2007)]{hayden_black_2007}
Patrick Hayden and John Preskill.
\newblock Black holes as mirrors: Quantum information in random subsystems.
\newblock \emph{Journal of High Energy Physics}, 2007\penalty0 (09):\penalty0
  120, September 2007.
\newblock ISSN 1126-6708.
\newblock \doi{10.1088/1126-6708/2007/09/120}.

\setcounter{firstbib}{\value{NAT@ctr}}
\end{thebibliography}

\begin{thebibliography}{99}
\providecommand{\natexlab}[1]{#1}
\providecommand{\url}[1]{\texttt{#1}}
\expandafter\ifx\csname urlstyle\endcsname\relax
  \providecommand{\doi}[1]{doi: #1}\else
  \providecommand{\doi}{doi: \begingroup \urlstyle{rm}\Url}\fi

\setcounter{enumiv}{\value{firstbib}}
\setcounter{NAT@ctr}{\value{firstbib}}

\bibitem[Polkovnikov(2010)]{polkovnikov_phase_2010}
Anatoli Polkovnikov.
\newblock Phase space representation of quantum dynamics.
\newblock \emph{Annals of Physics}, 325\penalty0 (8):\penalty0 1790--1852,
  August 2010.
\newblock ISSN 0003-4916.
\newblock \doi{10.1016/j.aop.2010.02.006}.

\bibitem[Huber et~al.(2022)Huber, Rey, and Rabl]{huber_realistic_2022}
Julian Huber, Ana~Maria Rey, and Peter Rabl.
\newblock Realistic simulations of spin squeezing and cooperative coupling
  effects in large ensembles of interacting two-level systems.
\newblock \emph{Physical Review A}, 105\penalty0 (1):\penalty0 013716, January
  2022.
\newblock \doi{10.1103/PhysRevA.105.013716}.

\end{thebibliography}
\end{document}